\documentclass[natbib]{svjour3}                     
\smartqed  %
\usepackage{graphicx}
\usepackage{latexsym}			%
\usepackage{amsmath} 			%
\usepackage{amssymb} 
\newfont{\sfsl}{cmssqi8 scaled 1200}
\newfont{\sfslp}{cmssqi8 scaled 1250}
\newfont{\sfsls}{cmssqi8 scaled 900}
\newfont{\sfsln}{cmssqi8 scaled 1350}
\newfont{\sfslz}{cmssqi8 scaled 1500}
\newfont{\sfslzz}{cmssqi8 scaled 1150}
\newfont{\sfslms}{cmssqi8 scaled 1000}	
\newfont{\sfsla}{cmssqi8 scaled 3000}
\newfont{\sfslb}{cmssqi8 scaled 2200}

\newcommand{\dif}{{\rm d}}

\newcommand{\ro}{{\it ROSAT}}
\newcommand{\ps}{{\rm PSPC}}

\newcommand{\as}{{\it ASCA}}
\newcommand{\sax}{{\it BeppoSAX}}
\newcommand{\xmm}{{\it XMM-Newton}}

\newcommand{\cha}{{\it Chandra}}
\newcommand{\suz}{{\it Suzaku}}
\newcommand{\swi}{{\it Swift}}

\newcommand{\athena}{{\it Athena}}

\newcommand{\rosi}{{\it eROSITA}}

\newcommand{\pla}{{\it Planck}}

\newcommand{\om}{\Omega_{\rm m}}

\newcommand{\ol}{\Omega_{\Lambda}}

\newcommand{\roc}{\rho_{\rm c}}

\newcommand{\sx}{S_{\rm X}}

\newcommand{\mpr}{m_{\rm p}}
\newcommand{\mel}{m_{\rm e}}

\newcommand{\rog}{\rho_{\rm gas}}

\newcommand{\rot}{\rho_{\rm tot}}
\newcommand{\fg}{f_{\rm gas}}

\newcommand{\rv}{r_{\rm vir}}

\newcommand{\nh}{n_{\rm H}}

\newcommand{\nhy}{N_{\rm H}}

\newcommand{\nel}{n_{\rm e}}
\newcommand{\npr}{n_{\rm p}}

\newcommand{\pel}{P_{\rm e}}
\newcommand{\tg}{T_{\rm gas}}
\newcommand{\tx}{T_{\rm X}}

\newcommand{\tel}{T_{\rm e}}

\newcommand{\mt}{M_{\rm tot}}

\newcommand{\mg}{M_{\rm gas}}

\newcommand{\fgz}{f_{{\rm gas}, 200}}

\newcommand{\fgsm}{f_{{\rm gas}, 2500}}

\newcommand{\ekin}{E_{\rm kin}}
\newcommand{\epot}{E_{\rm pot}}

\newcommand{\kb}{k_{\rm B}}

\def\M500{M_{500}}
\def\R500{R_{500}}
\def\L500{L_{500}}

\def\T500{\theta_{\rm 500}}
\def\Y500{\theta_{\rm 500}}

\def\fg  {f_{\rm g}}

\def \Rv {r_{500}}

\def\Planck{{\it Planck}}

\newfont{\gwpfont}{cmssq8 scaled 1000}
\newcommand{\rexcess}{{\gwpfont REXCESS}}

\begin{document}

\title{Outskirts of Galaxy Clusters%
}

\author{Thomas H. Reiprich \and
        Kaustuv Basu \and
	Stefano Ettori \and
        Holger Israel \and
	Lorenzo Lovisari \and
	Silvano Molendi \and
	Etienne Pointecouteau \and
	Mauro Roncarelli
}

\institute{T.H. Reiprich \and K. Basu \and H. Israel \and
L. Lovisari \at 
Argelander Institute for Astronomy,
Bonn University,
Auf dem H\"ugel 71,
53121 Bonn,
Germany\\
\email{reiprich@astro.uni-bonn.de, kbasu@astro.uni-bonn.de,
hisrael@astro.uni-bonn.de, lorenzo@astro.uni-bonn.de}           %
	      \and
S. Ettori \at
INAF-Osservatorio Astronomico, via Ranzani 1, 40127 Bologna, Italy \\
INFN, Sezione di Bologna, viale Berti Pichat 6/2, 40127 Bologna,
Italy\\
\email{stefano.ettori@oabo.inaf.it}
           \and
S. Molendi \at 
INAF-IASF, via Bassini 15, 20133, Milan, Italy\\
\email{silvano@iasf-milano.inaf.it}
           \and
E. Pointecouteau \at
Universit\'e de Toulouse, UPS-Observatoire Midi-Pyr\'en\'ees, IRAP,
31400, Toulouse, France; CNRS, Institut de Recherche en Astrophysique
et Plan\'etologie, 9 Avenue du Colonel Roche, BP 44346, 31028, Toulouse
Cedex 4, France\\
\email{etienne.pointecouteau@irap.omp.eu}
           \and
M. Roncarelli \at
Dipartimento di Astronomia, Universit\`a di Bologna, via Ranzani
  1, I-40127 Bologna, Italy\\
\email{mauro.roncarelli@unibo.it}
}

\date{Received: 2012-12-10 / Accepted: 2013-01-24}

\maketitle

\begin{abstract}
Until recently, only about 10\% of the total intracluster gas volume
had been studied with high accuracy, leaving a vast region essentially
unexplored. This is now changing and a wide area of hot gas physics
and chemistry awaits discovery in galaxy cluster outskirts. Also,
robust large-scale total mass profiles and maps are within reach.
First observational and theoretical
results in this emerging field have been achieved in recent years with
sometimes surprising
findings. Here, we summarize
and illustrate the relevant underlying physical and 
chemical processes and review the recent progress in X-ray,
Sunyaev--Zel'dovich, and weak gravitational lensing observations of
cluster outskirts, including also brief discussions of technical
challenges and possible future improvements.
\end{abstract}

\section{Introduction}
\label{out:intro}
\begin{figure}[htb]
\includegraphics[angle=0,width=0.5\textwidth]{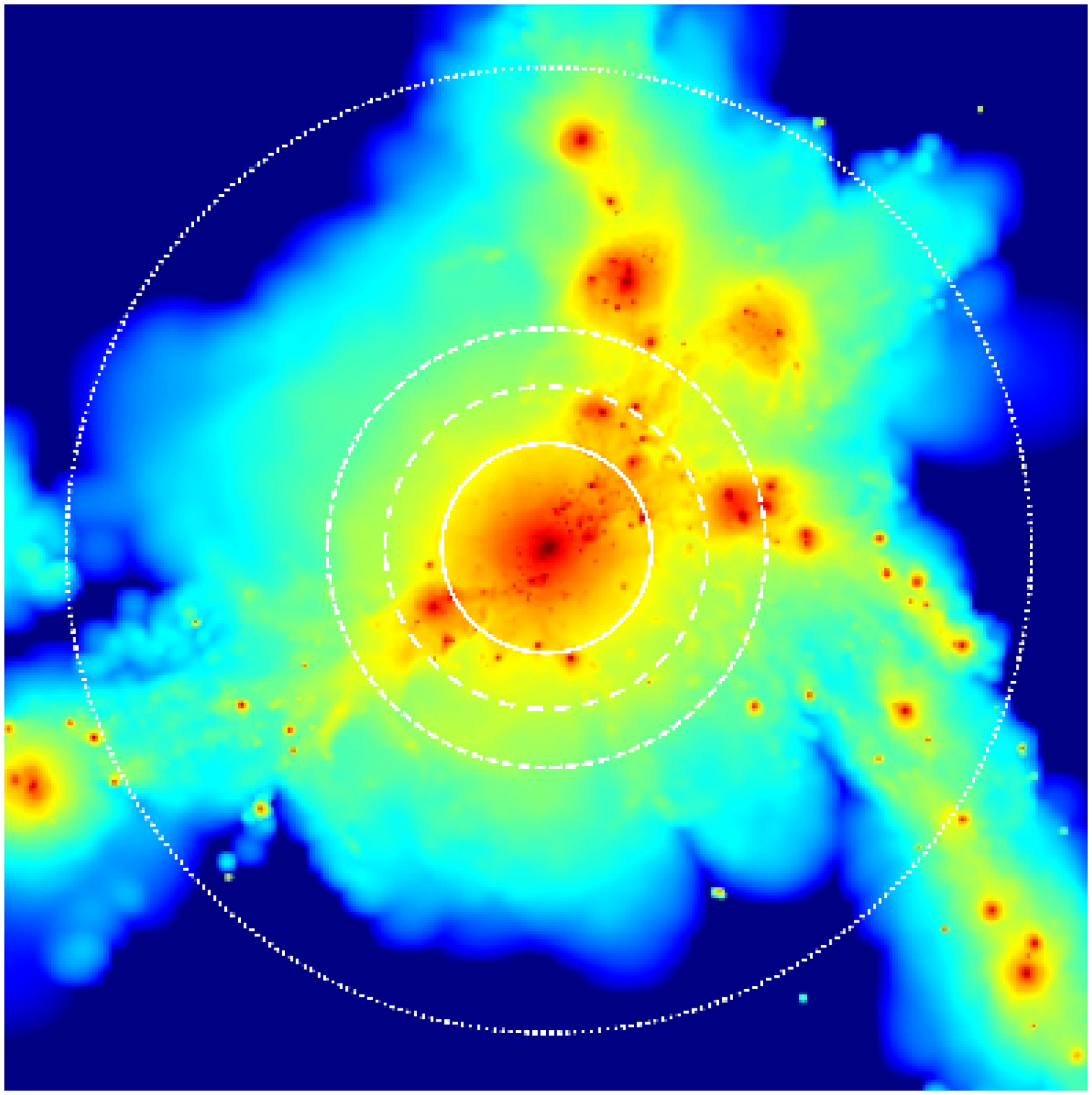}
\includegraphics[angle=0,width=0.5\textwidth]{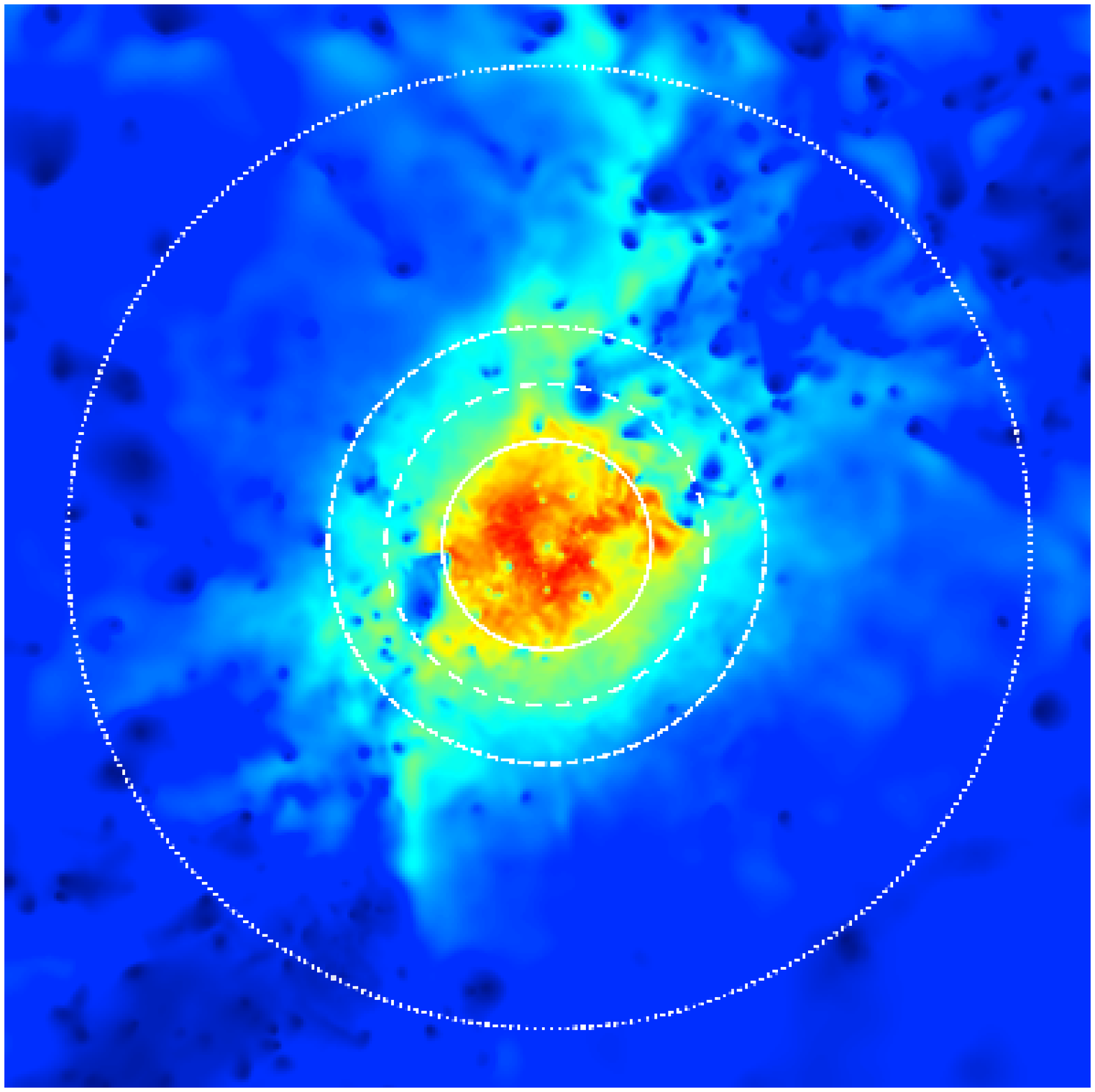}
\caption{Simulated galaxy cluster.
The white circles indicate $r_{500},\ r_{200},\ \rv,\ \mathrm{and}\ 3
\, r_{200}$ 
moving outwards, respectively (adapted from \citealt{red06}).
Left: X-ray surface brightness in the soft (0.5--2) keV band.
The color scale spans 16 orders of magnitude and has been chosen to
highlight cluster outskirts. 
Right: Temperature map on a linear scale from 0 keV (blue) to 11 keV
(red).
}
\label{out:physics:structure:simu}
\end{figure}

A plethora of physical effects is believed to be acting in the
outskirts of galaxy clusters, which ebbed away long ago in more
central regions.
This
includes, e.g., breakdown of equilibrium states like hydrostatic
equilibrium \citep[e.g.,][]{nvk07}, thermal equilibrium and
equipartition \citep[e.g.,][]{fl97}, and ionization equilibrium
\citep[e.g.,][]{wsj11}. It is also in the
outskirts, where structure formation effects should be widespread,
resulting, e.g., in multitemperature structure and a clumpy gas
distribution (Fig.~\ref{out:physics:structure:simu}).
Moreover, the primary processes of intracluster medium (ICM) enrichment with heavy elements
\citep[e.g.,][]{sd08} may be identified by determining the metal
abundance up to the cluster outskirts.
Last not least, future measurements of the evolution of the cluster
mass function with $\sim$100,000 galaxy clusters detected with the
extended ROentgen Survey with an Imaging Telescope Array 
\citep[\rosi, e.g.,][]{pab10,ppr12,mpb12} 
will heavily rely on a detailed understanding of the cluster mass distribution. Therefore,
tracing this distribution out to large radii will be important for
using clusters as accurate cosmological tools.

If cluster outskirts are so interesting, why haven't they been
studied extensively with observations and simulations already long ago?
In fact, we have only really seen the tip of the iceberg of the
ICM up to now; i.e., the relatively dense central
regions of galaxy clusters, the inner $\sim$10\% in terms of volume.
The reason is, of course, that robust observations and realistic
simulations are challenging in cluster outskirts.

Why is that? The difficulties differ depending on the waveband used for
cluster outskirt observations. For instance, the X-ray surface
brightness drops below various fore- and background components at large
radii, Sunyaev--Zel'dovich (SZ) effect measurements are also less
sensitive in cluster outskirts where the gas pressure is low, and the weak
gravitational lensing signal interpretation is increasingly plagued by projection effects. Naively,
simulations should be done most easily in outskirts because there the
least resolution might be required. While this may be true for dark
matter only simulations, it is not so simple if gas physics is
included, e.g., the cooling of infalling subclumps.

If it is so difficult, why has interest been rising in recent years?
This is certainly mostly due to technical advances in observational and
theoretical techniques but possibly also to partially unexpected and
sometimes controversial inital results.

As is true for all articles in this review volume, we will put the
emphasis on the ICM and total mass properties. The member galaxy and
relativistic particle properties of cluster outskirts have 
been reviewed, e.g., in the Proceedings to the IAU colloquium
``Outskirts of Galaxy Clusters: Intense Life in the Suburbs''
\citep{d04} and \citet{bbr11}, respectively.
Other useful reviews mostly about the ICM properties of 
clusters include, e.g., \citet{s86,bg01,rbn02,vo05,a05,n05,b06,bk09,aem11}.

This article is organized in 7 Sections. Section~\ref{out:out} contains our
definition of cluster outskirts, Section~\ref{out:mass} provides some basics
on cluster mass determination, Section~\ref{out:physics} includes a
summary of the status of ICM profiles as well as descriptions and
illustrations of physical effects relevant for cluster outskirts,
Section~\ref{out:chem} outlines chemistry aspects,
Section~\ref{out:tech} summarizes technical
considerations for X-ray, SZ, and weak lensing measurements, and
Section~\ref{out:outlook} gives a brief outlook.

\section{Where are the ``cluster outskirts''?}
\label{out:out}
Let us define, which radial range we consider as ``cluster
outskirts.'' Readers not interested in more details on the radial ranges
can skip this section and just take note of our subjective choice:
\begin{equation}
 r_{500}< {\rm cluster\ outskirts} <3r_{200}\,,
\end{equation}
where $r_{500}$ (defined below) used to be the observational limit for
X-ray temperature measurements and the range up to
$3r_{200}$
captures most
of the interesting physics and chemistry before clearly entering the
regime of the warm-hot intergalactic medium (WHIM,
Fig.~\ref{out:physics:structure:simu}). 
This range also includes (i) the turn
around radius, $r_{\mathrm{turn}}=2\rv$, from the spherical collapse
model \citep[e.g.,][]{ll00}, (ii) part of the infall region where caustics in galaxy
redshift space are observed, several Mpc \citep[e.g.,][]{d99}, (iii)
much of the radial range where accretion shocks might be expected,
(1--3)$\rv$ \citep[e.g.,][]{mhh09}, and (iv) the region where the
two-halo term starts dominating over the one-halo term in the matter
power spectrum, few Mpc \citep[e.g.,][]{cos02}. 

A theoretical recipe that can be used to define a cluster ``border,''
``boundary,'' or at least a ``characteristic'' radius is the spherical
collapse model \citep[e.g.,][]{at10}.
Based on this very idealistic model, a \emph{virial radius}, $\rv$,
separating the virialized cluster region from the outer ``infall''
region, can be obtained by requiring the mean total mass density of a
cluster, $\langle\rot\rangle$, to fulfill 
\begin{equation}
 \langle\rot\rangle(<\rv)\equiv\frac{3\mt(<\rv)}{4\pi\rv^3}=\Delta_{\rm
 c}^{\rm vir}(z)\roc(z)\,,
\end{equation}
where $\roc(z)$ is the critical density of the Universe at redshift
$z$.\footnote{Some authors use the mean matter density of the Universe,
$\bar{\rho}_{\rm m}(z)=\om(z)\roc(z)$, instead of the
critical density for their overdensity definition.}
The virial overdensity, $\Delta_{\rm c}^{\rm vir}(z)$, is a function
of cosmology and redshift, in general \citep[e.g.,][]{ks96,wk03}. E.g., for
a flat Universe with mean normalized matter density $\om=1$, $\Delta_{\rm c}^{\rm
vir}=(6\pi)^2/2\approx 178$ (at any $z$). Since $178\approx200$, 
$r_{\Delta_{\rm c}=200}\equiv r_{200}$ is sometimes used as a rather
crude approximation to the virial radius ($r_{200}\approx\rv$). On the
other hand, using a cosmology other than Einstein--de Sitter,
$\Delta_{\rm c}^{\rm vir}$ varies. E.g., for $\om=0.3$ and $\ol=0.7$,
one finds $\Delta_{\rm c}^{\rm vir}(z=0)\approx 101$, $\Delta_{\rm
c}^{\rm vir}(z=0.5)\approx 138$, and $\Delta_{\rm c}^{\rm
vir}(z=1)\approx 157$ \citep[using eq.~6 of][]{bn98}.

Another possibility to define a virial radius (at least in
simulations) is to use the region within which the condition of virial
equilibrium ($2\ekin=-\epot$) is satisfied.

The virial mass is also the one that should be measured when
comparing observational cluster mass functions to the
Press--Schechter (\citeyear{ps74}) mass function.
However, when it became clear \citep[e.g.,][]{gbq99} that
semi-analytic recipes for the mass function, like Press--Schechter,
are not accurate enough for modern measurements, authors shifted to using parametrized fits to
mass functions as obtained from numerical $N$-body simulations
\citep[e.g.,][]{jfw01,tkk08} to compare observations to predictions
(at the expense of loosing any analytic understanding of the mass
function, of course).
This removed the need to use virial masses for the comparison and it
has become common practice to simply use a fixed value for
$\Delta_{\rm c}$ in both observations and simulations\footnote{Note
that this definition also leads to a funny effect when comparing 
cluster mass functions at different redshifts: assume an unrealistic
Universe without evolution of clusters and their number density. For
instance, at $z=0$ and at $z=1$ we would then have the same number
of clusters (with identical distributions of physical mass profiles)
per comoving volume. Now, the measured $M_{200}$ of two clusters with 
identical mass profiles at both redshifts differ, the $M_{200}$ of
the higher redshift cluster being smaller because
$\roc(z=1)=E^2(z=1)\roc(z=0)\approx 8\roc(z=0)$ for a ``concordance cosmology'' and,
therefore, the mass profile gets integrated only to a much smaller physical
radius for the higher redshift cluster. Plotting the mass functions at
$z=0$ and $z=1$ would then result in a lower number density at higher
redshift. This effect is further enhanced when the mean density is
used for the overdensity definition instead of the critical density or
if the virial overdensity is used.  So, clearly, the definition of
outer radius has a strong effect on the perceived evolution of the
mass function.
Since the choice of $\Delta(z)\rho(z)$ is arbitrary (it just has to
be consistent between observations and predictions) and 
if one wanted to appreciate the pure number density evolution of the
mass function from a plot more directly one could, e.g., use masses
defined with a fixed overdensity with respect to the critical density
at $z=0$ for all clusters; i.e., make both $\Delta$ and $\rho$
redshift independent.
Note that this would still ensure that, at a given redshift, low mass
clusters would be treated in a way that allows comparison to high
mass clusters, which would be less obvious if a metric radius
(e.g., the Abell radius) was used.}; i.e.,
\begin{equation}
 \frac{3\mt(<r_{\Delta_{\rm c}})}{4\pi r_{\Delta_{\rm
 c}}^3}=\Delta_{\rm c}\roc(z)\,.
\label{out:eq:out:o_radius}
\end{equation}

Typical overdensities used in the literature include $\Delta_{\rm
c}$ = 100, 180, 200, 500, 666, 1000, and 2500.
Assuming an NFW \citep{nfw97} profile with concentration,
$c\equiv r_{200}/r_{\rm s}=4$, one finds
$\rv(z=0)\approx r_{100}\approx 1.36 r_{200}$, $r_{180}\approx 1.05
r_{200}$, $r_{500}\approx 0.65 r_{200}$, $r_{666}\approx 0.57
r_{200}$, $r_{1000}\approx 0.46 r_{200}$, and $r_{2500}\approx 0.28
r_{200}$.
While it is obvious that using a fixed value for $\Delta_{\rm c}$
makes things simpler, this choice also requires picking a ``magic'' number: which
overdensity to pick?, which is the best radius, compromising
between simulations and obervations?
It appears that, currently, a good choice would be in the range
$500\leq\Delta_{\rm c}\leq1000$, where the lower limit comes from
observations and the upper limit from simulations.
An interesting number is then also the ratio of volumes within
overdensities 500 and 100:
\begin{equation}
 \left[\frac{r_{500}}{\rv(z=0)}\right]^3\approx 0.1\,,
\end{equation}
which implies that measurements
limited to $r_{500}$ explore only about 10\% of the total cluster
volume!

Due to their high particle backgrounds, \cha\ and \xmm\ are basically
limited to $\lesssim$$r_{500}$ for robust gas temperature
measurements. As we will see in Section~\ref{out:physics:Tr}, \suz\
now routinely reaches $\sim$$r_{200}$. \swi\ may soon follow for a few
bright clusters. ROentgenSATellit (\ro) and SZ (stacking) observations constrain well
gas density and pressure out to $\sim$$r_{200}$, respectively
(Sections~\ref{sec:sb} and \ref{out:physics:P}).
Currently, we cannot observationally reach the outer
border of our definition of cluster outskirts, $3r_{200}$, leaving
ample discovery space for the future.

\section{Mass}
\label{out:mass}
In the contributions by Ettori et al.\ and Hoekstra et al.\ of this
volume, detailed reviews 
on cluster mass reconstruction are provided. Here, we summarize some
basics that are important for our discussion of cluster outskirts.
\subsection{Total mass inferred from ICM properties}
\label{out:mass:X}
The total mass of galaxy clusters can be determined by measuring
ICM properties, like density, temperature, and pressure.
We will see in
Section~\ref{out:physics} that a clear understanding of the gas
physics is required for accurate mass determinations. In
Section~\ref{out:tech}, examples are given that illustrate technical
challenges that need to be overcome to understand gas physics in
cluster outskirts.

Under the assumption\footnote{Other, mostly minor,
assumptions that we will not discuss include:
gravitation is the only external field (e.g., no magnetic field),
clusters are spherically symmetric (e.g., do not rotate),
no (pressure supplied by) relativistic particles,
$\mu$ is independent of $r$ (e.g., negligible
helium sedimentation, constant metallicity),
Newtonian description of gravity is adequate (e.g., no relativistic
corrections), the
effect of a cosmological constant (dark energy) is negligible.}
that the ICM is in hydrostatic equilibrium with the gravitational
potential, the integrated total mass profile, $\mt(<r)$, is given by
\begin{equation}
 \frac{1}{\rog}\frac{\dif P}{\dif r}=-\frac{G\mt(<r)}{r^2}\,,
\label{out:eq:mass:X:hydro1}
\end{equation}
where $P$ is the gas pressure, $\rog$ its density, and $G$ the
gravitational constant. Applying the ideal gas equation,
$P=\frac{\kb}{\mu\mpr} \rog\tg$, results in 
\begin{equation}
 \mt(<r)=-\frac{\kb\tg r}{G\mu\mpr}\left(\frac{\dif \ln\rog}{\dif \ln
 r}+\frac{\dif \ln\tg}{\dif \ln r}\right)\,,
\label{out:eq:mass:X:hydro2}
\end{equation}
where $\mu\approx 0.6$ (Section~\ref{out:physics:helium}) is the mean
particle weight in units of the 
proton mass, $\mpr$, and $\kb$ is Boltzmann's constant. So, the total
mass within a given radius depends on the gas temperature at this
radius, as well as the temperature gradient, and the gas density
gradient. Note there is no dependence on the absolute value of the gas
density, only on its gradient.

\subsubsection{X-ray measurements}
\label{out:mass:X:X}
\begin{figure}[htb]
\center
\includegraphics[angle=270,width=0.9\textwidth]{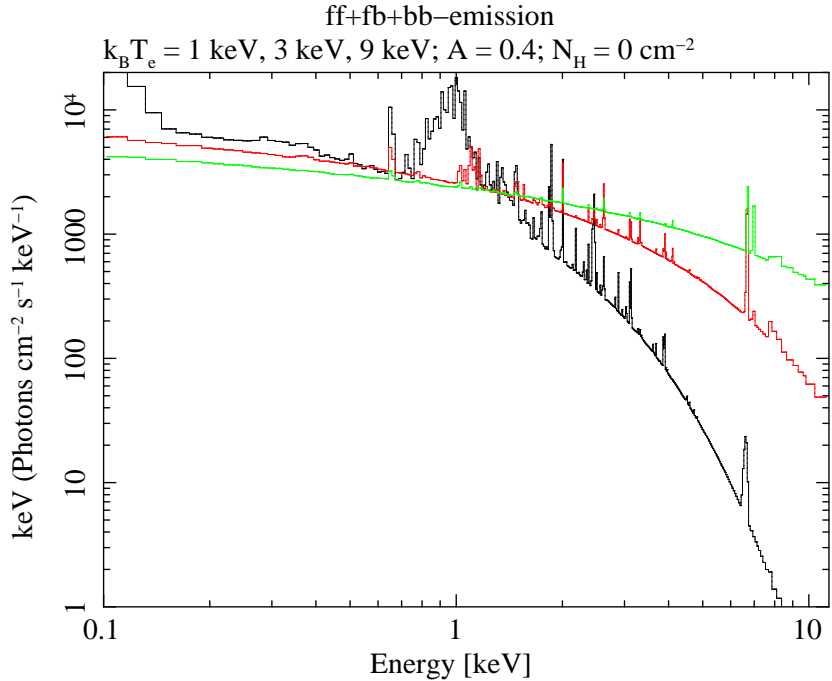}
\caption{Hot gas model X-ray spectra including free-free, free-bound,
and bound-bound emission as a function of $\kb\tel$ (black: 1 keV,
red: 3 keV, green: 9 keV). The emission measure was
kept fixed for all spectra; i.e., all spectra assume the same electron
density distribution. Notice how the exponential bremsstrahlung cutoff
shifts to higher energies for higher temperatures, as expected from
(\ref{out:eq:ff-emiss}). Note also the Fe L
and K shell emission line complexes at $\sim$1 and $\sim$6 keV,
respectively. The line strength depends on metallicity (here assumed
to be 0.4 solar) \emph{and} temperature. For hot clusters
($\kb\tel > 2$ keV), the
emission in a soft energy band, e.g.\ (0.5--2.0) keV, is almost
independent of $\tel$.}
\label{out:mass:apec}
\end{figure}

The hot ICM is collisionally highly ionized and mostly optically thin.
Using X-rays, the gas density and temperature profiles can be
determined. At temperatures $\kb\tel\gtrsim$ 2 keV\footnote{$\tel$
is often used synonymously to $\kb\tel \Rightarrow 1$ keV $\approx
1.16\times 10^{7}$K; X-ray photon energies are also typically 
expressed in keV;
1 keV $\approx$ 2.42 $\times 10^{17}$ Hz $\approx$ 12.4 \AA.}
and typical ICM
metallicities (0.1--1 solar) thermal bremsstrahlung (free-free)
emission is the dominant emission process. The emissivity; i.e., the
energy emitted per time and volume, at frequency $\nu$ is given in
this case by 
\begin{equation}
 \epsilon_\nu^{\rm ff}\propto \nel^2\tel^{-\frac{1}{2}}
 e^{-\frac{h\nu}{\kb\tel}}\,,
\label{out:eq:ff-emiss}
\end{equation}
where $\nel$ ($\propto\rog$) and $\tel$ denote electron number density and
temperature, respectively. So, fitting a model to a measured X-ray
spectrum yields density and temperature of the hot electrons.
At lower temperatures ($\kb\tel\lesssim$ 2 keV), line emission becomes
important or even dominant and serves as an additional temperature
discriminator (Fig.~\ref{out:mass:apec}). Note also that the
abundances of heavy elements and the cluster redshift can be
constrained by modelling the line emission.

So, to determine the total mass out to large cluster radii, one needs
to measure gas density and temperature profiles in low surface
brightness outer regions. There, not only are the measurements
themselves quite challenging but also several physical effects may
become important that can usually be ignored in inner parts. Both
issues will be discussed in some detail in this review
(Sections~\ref{out:tech} and \ref{out:physics}, respectively).

\begin{figure}[htb]
\hspace{1cm}
\includegraphics[angle=0,width=0.8\textwidth]{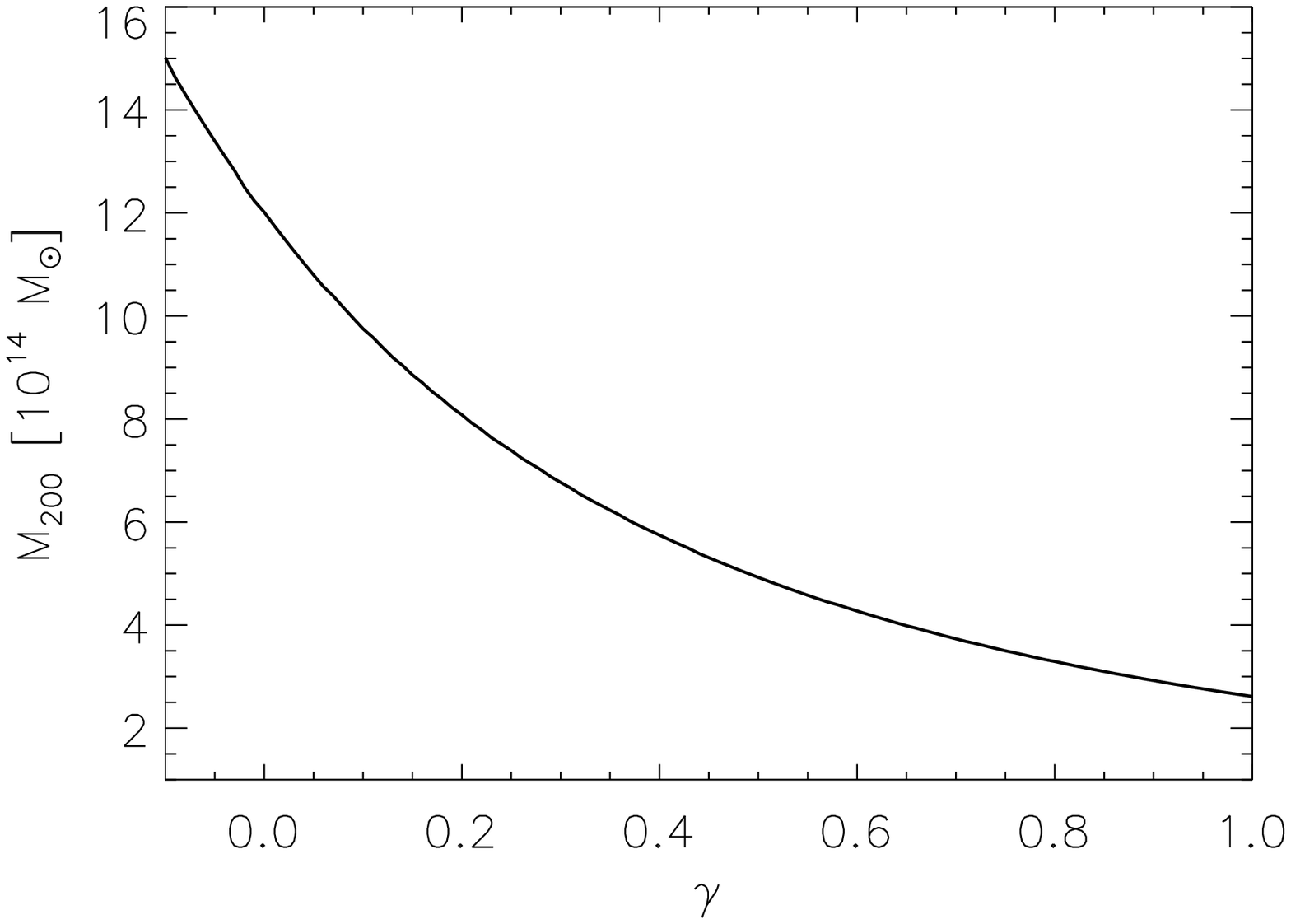}
\vspace{0.5cm}
\caption{Change of total mass depending on the steepness of the
temperature profile, $\tg\propto r^{-\gamma}$, for some fiducial ICM
parameter values (see text). The steeper the profile, the lower the
inferred mass.}
\label{out:mass:M-Tr}
\end{figure}

As we will see, gas temperatures typically decline with radius in the
outer regions of clusters. Equation (\ref{out:eq:mass:X:hydro2}) shows
that both the absolute value as well as the gradient of the
temperature at a given radius contribute, and they work in opposite
directions: for a declining temperature,
the former term decreases the total mass while the latter increases
it. It will be of interest in the course of this article, which term
usually dominates. To illustrate this, we show in
Fig.~\ref{out:mass:M-Tr} how $M_{200}$ changes depending on the slope
of the temperature profile, for a simple model cluster with a density
profile following a single beta model with $\beta=2/3$ and a core
radius of 150 kpc, and a temperature profile
$T(r)=T_0(r/r_{\mathrm{cut}})^{-\gamma}$ with $T_0=6$ keV and
$r_{\mathrm{cut}}=400$ kpc. One notes immediately two things: first,
the absolute value of the temperature at a given radius is much more 
important than its gradient because the steeper the temperature
profile the lower the total mass, and, second, the temperature
profile can have a significant impact on the total mass determination
even if its gradient is much smaller than the gas density gradient.
The density gradient effect is clear from (\ref{out:mass:M-Tr}): the
steeper the profile, the larger the total mass.
Recall in this context that convection will set in if the
gradient of specific entropy
becomes negative, so hydrostatic equilibrium is likely not a good
assumption if  $-\dif\ln\tg/\dif r\gtrsim -2/3\, \dif\ln\rog/\dif r$
(Section~\ref{out:physics:hydro}).

\subsubsection{SZ measurements}

The hot ICM electrons emitting in X-rays also change the intensity
of the cosmic microwave background (CMB) radiation via inverse Compton
scattering. The characteristic features of this spectral distortion of
the CMB were predicted by R. Sunyaev and Y. Zel'dovich shortly after
the discovery of X-rays from clusters, and is named after them
\citep{sz72}. The distinguishing feature of this effect is a decrement
of CMB intensity below $\sim$220 GHz, where clusters appear as a dark
spot or a ``hole" in the microwave sky, and an increment above 220
GHz. This effect is more precisely called the {\it thermal}
Sunyaev-Zel'dovich (tSZ) effect, to distinguish it from the
scattering signal caused by the bulk motion of the intracluster gas
(the kinematic Sunyaev-Zel'dovich, kSZ, effect). The
latter has more than an order of magnitude lower amplitude than the
thermal effect, and for the rest of the discussion we will
specifically focus on the thermal SZ effect only. 

Excellent reviews for the SZ effect and its cosmological applications
are given, e.g., by \citet{birkinshaw99} and \citet{carlstrom02}. Recent
advances in detector technology have made the first blind detection
of SZ clusters possible \citep{staniszewski09}, and three large-scale
experiments are currently in operation which are providing many more
SZ selected clusters out to $z=1$ and beyond: the South Pole Telescope
\citep{vanderlinde10}, the Atacama Cosmology Telescope
\citep{marriage11} and the \pla\ satellite \citep{planckESZ}.

The signal of the SZ effect is directly proportional to the integrated
pressure of the intracluster gas along the line-of-sight, which is
measured as the {\it Comptonization parameter, y}. The change in the
background CMB intensity is thus $\Delta I_{\mathrm{CMB}} /
I_{\mathrm{CMB}} = f(\nu) y$, where $f(\nu)$ is the spectral shape
function, and 
\begin{equation}
y = \int \frac{\kb \tel}{\mel c^2} \sigma_{\rm T} \nel \dif l\,.
\label{eq:sz}
\end{equation}
Here $\mel$ is the electron mass and $\sigma_{\rm T}$ is the Thomson
scattering cross-section. The integration is along the line-of-sight
path length $\dif l$. For typical ICM temperatures
and densities,
the relative change in CMB intensity is small:
$\Delta I_{\mathrm{CMB}} / I_{\mathrm{CMB}} \sim 10^{-4}-10^{-5}$. The
major advantage of the SZ effect comes from its redshift independence,
since the signal is the result of scattering of the background CMB
photons, and both scattered and un-scattered photons redshift
together. This puts the SZ effect in contrast to all other
astrophysical signals, for example the $(1+z)^4$ dimming of the X-ray
surface brightness (eq.~\ref{out:eq:tech:X:SB}). In practice, however, this redshift independence
is currently not fully exploitable due to finite beam sizes. Another major
advantage is the linear dependence of the SZ signal on electron
density, as opposed to the $\nel^2$ dependence of X-ray brightness,
which potentially makes it more suitable to study the low density
outskirt environments.

SZ measurements can be used in at least two different ways to
determine the total mass from ICM properties. Both methods, however,
do require additional information from X-rays since constraining the
SZ observable, pressure integrated along the sight line, alone is
insufficient to apply the hydrostatic equation
(\ref{out:eq:mass:X:hydro1}).

The first method aims at directly measuring the cluster pressure
profile. By adding the gas density profile from X-ray observations,
the hydrostatic equation can be applied.
In the second method, density and temperature profiles are determined
simultaneously from joint X-ray/SZ modelling. More details of
both methods are described in Section~\ref{out:tech:SZ}.
See also \citet[][in this volume]{lms13} for a review on combining
X-ray, SZ, and gravitational lensing measurements to constrain the
three-dimensional shape of clusters. 

\subsection{Total mass inferred from weak gravitational lensing}
\label{out:mass:WL}
To first approximation, gas physics can be ignored for weak lensing
mass reconstructions. This simplifies the situation considerably if
one is only interested in cluster mass.

Weak gravitational lensing offers an alternative route to measuring cluster
mass profiles, independent of the physical state and nature (dark or luminous,
baryonic or non-baryonic) of the matter. One exploits the spatial
correlation of weak shape distortions of background galaxies induced by a
cluster's gravitational potential.
The weak lensing observable, the
so-called reduced shear $g(\pmb{\theta})$ as a function of the lens-plane 
position $\pmb{\theta}$ is connected to the projected surface mass 
density\footnote{Defined as $\kappa\!=\!\Sigma/\Sigma_{\mathrm{crit}}$ in terms
of the critical surface mass density.} $\kappa$ via a non-local
relation \citep[e.g.,][]{ks93}.
Cluster masses can be inferred from the reduced shear profile $g(\pmb{\theta})$
by either fitting with a profile 
\citep[e.g.,][NFW]{1996ApJ...462..563N,1996A&A...313..697B,2000ApJ...534...34W} 
function or by directly inverting the shear--mass problem.
An early direct method, the aperture densitometry or $\zeta$-statistics
\citep{1994ApJ...437...56F}, spawned the development of the aperture mass
estimator \citep{1996MNRAS.283..837S}, which is mainly used to detect mass
overdensities via weak lensing. \citet{2001A&A...374..740S} developed a
mass reconstruction algorithm computing a two-dimensional convergence map from
an input shear catalogue. \citet{2008ApJ...684..177U} introduced a maximum 
entropy method
tackling the same problem from a Bayesian viewpoint and
present mass profiles for a well-studied lensing cluster, Abell 1689, using
different lensing methods.

While weak lensing shear profiles can be measured as far out as the field-of-view of the
camera permits, the cluster signal slowly sinks into the cosmic-shear background
caused by lensing due to uncorrelated large-scale structure. We address this
topic in greater detail in Section~\ref{out:tech:WL}.
As the mass enclosed within a sphere described by an NFW profile diverges
logarithmically, \citet{2009JCAP...01..015B} introduced a smoothed cut-off at
large radii. \citet{2011MNRAS.414.1851O} provide the corresponding lensing
profile which they find to give a better representation of the cluster shear
obtained by ray tracing through an $N$-body numerical simulation. 

A further practical limitation to the precision of weak lensing mass
profiles arises from the considerable intrinsic and observational
scatter in galaxy ellipticities, which dominates over the shear signal
outside a certain radius depending on both the cluster mass and the
lensing efficiency (e.g., Hoekstra et al., this volume).

\subsection{Total mass inferred from galaxy velocities}
\label{out:mass:caustics}
While this review focusses on cluster outskirts mass estimates through
ICM and weak lensing measurements, a tremendous amount of work has
been done using galaxy velocites. Indeed, the first robust hints on
the existence of dark matter are due to them \citep{z33}. As a simple
example, assuming virial equilibrium, the total cluster mass can be
related to the radial galaxy velocity dispersion through
$\mt\approx r\sigma^2/G$. Masses have been estimated for large
samples of galaxy clusters through galaxy velocites, resulting in
cosmological constraints, scaling relations etc.\
\citep[e.g.,][]{bgg93,gbg98,bgc99,zac11}.

Of particular interest for cluster outskirts is the so-called caustics
method \citep[e.g.,][]{dg97,d99}, which has been used to infer cluster
masses out to large radii without equilibrium assumptions
\citep[e.g.,][]{rgk03,rgd12,rd06}. The method is based on the measurement of
sharp, trumpet-like features in redshift space as a function of
cluster centric distance in cluster infall regions
\citep[e.g., Fig.~5 in][]{k87}. The amplitude of these ``caustics''
depends on the escape velocity and, therefore, the mass.

\section{Gas physics}
\label{out:physics}
In this Section, we discuss several physical effects that may
influence the uncertainty of the X-ray mass determination in cluster
outskirts.
While observations and simulations are always discussed in the
following when relevant, we summarize the status of ICM
density, temperature, pressure, and entropy profiles as well as the gas mass
fraction in the first Section (\ref{out:physics:profiles}).

\subsection{Overview of ICM properties in cluster outskirts}
\label{out:physics:profiles}

\subsubsection{Surface brightness and gas density profiles}
\label{sec:sb}

As has been summarized recently by \citet{em11}, the X-ray surface
brightness is a quantity much easier to characterize than 
the temperature and it is rich in physical information being proportional
to the emission measure, i.e.\ to the square of the gas density, of the emitting source.
Thanks to its large field-of-view and low instrumental background, 
\ro\ \ps\ is still the main instrument for providing robust constraints
on the X-ray surface brightness profile of galaxy clusters over a 
significant fraction of the virial radius 
\citep[e.g.,][]{vfj99,ne05,eve12}.

\citet{vfj99} found that a $\beta$-model with $\beta=0.65$--$0.85$
(i.e., a power-law slope in the range $-2.9$ to $-4.1$; from $\sx(r) \approx r^{2 (0.5 -3\beta)} 
= r^{1 - 6 \beta}$)
described well the surface brightness profiles, $\sx(r)$, in the range (0.3--1)
$r_{180}$ of 39 massive local galaxy clusters observed with \ro\ \ps\ 
in the soft X-ray band, (0.5--2) keV.
\citet{ne05} found that the stacked profiles of a few massive nearby systems
located in regions of low ($<$$6 \times 10^{20}$ cm$^{-3}$) Galactic absorption
observed by \ro\ \ps\ provide values of $\beta$ around $0.8$ at
$r_{200}$, with a power-law slope that increases from $-3$, when the fit is
done over the radial range (0.1--1) $r_{200}$, to $-5.7^{+1.5}_{-1.2}$ over
(0.7--1.2) $r_{200}$.

\begin{figure}[ht]
\hspace*{-0.7cm}
     \includegraphics[height=.22\textheight]{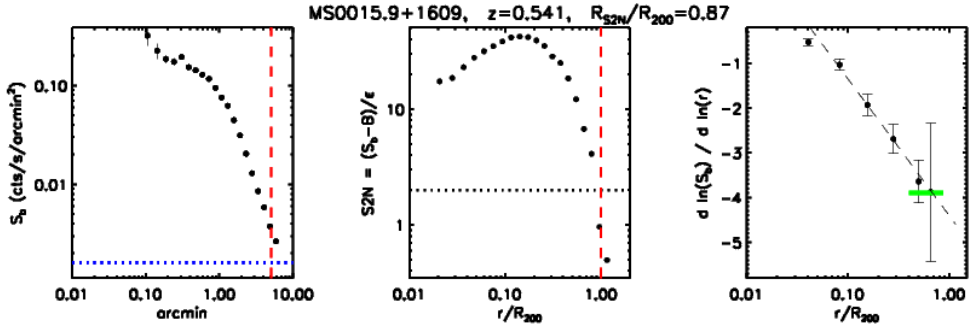}
\caption{
From left to right:
(i) \cha\ surface brightness profile of MS0015.9 with the fitted background
({\it horizontal dotted line}) and the radius $r_{200}$
({\it vertical dashed line}); (ii) signal-to-noise profile, evaluated
as $S2N = (\sx - B)/ \epsilon$ as function of $r / r_{200}$,
where the error $\epsilon$ is the sum in quadrature of the Poissonian error
in the radial counts and the uncertainties in the fitted background, $B$,
defined by considering
a region far from the X-ray center that covered a significant portion
of the exposed CCD with negligible cluster emission;
(iii) the best-fit values of the slope of the
surface brightness profiles as a function of $r / r_{200}$.
The thick horizontal solid line indicates the slope evaluated between
$0.4$ $r_{200}$ and $R_{S2N}$ with a minimum of 3 radial bins.
The dashed line indicates the best-fit of $d \ln (\sx) / d \ln (r/r_{200})$
with the functional form $s_0 +s_1 \ln (r/r_{200})$ over the radial range
$0.1$ $r_{200}$--$R_{S2N}$ (reprinted with permission from
\citealt{eb09}).
} \label{fig:sb_obs} \end{figure}

\citet{eb09} studied the X-ray surface brightness profiles
at $r > r_{500}$ of 11 objects extracted from
a sample of hot ($\tg > 3$ keV), high-redshift ($0.3 < z < 1.3$)
galaxy clusters observed with \cha. 
They performed a linear least-squares fit between the logarithmic
values of the radial bins and the background-subtracted X-ray surface
brightness (Fig.~\ref{fig:sb_obs}). 
Overall, the error-weighted mean slope is $-2.91$ (with a standard deviation
in the distribution of $0.46$) at $r > 0.2$ $r_{200}$ and $-3.59$ $(0.75)$
at $r > 0.4$ $r_{200}$.
For the only 3 objects for which a fit between $0.5$ $r_{200}$ and
$R_{S2N}$,
the maximum radius out to which the cluster surface brightness could
be measured with a signal-to-noise ratio of at least 2,
was possible, they measured a further steepening of the profiles, with a
mean slope of $-4.43$ and a standard deviation of $0.83$.
They also fitted linearly the derivative of the logarithmic $\sx(r)$
over the radial range $0.1$ $r_{200}$--$R_{S2N}$, excluding in this way the
influence of the core emission.
The average (and standard deviation $\sigma$) values of the extrapolated slopes are
then $-3.15$ $(0.46)$, $-3.86$ $(0.70)$, and $-4.31$ $(0.87)$
at $0.4 r_{200}$, $0.7 r_{200}$ and $r_{200}$, respectively.
These values are comparable to what has been obtained in previous analyses of local systems
through \ro\ \ps\ exposures and are supported from the studies of the plasma's properties
in the outer regions of hydrodynamically simulated X-ray emitting galaxy clusters 
\citep[e.g.,][]{red06,nl11,vre11}.

Indeed, modelling cluster outskirts with hydrodynamical simulations is generally considered an 
easier task with respect to cluster cores. The reason is that any form of feedback effect is 
usually connected with active galactic nuclei (AGN) and star-formation
and, therefore, with gas density, thus leaving the  
external regions more dominated by gravitational collapse.
In fact, it has been shown with hydrodynamical simulations that different physical prescriptions 
have a small impact on clusters' profiles in the outskirts. 
\citet[see also Fig.~\ref{out:physics:structure:simu}]{red06} studied the behavior of the profiles of density, 
temperature and surface brightness in a sample of 9 galaxy groups and clusters simulated with 
the Lagrangian code GADGET-2, with non-radiative physics and with several treatments of cooling, 
star-formation, feedback, viscosity, and thermal conduction. They found a clear steepening of all 
the profiles around the virial radius with very small differences due to the physical models.
In detail, the slope of the soft, (0.5--2) keV, X-ray surface
brightness changes from about $-3.2$ 
close to the center to about $-5.5$ in the outskirts for high-mass
objects, with a slope of $-4, -4.5, -5.2$ when estimated in the radial
range  (0.3--1.2) $r_{200}$, (0.7--1.2) $r_{200}$, (1.2--2.7)
$r_{200}$, respectively,  
in excellent agreement with the observational constraints.
Moreover, on the scale of galaxy groups ($M_{200} \sim 10^{14} M_\odot$),
the steepening is much more prominent, with slopes that vary from 
about $-2.5$ to $-7.5$.
A similar behavior is measured in the profiles of the gas density and temperature.
These evidences can be explained as due to the fact that, in the present scenario
for structure formation, galaxy groups are dynamically older and more relaxed 
than clusters, as also suggested from the observed and simulated behavior
of the concentration--mass relation
\citep[e.g.,][]{ngb07,bgh07,egl10}.

However, simulations highlight that the accretion pattern in the
outskirts is indeed complicated. Non-adiabatic effects like turbulence
and shocks are frequent in unrelaxed systems and can influence the
pressure profiles. Even if these processes are not connected to any
feedback  
of the star-forming phase, they often require special treatment in the simulations because they act at 
small scales that are difficult to reach in cosmological simulations.
Recently, \citet{vbk09} ran simulations with the Eulerian adaptive mesh
refinement (AMR) code ENZO including the implementation of a new 
sub-grid refining scheme, allowing them to study the velocity pattern of the ICM in simulated 
clusters. Their results showed that the kinetic and turbulent energy
associated with the ICM 
account for 5--25 per cent of the total thermal energy inside $\rv$.
\citet{bso10} found a remarkable agreement between the temperature,
density, and 
entropy profiles of 24 mostly substructure-free massive clusters simulated with the ENZO code and the
published \suz\ results, implying that (i) the simplest adiabatic gas physics is adequate to model the 
cluster outskirts without requiring other mechanisms (e.g., non-gravitational heating, cooling, magnetic fields, or cosmic rays),
and (ii) the outer regions of the ICM are not in hydrostatic equilibrium. 

Inhomogeneities in the X-ray emission due to random density fluctuations are expected.
Using simulated clusters, \citet{mem99} measured a mean mass-weighted clumping factor 
$C = \langle\rog^2\rangle/\langle\rog\rangle^2$ between 1.3 and 1.4
within a density contrast of 500. If clumping is ignored, the gas
density, and therefore, the gas mass, gets overestimated by $\sqrt{C}$.
\citet{nl11} analysed 16 simulated clusters and computed their clumpiness 
profiles with and without radiative cooling. They found that, at $r_{200}$, $C$ goes from
1.3 in the former to 2 in the latter ones.
The non-radiative clumping factor is higher because radiative cooling
removes gas from the hot $T>10^6$ K X-ray emitting phase in the
simulations. This shows that adding gas physics has quite a
significant effect in cluster outskirts. The authors suggest that,
since hydrodynamical simulations suffer of a form of overcooling
problem, the true result is likely to be between these values.
Overall, they concluded that gas clumping leads to an overestimate of the observed gas density 
and causes flattening of the entropy profile, as suggested
qualitatively from recent \suz\ observations 
\citep[e.g.,][]{hhs10,sam11}.

\begin{figure}[th]
\hbox{
\hspace*{-1cm}
\includegraphics[height=.45\textheight]{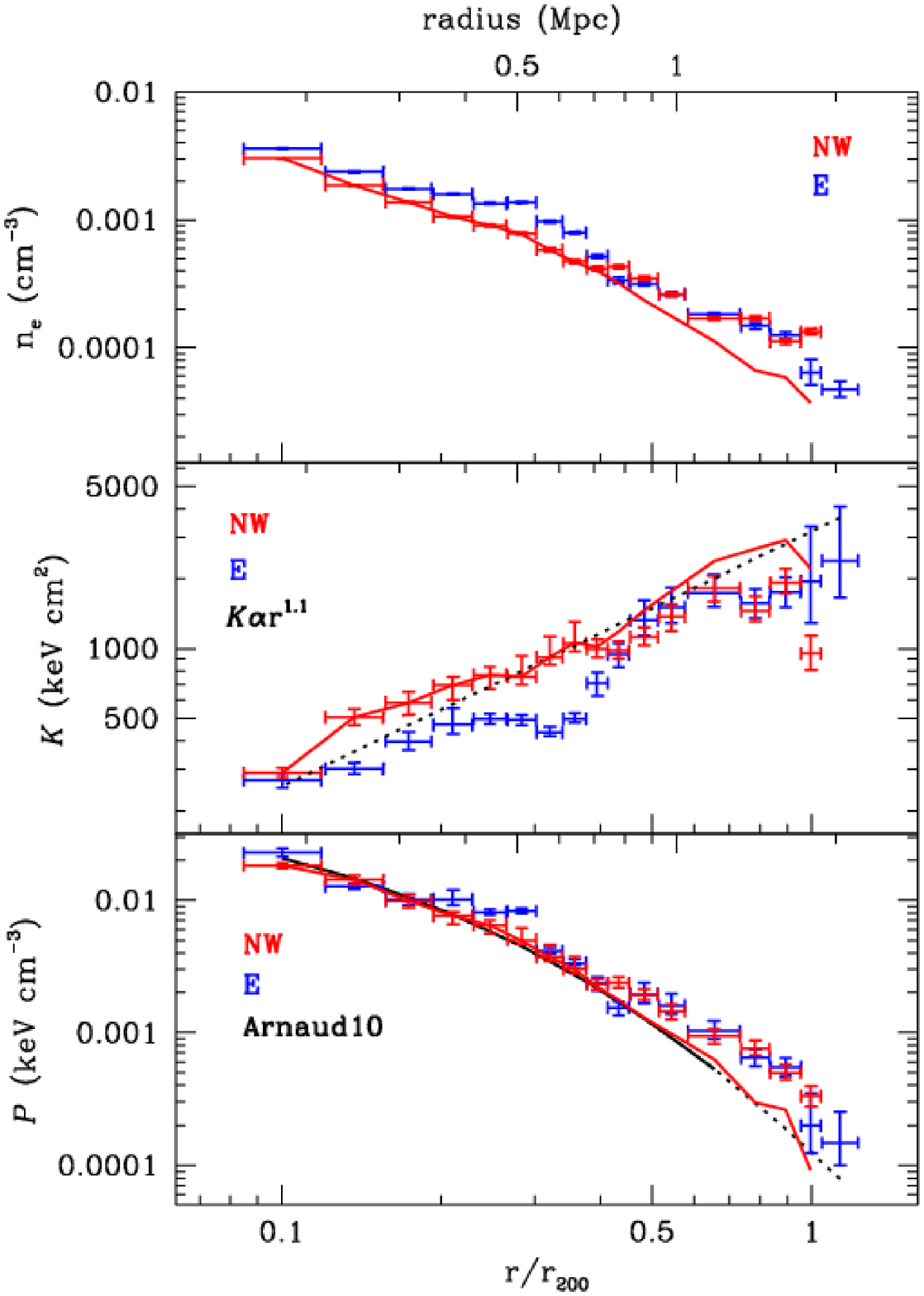}
\includegraphics[height=.45\textheight]{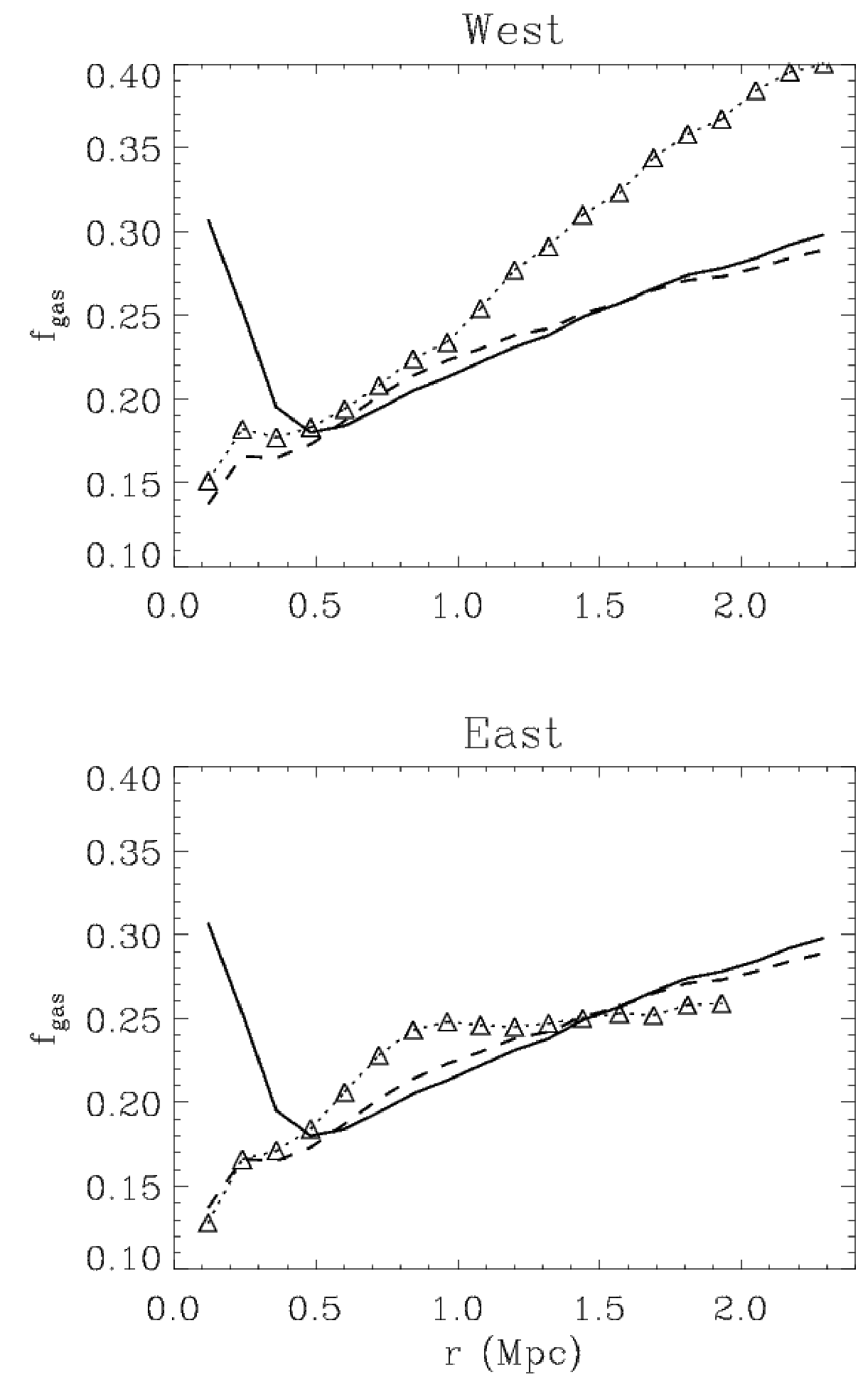}
}
\caption{Left (from \citealt{sam11}, reprinted with permission from
AAAS): 
Deprojected electron density, entropy ($K$), and pressure profiles
towards the northwest (red data points) and east (blue data points) of
the Perseus cluster, measured with \suz.
The red line shows the northwestern profiles corrected for clumping.
The expected entropy profile from simulations of gravitational
collapse \citep{vo05} is a power-law with index 1.1 over-plotted  in
the entropy panel (black dotted line). The average profile of a sample
of clusters previously studied with \xmm\ within $\sim$0.5 $r_{200}$
\citep{app10} is shown in the pressure panel (solid black curve;  its extrapolation to $r_{200}$ is shown with a dotted black line).
Right (from \citealt{efw98}, reprinted with permission): Gas mass fraction profiles obtained from
the deprojection of the \ro\ \ps\ surface brightness profiles in
different azimuthal sectors by assuming a Navarro-Frenk-White (NFW) dark
matter density profile (triangles joined by a dotted line). The gas
fraction measured from the azimuthally averaged profile, both using an
isothermal profile (solid line) and an NFW profile (dashed line), is
shown. $r_{200} = 2.7$ Mpc in this analysis. An Einstein-de Sitter
Universe with $H_0=50$ km s$^{-1}$ Mpc$^{-1}$ is here adopted.   
} \label{fig:perseus} \end{figure}

\suz\ has indeed improved the observational characterization of the
faint emission from cluster outskirts, providing the first
spectroscopic constraints on it (Section~\ref{out:physics:Tr}). 
\citet{sam11} have resolved the baryonic and metal content
of the ICM in the Perseus cluster out to $r_{200} \sim 1.8
h_{70}^{-1}$ Mpc, estimating a clear excess of the gas fraction with
respect to the expected cosmic value along the northwest axis, and
confirming some preliminary evidence from \ro\ \ps\ (\citealt{efw98},
where the gas density is resolved at $r_{200}$ but an isothermal
gas is assumed due to limited spectral capability of the \ps; see
Fig.~\ref{fig:perseus}). They suggest that clumpiness of the ICM on
the order of $C\approx$ 4--16 over the radial range (0.7--1) $r_{200}$
is required to properly reconcile the expected and measured gas mass
fraction; i.e., significantly higher $C$ than indicated by the
simulations described above.

A more observable quantity is the estimate of the azimuthal scatter along the cluster radius
$S(r) = \sqrt{1/N \, \sum_i \frac{ \left[ y_i(r) -Y(r) \right]^2 }{Y(r)^2}}$, where $y_i(r)$ is the 
radial profile of a given quantity, taken inside a given $i-$sector, and $Y(r)$ is the 
average profile taken from all the cluster volume. 
Contributions to this scatter are expected from both an intrinsic deviation with respect to
the spherical symmetry and, most importantly, to the presence of filaments of X-ray emission
associated to any preferential direction in the accretion pattern of the ICM.

Two sets of high-resolution cosmological re-simulations obtained with the codes ENZO and GADGET2 
are used in \citet{vre11} to show that, in general, the azimuthal scatter in the radial 
profiles of X-ray luminous galaxy clusters is about 10 per cent for
gas density, temperature, and 
entropy inside $r_{200}$, and 25 per cent for X-ray luminosity for the same volume. 
These values generally double approaching 2 $r_{200}$ from the cluster
center, and are found to be 
higher (by $\sim$20--40 per cent) in the case of perturbed systems. These results suggest the 
possibility to interpret the large azimuthal scatter of observables as
estimated from, e.g., \suz\ with the present simulated data.

\begin{figure}[th]
\hbox{
\hspace*{-1cm}
\includegraphics[height=.33\textwidth]{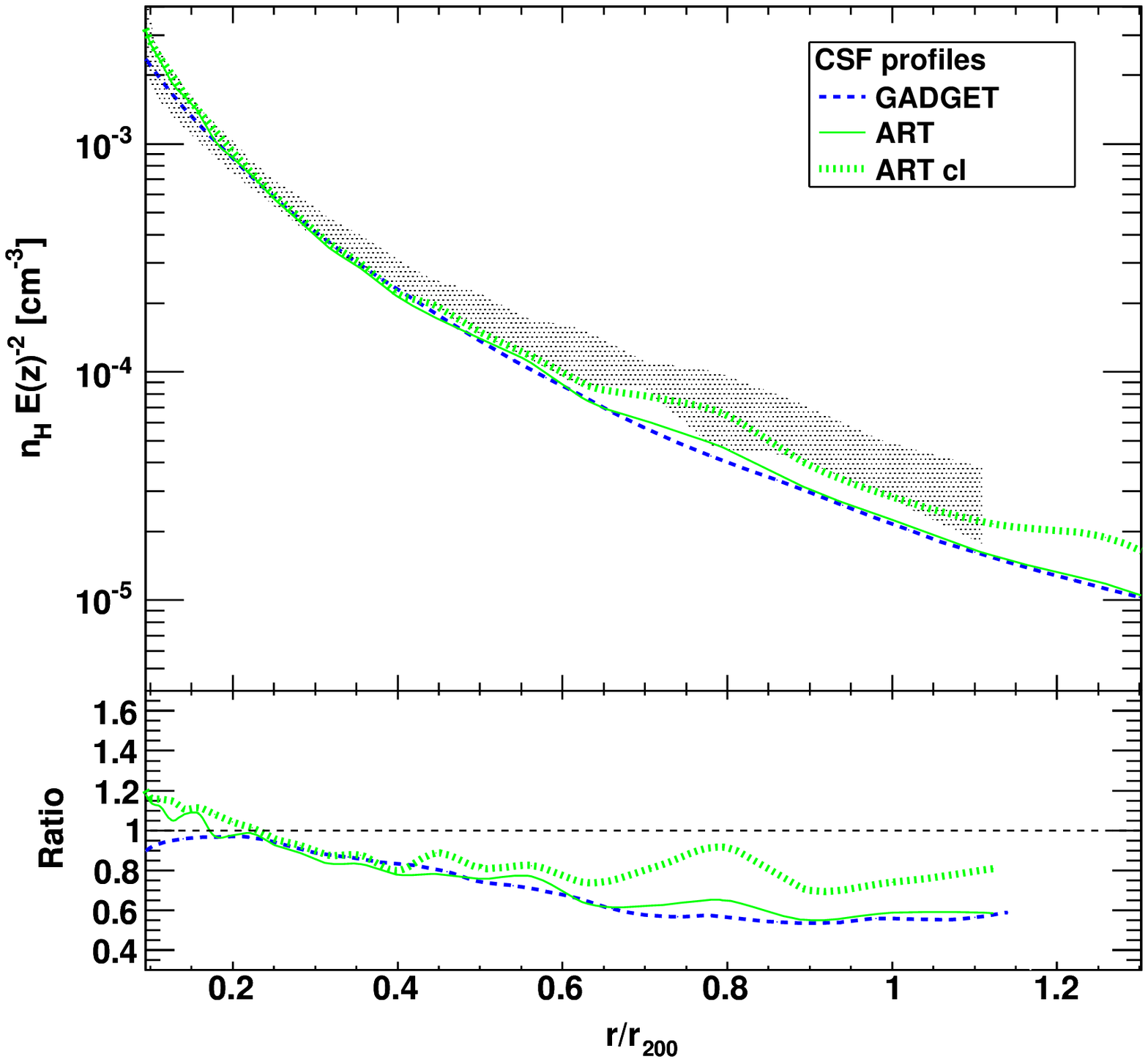}
\includegraphics[height=.33\textwidth]{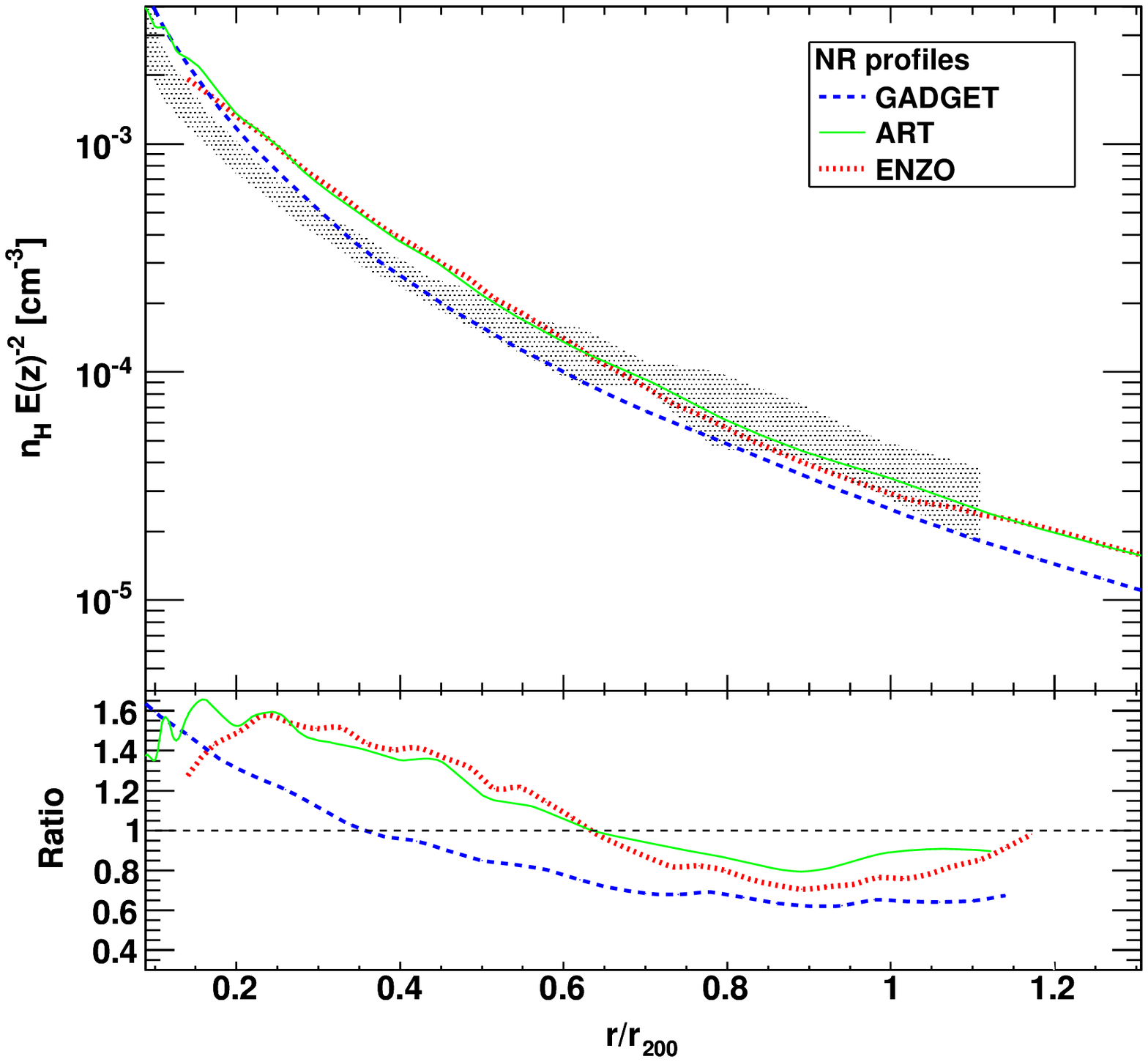}
\includegraphics[height=.33\textwidth]{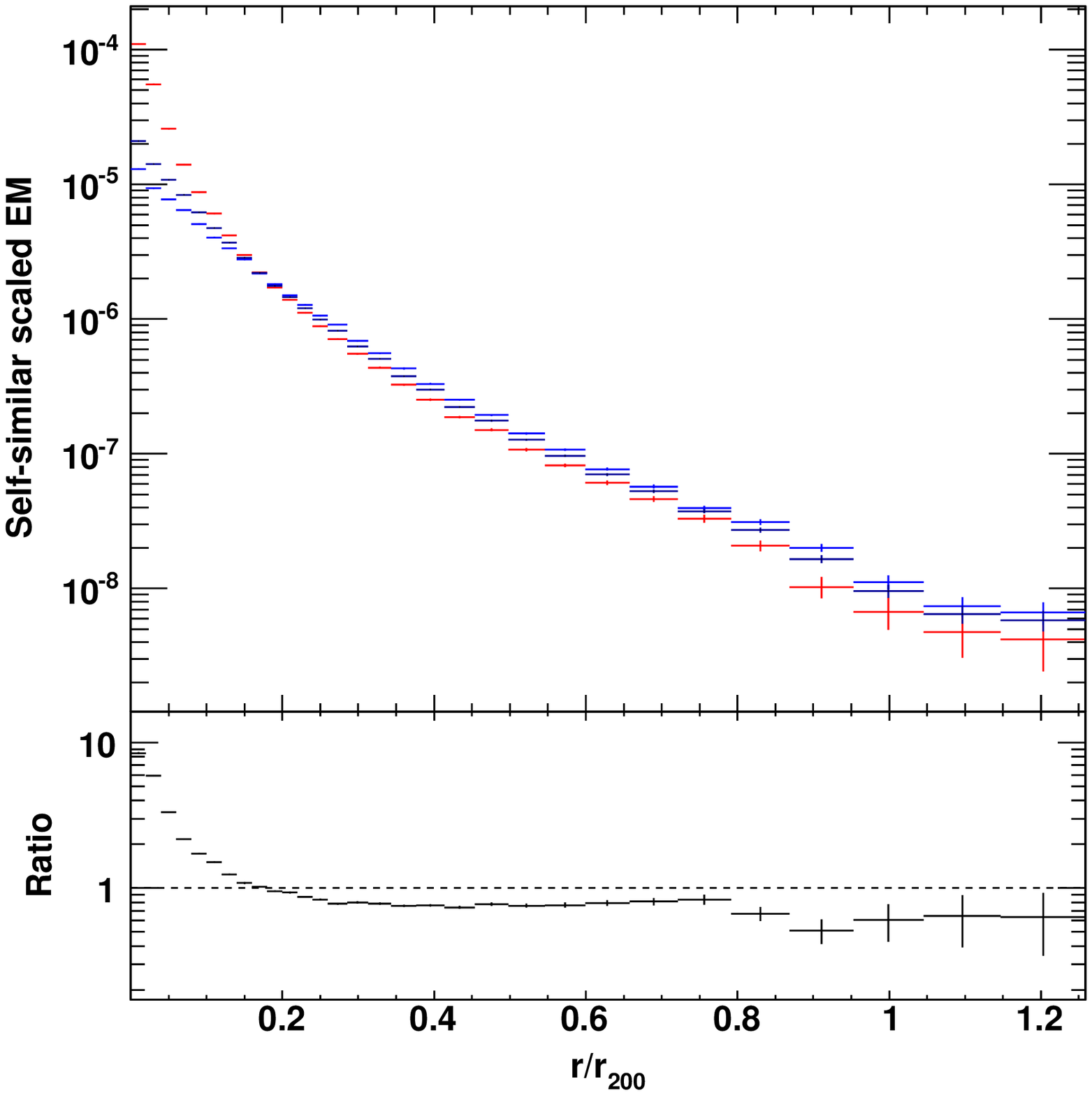}
}
\caption{Comparison between the mean \ro\ density profile for a sample
of 31 nearby X-ray bright 
galaxy clusters and numerical simulations both including cooling and
star formation (left) and with non-radiative 
physics (center).
Right: Stacked emission measure profile (in units
of cm$^{-6}$ Mpc) for the entire sample (black),  
the cool-core (CC, red) and non-cool-core (NCC, blue) systems. The bottom panel
shows the ratio between the CC and NCC populations. 
Reprinted with permission from \citet{eve12}.
} \label{fig:out:physics:SB:Eckert}
\end{figure}
In \citet{eve12}, a stacking analysis of
the gas density profiles
in a local ($z =$ 0.04--0.2) sample of 31 galaxy clusters observed with
\ro\ \ps\ is presented (Fig.~\ref{fig:out:physics:SB:Eckert}).
They observe a steepening of the density profiles beyond $\sim$$r_{500}$. 
They also report the high-confidence detection of a systematic difference
between cool-core and
non-cool-core clusters beyond $\sim$0.3 $r_{200}$, which is explained by a
different distribution 
of the gas in the two classes. Beyond $r_{500}$, galaxy clusters deviate
significantly from 
spherical symmetry, with only little differences between relaxed and
disturbed systems. 
The observed and predicted (from numerical simulations) scatter profiles are
in good agreement 
when the 1\% densest clumps are filtered out in the simulations.
While all the different sets of simulations used by them, especially
beyond $\sim$0.7 $r_{200}$, show a relatively good agreement, 
they seem all to predict steeper profiles than the observed one from
the \ps, in particular in the radial range
between 0.4 and 0.65 $r_{200}$. Approaching
$r_{200}$, the slope increases up to  $\beta\approx 1$ both in
simulated and observed profiles.
\citeauthor{eve12} also conclude that a slightly better agreement in terms of
shape of the gas density profile 
is found when a treatment of the observational effects of gas clumping is
adopted (as, e.g., in \citealt{nl11}).

At very large radii up to $\rv$, \suz\ data seem to indicate a
flattening of the gas density profile \citep[e.g.,][]{kou10}. Recall
(eq.~\ref{out:eq:mass:X:hydro2}) that the gas density gradient enters
linearly in the total mass determination. A flatter profile, therefore,
results in a lower total mass estimate (and a larger gas mass
estimate).

\subsubsection{Temperature profiles}
\label{out:physics:Tr}
We have seen (e.g., Fig.~\ref{out:mass:M-Tr}) that the inferred total
cluster mass strongly depends on the measured temperature profile.
This is mostly because $\tg(r)$ enters linearly in the hydrostatic
equation (\ref{out:eq:mass:X:hydro2}), therefore, a 20\% uncertainty
in $\tg(r)$ results in a mass uncertainty contribution of 20\% if a
fixed radius is chosen ($\sim$30\% if the mass is determined within an
overdensity radius). 

Despite the poor, energy-dependent point-spread-function (PSF), the
Advanced Satellite for Cosmology and Astrophysics (\as) 
provided temperature profiles to large radius for some clusters
\citep[e.g.,][]{mmi96,f97,mfs98,w00}, as well as \sax\
\citep[e.g.,][]{ibe99,ib00,dm02}.
Due to their high particle backgrounds,
\xmm\ and \cha\ usually cannot robustly reach the cluster outskirts
\citep[e.g.,][]{asf01,zfb04,vmm05,app05,kv05,pjk05,pbc07,smk08,lm08},
apart from a few special, e.g., very bright, low temperature, systems 
\citep[e.g.,][]{uws11,blm12}.

The breakthrough came recently after the launch of \suz, a satellite in
low-Earth orbit and with short focal length \citep{mbi07}, resulting
in a low and stable particle background.
The first \suz\ temperature measurements reaching beyond the \xmm\ and
\cha\ limit were published by
\citet{fth07,rhz08,gfs08,bms09}. The latter three are all based on
relaxed cool-core clusters, while \citet{fth07} targeted the
compressed and heated interaction region between A399 and A401 with
the primary goal to constrain the metallicity
(Section~\ref{out:chem}).
While initially there were sometimes difficulties to properly account
for all fore- and background components (see the technical
Section~\ref{out:tech:X:bkg} for details on these components),
especially for clusters at low Galactic latitude \citep{gfs08,emg11},
more elaborate robust analyses are now routinely being performed
\citep[e.g.,][]{waf12}. 

The $\sim$100 refereed articles that are turned up by
ADS\footnote{Astrophysics Data System,
http://adsabs.harvard.edu/abstract\_service.html\,.} when 
searching for ``Suzaku'' and ``cluster'' in the abstract have already
received $>$1,000 citations at the time of writing (December 2012), demonstrating the large interest in \suz\
cluster studies. The six most highly cited references of these all
deal with cluster outskirts, which shows that this interest is driven
particularly by this subject.

\begin{figure}[htb]
\includegraphics[angle=0,width=1.0\textwidth]{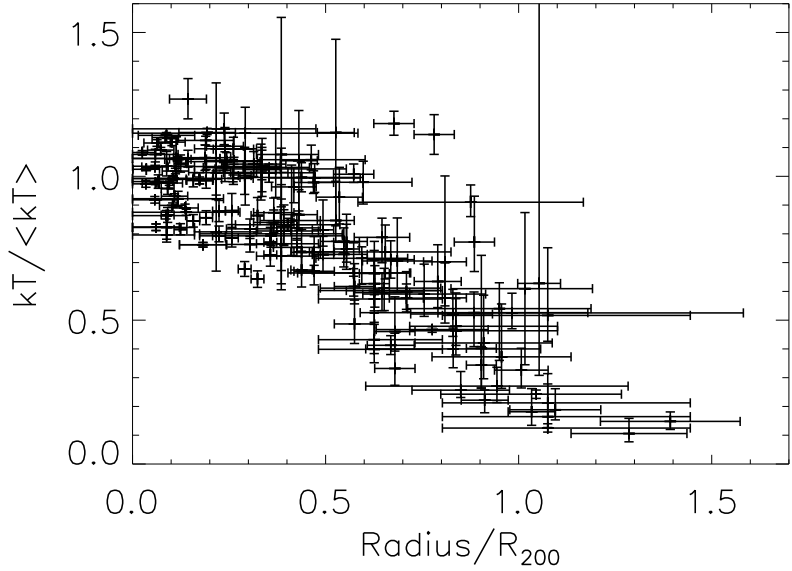}
\caption{All published  \suz\ temperature profiles that reach cluster
outskirts. The references are listed in the text.}  
\label{out:physics:Tr:Suz1}
\end{figure}
In Fig.~\ref{out:physics:Tr:Suz1} all currently available \suz\
cluster temperature profiles are shown that reach beyond about 1/2
$r_{200}$ (to our knowledge; see also \citealt{ahi11} for an earlier
compilation).%
\footnote{The data were thankfully provided electronically by
H. Akamatsu (A2142, \citealt{ahi11}, as well as A3667, A3376, CIZA2242.8--5301, and
ZwCl2341.1--0000, \citealt{ak12};
A. Fabian and S. Walker (A2029, \citealt{wfs12}, and 
the reanalysis of PKS0745-191, \citealt{waf12}, updated from the
inital results of \citealt{gfs08});
A. Hoshino \citep[A1413,][]{hhs10};
P. Humphrey \citep[RXJ1159+5531,][]{hbb12};
M. Kawaharada \citep[A1689,][]{kou10};
E. Miller \citep[A1795,][]{bms09};
T. Reiprich \citep[A2204,][]{rhz08};
K. Sato \citep[A2811, A2804, and A2801,][]{sk10};
T. Sato (Coma, \citealt{smo11}, and Hydra~A, \citealt{ssm12});
A. Simionescu \citep[Perseus,][]{sam11}.
Note that for several clusters more than one data set is shown, each
covering a different azimuthal direction within a cluster. 
After acceptance of this review, two other \suz\ temperature profile
papers appeared, which are not included in the compilation above
(A1835, \citealt{imo13}, and Coma, \citealt{swu13}).}
The purpose of this compilation is to test for similarities of the
temperature profiles and, if similar, to determine the general trend
and compare to predicted profiles.
Before proceeding however, a few words of 
caution are in order: While the \citet{app05} $\langle
\kb T\rangle$--$r_{200}$ relation (for their hot cluster subsample,
assuming a flat Universe with $H_0 = 71$ km/s/Mpc and $\om = 0.27$)
was used homogeneously for all clusters for the radial scaling, 
the shown data are inhomogeneous in several other aspects. These
include, e.g.,
background subtraction, PSF correction, deprojection, $\langle
\kb T\rangle$ determination, and radial bin center calculation, so this
will cause increased dispersion in the profiles. Also, both axes are
not completely independent since both are scaled by $\langle
\kb T\rangle$ ($\langle \kb T\rangle^{1/2}$ for $r_{200}$), which 
further increases dispersion (due to the negative slope of the
profiles). So, even if the true scaled cluster
temperature profiles were perfectly self-similar, we expect to see
dispersion in Fig.~\ref{out:physics:Tr:Suz1}. 

The vertical bars shown are the 68\% confidence level statistical
uncertainties. Some authors provided also total systematic
uncertainties. Typically, in the outer parts, they are of roughly
similar size as the statistical errors. The horizontal bars
indicate the radial range used for accumulating the spectra.

\begin{figure}[thb]
\includegraphics[angle=0,width=0.5\textwidth]{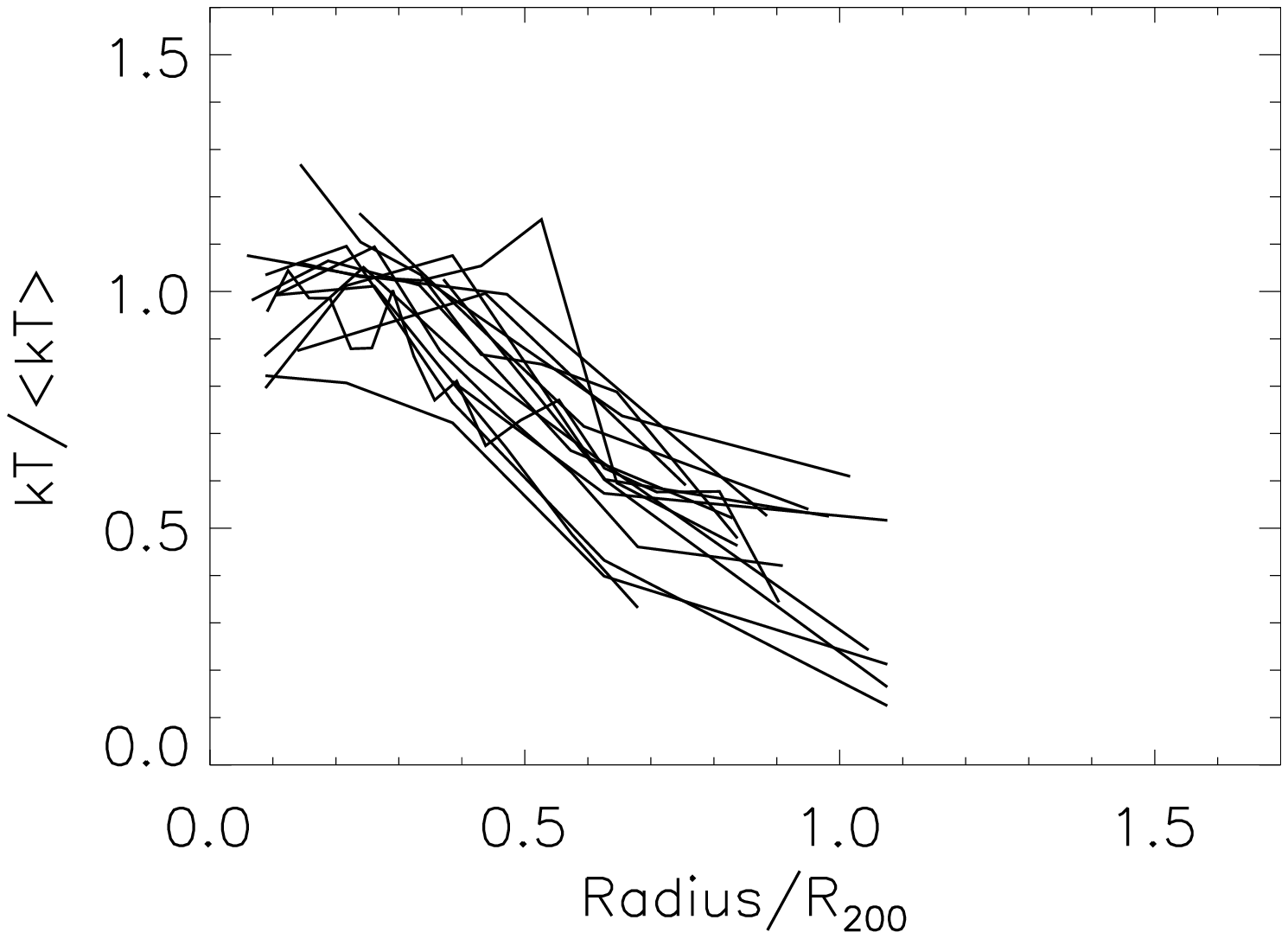}
\includegraphics[angle=0,width=0.5\textwidth]{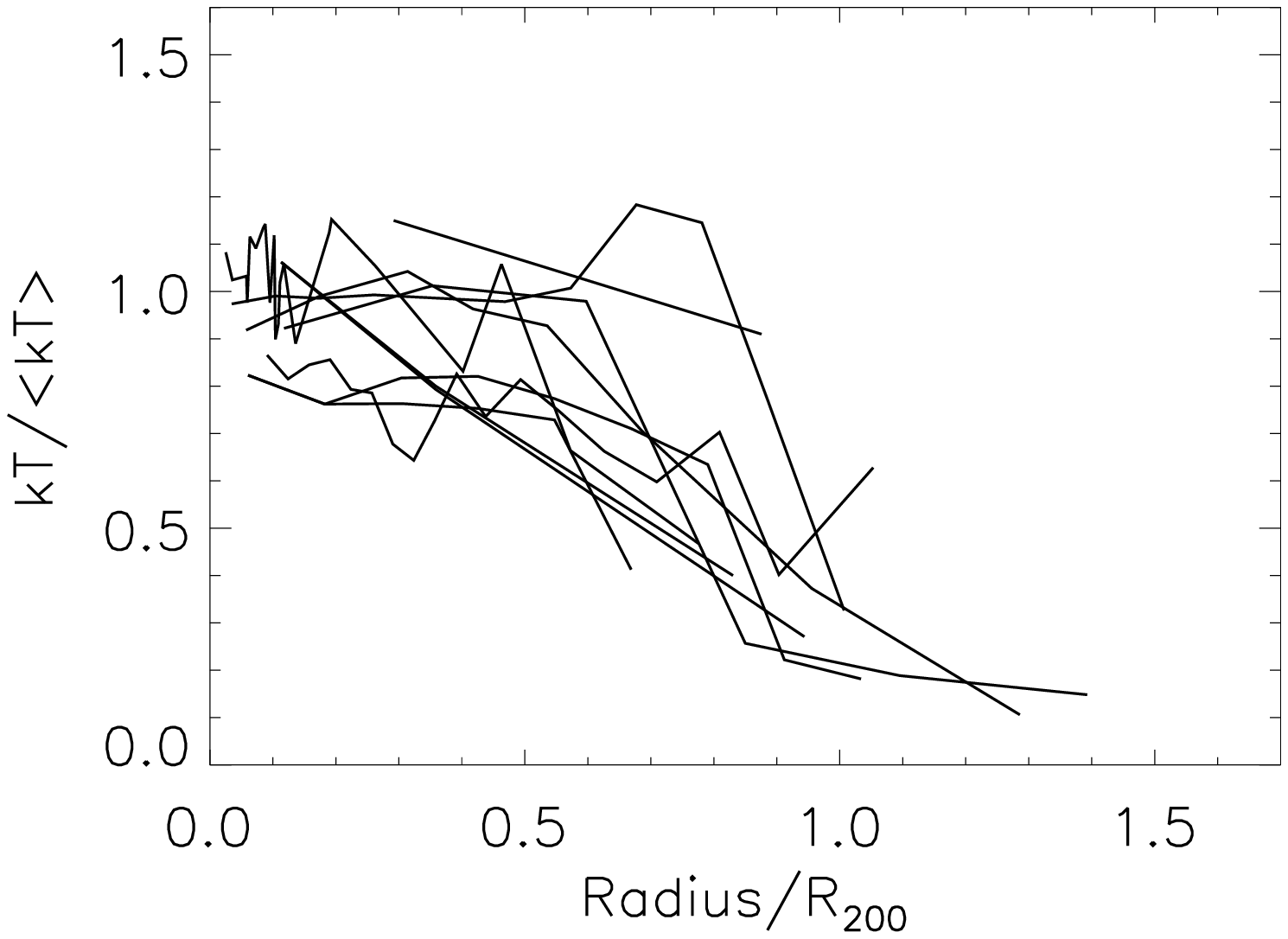}
\caption{\suz\ temperature profiles of clusters flagged as relaxed
(left) and merging (right).
}
\label{out:physics:Tr:Suz1a}
\end{figure}

\begin{figure}[thb]
\includegraphics[angle=0,width=0.5\textwidth]{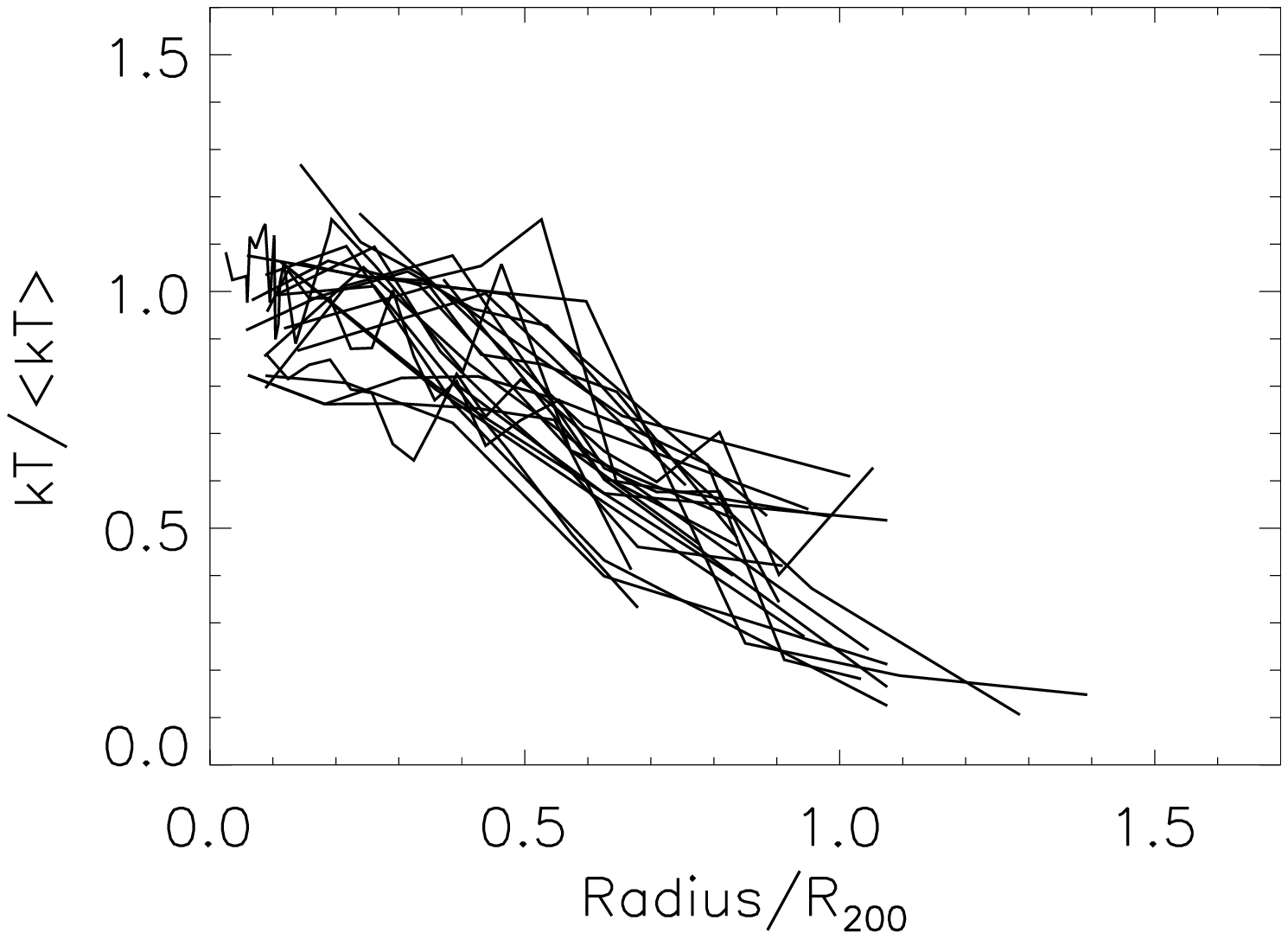}
\includegraphics[angle=0,width=0.5\textwidth]{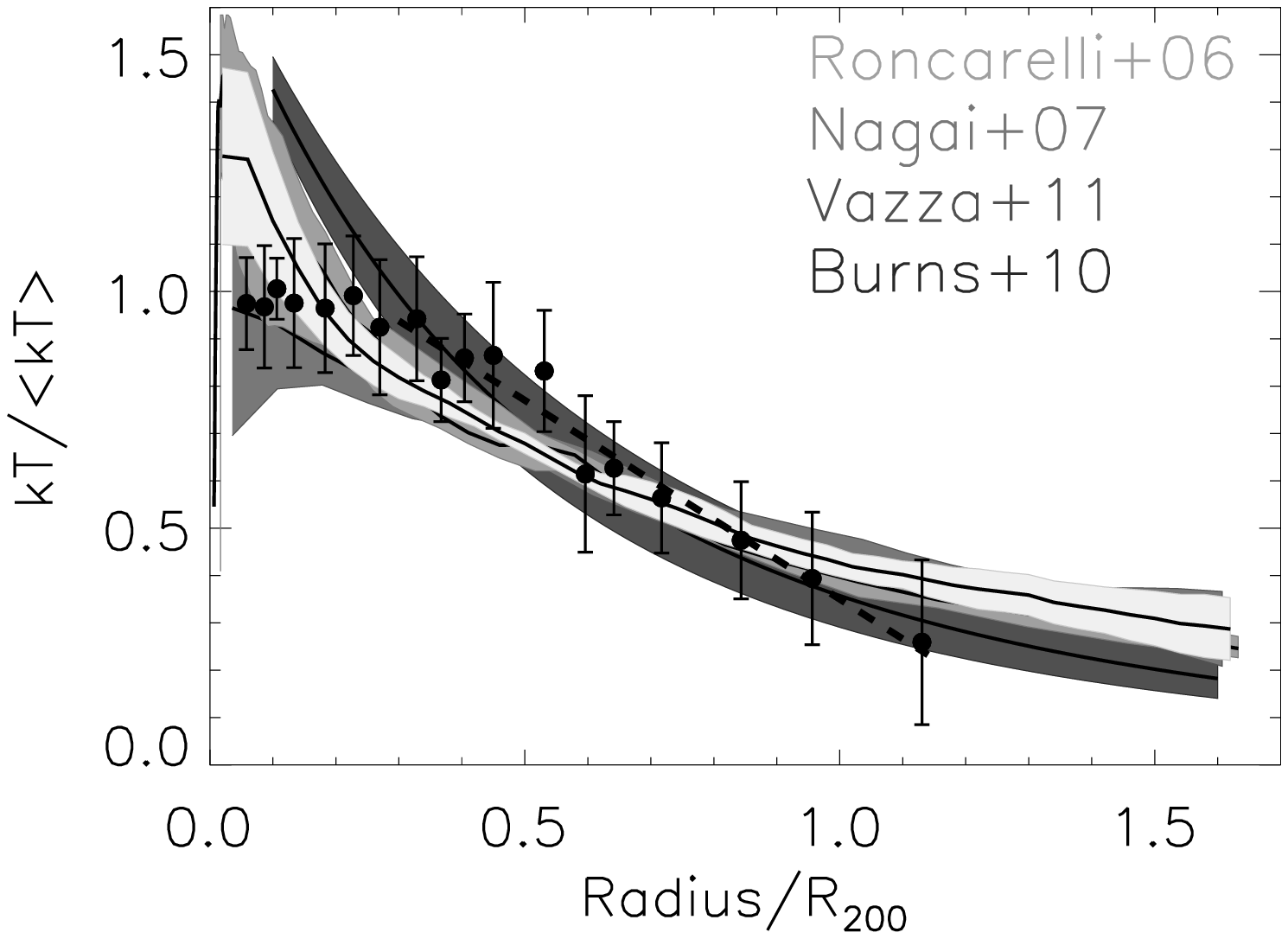}
\caption{All \suz\ temperature profiles, excluding A3376 and A2804 (see
text; left: individual profiles, right: binned average profile).
162 temperature measurements are binned into 18 radial bins,
each containing 9 temperature measurements. Error bars denote the
standard deviation of temperature measurements in each bin.
Also shown are (partially renormalized, see text) average temperature
profiles from numerically simulated clusters from
\citet{red06,nkv07,bso10,vre11}. The dashed line represents a simple
linear fit to the data points beyond 0.3$r_{200}$.
}
\label{out:physics:Tr:Suz2}
\end{figure}

The authors were asked to flag their clusters or azimuthal directions
as either ``relaxed'' or ``merging.'' In
Fig.~\ref{out:physics:Tr:Suz1a}  both sets of profiles are compared
(without error bars, for clarity). While there appears to be more
scatter in the profiles of the clusters flagged as merging, this is
mostly due to A3376 and A2804. In the merging cluster A3376,
\citet{atn12} clearly identified a shock front, which can be
appreciated in Fig.~\ref{out:physics:Tr:Suz1a} (right) as the profile with a
maximum around (0.7--0.8) $r_{200}$. A2804 is a group that lies
between two hotter clusters \citep{sk10}, which seems to cause an
untypically flat temperature profile in the outer parts (second
highest relative temperature at $\sim$0.8 $r_{200}$). Overall, the
trends of the relaxed and merging profiles are rather similar and in
Fig.~\ref{out:physics:Tr:Suz2} all profiles are therefore combined,
excluding only those of A3376 and A2804.

In the inner region ($\lesssim$\,0.3 $r_{200}$) the profiles appear flat.
This is due to the large spread and radially varying slopes of central
temperature profiles as observed with \cha\ and \xmm\
\citep[e.g.,][]{vmm05,pbc07,hmr09} combined with \suz's broad PSF.
Beyond this central region, temperatures systematically decline by
a factor of about three out to $r_{200}$ and slightly beyond.

The \suz\ average profile is compared to profiles predicted by
$N$-body plus hydrodynamic simulations
\citep{red06,nkv07,bso10,vre11}, employing different
numerical algorithms and incorporating different physics
(Fig.~\ref{out:physics:Tr:Suz2}, right).
Before proceeding, it needs to be stressed that this comparison should
not be overinterpreted. This is, on the one hand, because the
simulations vary in several aspects, e.g.,
they employ different temperature definitions. Moreover, all
simulations have difficulties reproducing the temperature
structure in cluster cores as observed by \cha\ and \xmm. Results in
cores depend strongly on the
additional physics put in. The point is now that the profiles in
Fig.~\ref{out:physics:Tr:Suz2} are \emph{normalized} by some average
temperature ($\langle \kb T\rangle$), which depends also on the
temperature structure in the core. We have,
therefore, increased $\langle \kb T\rangle$ by 50\% for the
\citet{red06} and \citet{vre11} profiles in order to roughly renormalize
them to fit the cluster outskirts, as many authors find that employing
different physics recipes does not affect much the outer
temperature profiles. Without this rescaling, the \citeauthor{red06}
and \citeauthor{vre11} profiles would lie above the two other
simulated profiles.

In any case, the slopes of observed and simulated temperature
profiles appear consistent in the outskirts; if anything, then the
observed average temperature profile drops off slightly faster than
the simulated ones at the largest radii, as already noted in the very
first \suz\ temperature profile determination of a relaxed cluster
\citep{rhz08}. In Fig.~\ref{out:physics:Tr:Suz2} (right), this is illustrated by the
dashed line, which gives the best linear fit to the data points in the
range $0.3r_{200}<r<1.15r_{200}$ as $\kb T/\langle \kb T\rangle=
1.19-0.84r/r_{200}$.

It has already been mentioned that some authors provided more than one
\suz\ temperature profile for a given cluster by subdividing the
profile into azimuthal directions, sometimes finding significant
differences. For instance, \citet{kou10} found that in the northeastern
direction of Abell 1689 the temperature around the virial 
radius is about three times larger than temperatures in other
directions. Moreover, they found this enhanced temperature to be
correlated with a large-scale structure filament in the galaxy
distribution and argue that thermalization is faster in this overdense
infall region.

So, with \suz\ we have been moving forward, temperature measurements
out to $\sim$$r_{200}$ can be performed and we expect more progress
in the next few years through homogeneous sample studies with \suz\
and the upcoming \rosi\ and \emph{Astro-H}
\citep[e.g.,][]{tmk12}
instruments, the latter satellite also carrying a high spectral
resolution micro-calorimeter array \citep[e.g.,][]{mkb12}.
Nevertheless, we are still quite far away from the outer
border of cluster outskirts according to our definition (3$r_{200}$,
Section~\ref{out:out}). New X-ray missions with low particle
background, low soft proton contamination, good PSF for AGN removal,
large field-of-view, and large effective area will likely be required 
to reach this frontier.

There is another route to constraining temperature profiles in
cluster outskirts: the combination of X-ray surface brightness
measurements with SZ decrement profiles \citep[e.g.,][]{bnp09}. This
will be discussed in more detail in Section~\ref{out:tech:SZ}. Naively
speaking, the SZ effect depends on the gas density and the X-ray
emission on the square of the density, therefore, the SZ signal might
trace the low density cluster regions better than the X-ray signal.
However, the actual situation is slightly less straightforward since
the SZ signal also decreases linearly with decreasing temperature; i.e., in
cluster outskirts (Fig.~\ref{out:physics:Tr:Suz2}) while the soft X-ray
emission \emph{increases} with decreasing temperature for
$\kb\tel\lesssim 2$ keV due to line emission
(Fig.~\ref{out:mass:apec}). So, overall, the relative gain in
sensitivity of SZ measurements compared to X-ray measurements in
cluster outskirts is not as strong as naively implied by the
comparison of density dependencies.

\subsubsection{Pressure profiles}
\label{out:physics:P}
As the main balance for gravitation in massive halos, the distribution
of thermal pressure within the ICM is of particular interest. It has
been investigated via X-ray observations. Work based on \xmm\
observations of the \rexcess\ sample, a representative sample of nearby
clusters \citep{boe07,pra09}, has demonstrated that the scaled
distribution of the ICM pressure follows a  universal shape.
The observational constraints extend out to  a radius of $\Rv$.  
Within this radial range, the observed pressure profiles are
well-matched by predictions from various numerical simulations 
\citep{borgani04,nkv07,piffaretti08}. 
Without observational constraint beyond $\Rv$,
\citet{app10}
provided
a simple ``universal pressure profile'' parametrisation of a
generalized NFW (GNFW)
function  from the \xmm\ data and the aforementioned numerical
simulation (beyond $\Rv$). This work has been extended down to the
group regime by  \cite{sun11} from \cha\ data; see also, e.g.,
\citet{fpo07} for earlier work based on \xmm.

Baryons in the outskirts bear the signature of the continuous
three-dimensional non-spherical accretion from surrounding filaments.
In this sense, access to the level of thermal pressure in the cluster
outer parts provides a neat way to assess the virialization degree
achieved, the thermodynamical state of the (pre-shocked) in-falling
material \citep{voi02,voi03}, etc. 

Observational constraints on cluster outskirts are
sparse and difficult to gather, although these regions encompass
most of the cluster volume. X-ray observations have recently
provided a first insight out to $\sim$$r_{200}$ of the
physical properties of the gas (Sections~\ref{sec:sb} and \ref{out:physics:Tr}).
\begin{figure*}
\center
\includegraphics[scale=1.,angle=0,keepaspectratio,width=0.5\textwidth]{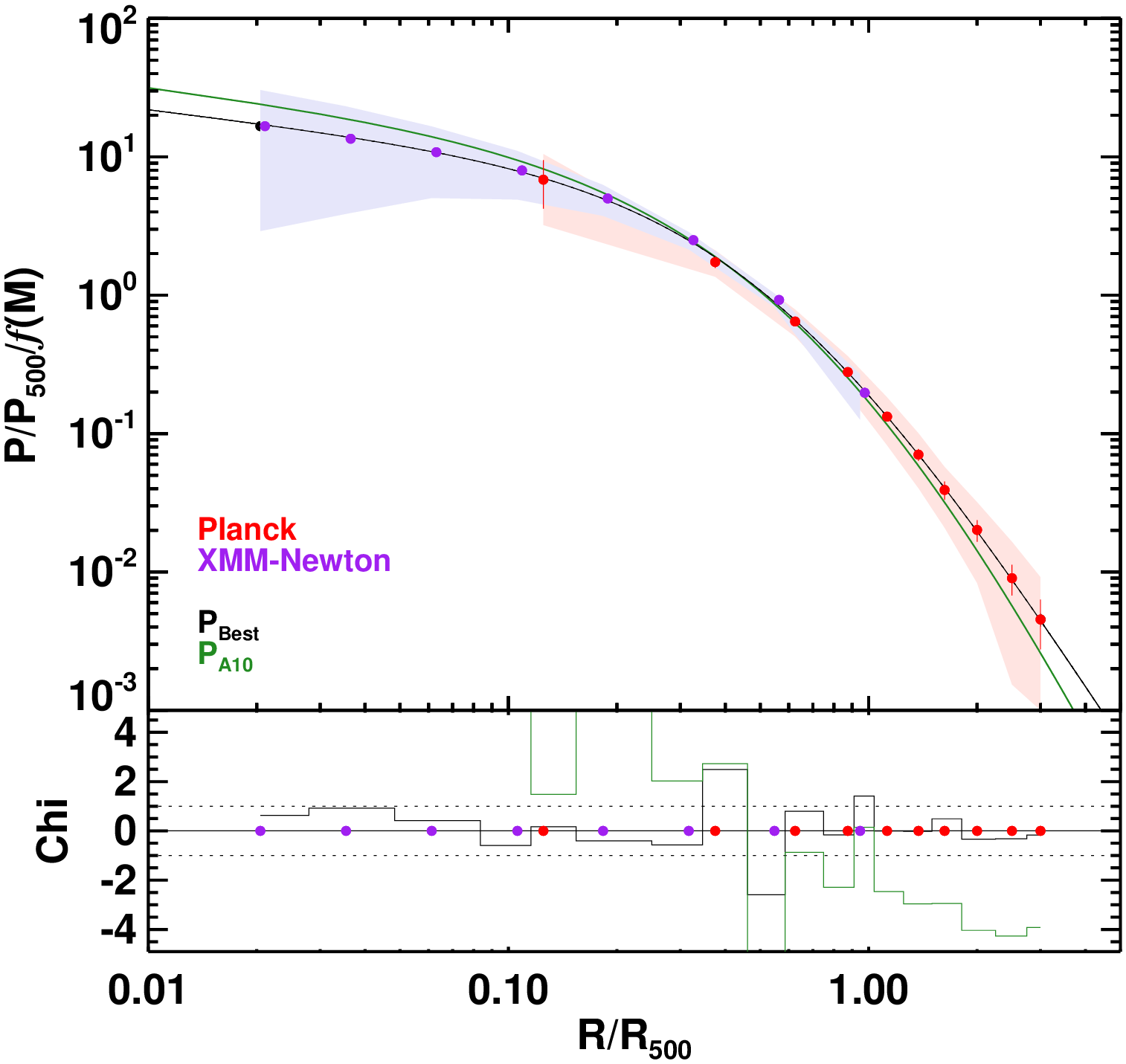}%
\includegraphics[scale=1.,angle=0,keepaspectratio,width=0.5\textwidth]{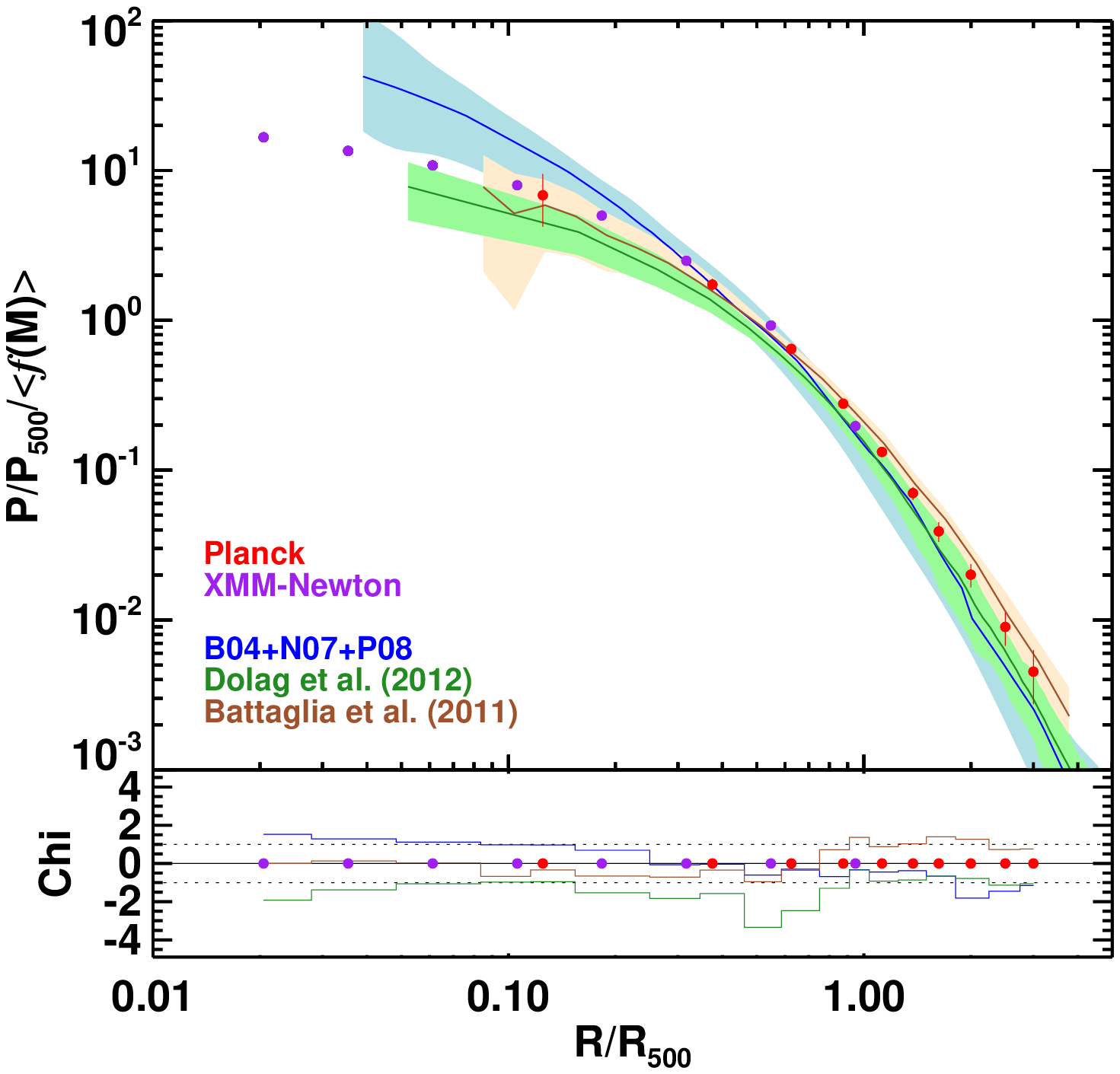}%
\caption{%
Both figures and caption reproduce
content from
\citet{paa13}, reprinted with permission.
Refer to this paper for a
description of the $f(M)$ factors.
Left: \pla\ pressure profile obtained from the 
average of the individual pressure profiles across the sample (red points) 
shown together with the stacked pressure profile derived from the \xmm\ data 
for the same sample (purple points). The dispersions about the SZ and X-ray 
profiles are depicted, respectively, by the red and purple shaded
areas. The best-fit GNFW profile is shown as a solid black line and
that of \citet{app10} as a solid 
green line. The lower panel shows the $\chi$ profile of these two
best-fit models 
taking into account the statistical errors and dispersion about the observed 
profile.
Right: Measured pressure profiles from \pla\ and \xmm. The three shaded areas mark the dispersions 
about the average profiles for three samples of simulated clusters: in blue, 
the simulations from \citet{borgani04}, \citet{nkv07} and \citet{piffaretti08},
which were used in 
\citet{app10} to derive the universal pressure profile together with the \rexcess\ data; 
in green, the simulated sample of clusters from
Dolag et al.\ (in prep.);
and in brown 
the simulated clusters from
\citet{bbp12}.
The corresponding average profiles for each set of 
simulations are plotted as solid lines using the same color scheme. In the 
bottom panel, the $\chi$ profiles between the observed profile and 
the simulated average profiles are presented, taking into account
their associated dispersion.  
}
\label{out:p:SZ}
\end{figure*}
The SZ effect has the potential to contribute greatly to the
discussion on cluster outskirts due to its linear dependence on
density and temperature. The radial pressure distributions of the first
SZ cluster samples have recently been presented based on observations
with, e.g., the SZ Array / Combined Array for Research in Millimeter-wave
Astronomy \citep[SZA/CARMA,][]{mbc09,bhb12},
APEX-SZ \citep{bnp09},
and the South Pole Telescope
\citep[SPT,][]{pba10}.
These studies confirmed that
the ICM  properties 
as seen by SZ and X-ray observations are consistent at least out to
$\Rv$.  
Noticeably, \citet{pba10} have obtained from 11 SPT clusters a
stacked SZ profile out to $\sim$(1.5--2)$\Rv$, where the shape of the
underlying pressure profile is compatible with the one given by
\citet{app10}.

A significant set of results has recently been published by
the \citet{paa13} based on the \pla\ nominal survey (i.e.,
14 months of survey).  
Making use of its full sky coverage over nine frequency bands
\citep{paa11I},
the \pla\ satellite maps all cluster
scales from its native resolution (5 to 10~arcmin at the SZ relevant
frequencies) to their outermost radii, even for nearby clusters. The
\Planck\ Collaboration adopted a statistical approach to derive a
stacked SZ profile from a sample of 62 nearby clusters detected in SZ
with high significance. These clusters were selected from the \Planck\
Early SZ (ESZ) sample and were previously used to investigate the total
integrated SZ flux and the SZ scaling relations 
\citep{paa11XI}. 

The statistical detection of the SZ signal extending out to 3$\Rv$
provides the first stringent observational constraints on the ICM
pressure
out to a density contrast of $\delta\,{\sim}\,50$--100.
Correcting for the instrument PSF and deprojecting the 2D profile
into a 3D one, the \pla\ collaboration derived the
underlying thermal pressure profile of the ICM. This observed
pressure profile is in excellent agreement with the one derived from
\xmm\ archive data for all 62 clusters, over the overlapping radial
range of (0.1--1)$\Rv$. The combined SZ and X-ray pressure profile
gives for the first time a comprehensive measurement of the
distribution of thermal pressure support in clusters from 0.01$\Rv$ out
to 3$\Rv$. Similarly to \citet{app10}, the \pla\ Collaboration has
derived an analytical representation assuming a GNFW profile
\citep[as formulated by][]{nkv07} with best-fit parameters 
$[P_0, c_{\rm 500}, \gamma, \alpha, \beta]= [6.41, 1.81, 0.31, 1.33,
4.13]$ (Fig.~\ref{out:p:SZ}), the outer slope ($\beta$) being slightly
shallower than the extrapolation based on simulations that was used by
\citet{app10}. Weak hints for even flatter outer slopes have been found
recently with Bolocam data of a sample of 45 clusters
\citep[][]{scm12}.
In the inner parts, there seems to be some tension between the \xmm\
data points and the \citet{app10} \xmm\ profile (although well within
the dispersion). As discussed in \citet{paa13}, this is likely caused
by differences in selection of the two samples (i.e., fraction of
dynamically perturbed versus relaxed clusters).

The observed joint \pla\ and \xmm\ pressure profile is also in
agreement within errors and dispersion over the whole radial range
with various sets of simulated clusters
(\citealt{borgani04,nkv07,piffaretti08,battaglia10}; Dolag et al.\ in
prep.). Within
the spread of predictions it matches best the numerical simulations
that implement AGN feedback, and it presents a slightly flatter
profile compared to most of the above theoretical expectations in the
outerparts.
As outlined in Section~\ref{out:intro}, the physics at play in
cluster outskirts is still to be understood. The constraints on gas
pressure brought from the \pla\ stacked SZ measurements have shed
light over a volume almost an order of magnitude larger than that
accessible from X-ray data of individual clusters alone. Further SZ
and X-ray measurements, especially for complete samples that are not
subject to possible archive biases, will continue to provide an
observational insight of cluster outskirts and serious constraints for
theoretical studies on issues such as gas clumping, departures from
hydrostatic equilibrium, contribution from non-thermal pressure, etc.,
as will be discussed in the following several Sections. SZ and X-ray
measurements complement each other nicely in this sense since they
primarily constrain different physical properties of the gas (SZ:
pressure, X-ray: density and temperature).

\subsubsection{Entropy profiles}
\label{out:physics:entropy}
\begin{figure}[thb]
\centering
\vspace{-1cm}
\hspace{-1cm}
\includegraphics[angle=0,width=0.54\textwidth]{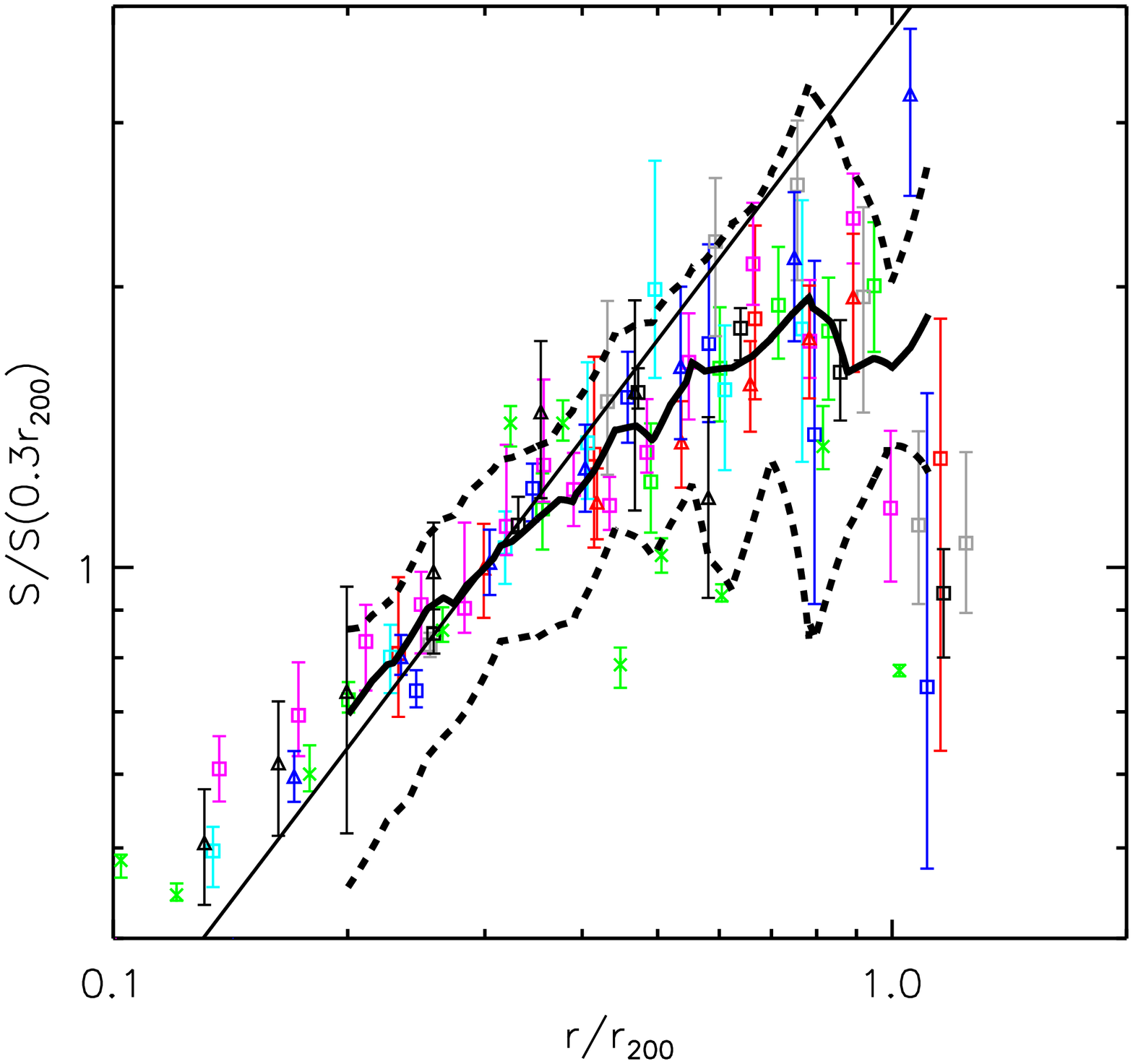}
\hspace{-0.4cm}
\includegraphics[angle=0,width=0.55\textwidth]{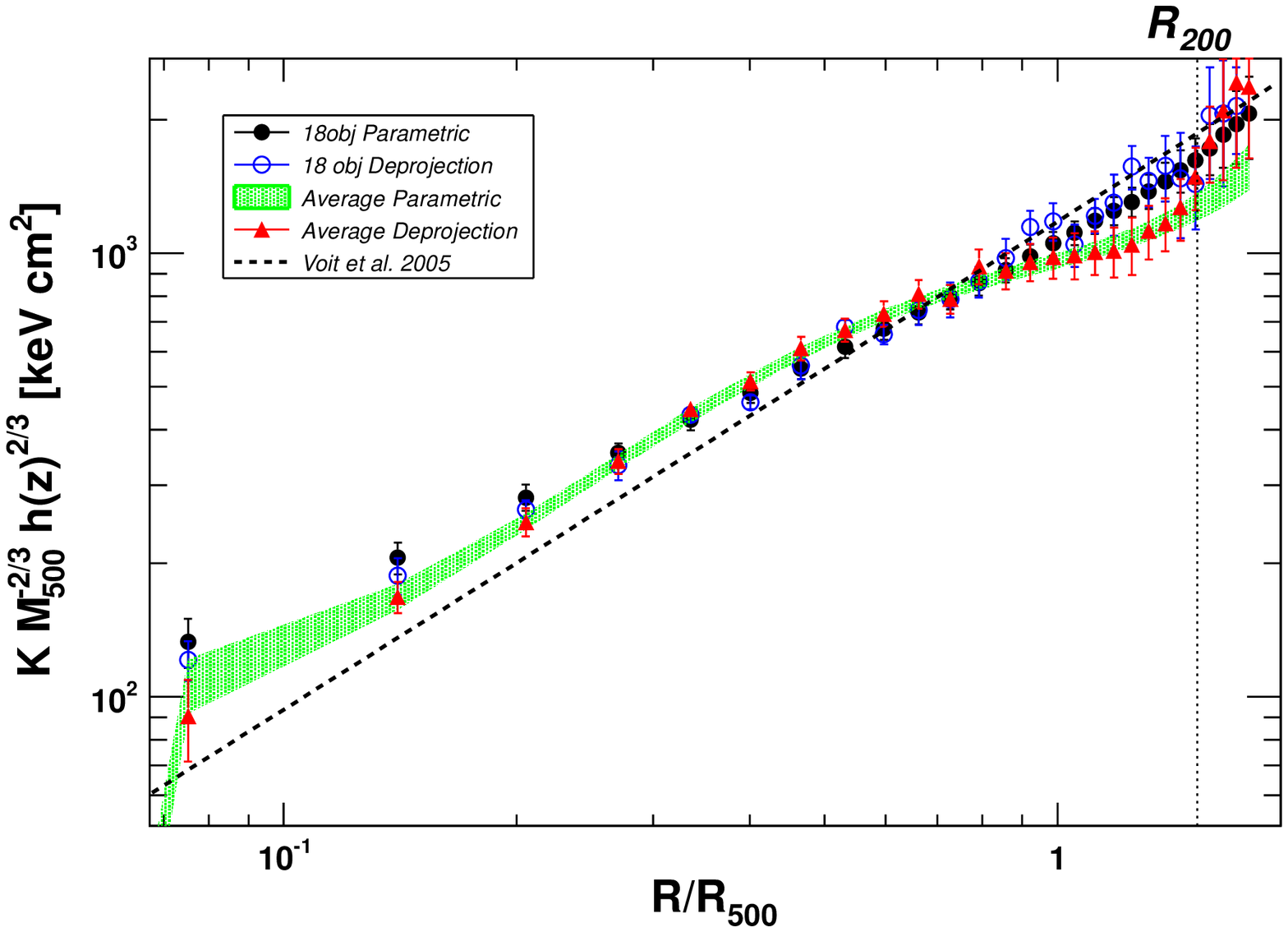}
\caption{Left: Scaled entropy profiles measured with \suz, apart from A1835
and A2204 (\cha, black boxes and triangles, respectively) and Virgo (\xmm, green crosses);
compiled by \citet[reproduced with permission]{wfs12c}. The black lines show the entropy profile and
scatter as derived by combining the \pla\ average pressure profile
(Fig.~\ref{out:p:SZ}) with the \ro\ average density profile
(Fig.~\ref{fig:out:physics:SB:Eckert}). The straight line represents an
$r^{1.1}$ power-law normalized to 1 at 0.3$r_{200}$.
Right: Scaled entropy profiles determined by \citet[reproduced with
permission]{emv13}. The straight dotted line again has a slope of
$r^{1.1}$ but is normalized using the definition of \citet{vo05}.
For the green band and the red triangles, the same data (average
\pla\ and \ro\ profiles) as on the left are used. The two symbols
represent different analysis methods (density determination through
fitting a function to the emission measure profile and through direct
deprojection, respectively). Also shown (circles) are the results from
determining individual entropy profiles for 18 clusters in common
between the \pla\ and \ro\ samples.
}
\label{out:entropy:walker}
\end{figure}
Often, ``entropy'' is defined in this field as $K\equiv\kb\tg/\nel^{2/3}$;
i.e., the discussion can be kept short here since it can basically be
derived by combining Section~\ref{sec:sb} (density profiles) either
with Section \ref{out:physics:Tr} (temperature profiles) or with
Section~\ref{out:physics:P} (pressure profiles, through
$K=\pel/\nel^{5/3}$). Moreover, several
aspects of entropy have already been addressed in Section
\ref{sec:sb}.

Typically, the temperature profiles observed with \suz\
in cluster outskirts tend to be fairly steep while the density
profiles are often found to be less steep than expected (in particular
for \suz). This results in a flatter slope compared to the expected
entropy profile gradient $K\propto r^{1.1}$ \citep[e.g.,][]{vo05}
but note also exceptions to this, e.g., the fossil group
RXJ1159+5531 \citep{hbb12}.
This is illustrated in Fig.~\ref{out:entropy:walker} where the \suz\
data on the left show a clear drop off at large radii while this is
much less obvious from the results on the right, possibly due to
systematic differences between \ro\ and \suz\ density profiles,
different chosen normalizations for the $r^{1.1}$ power-law, or
differences in sample selection.
Additionally, a possible
difference in the way how the very same \ro\ \citep{eve12} and \pla\
\citep{paa13} data are combined in both works is suggested by the
presence (absence) of non-monotonicity in the average profiles on the
left (right). Interpretations also vary; while  
\citet{emv13} argue that the well-known central entropy excess extends out
further than previously thought, \citet{wfs12c} suggest that clumping
in cluster outskirts could be one (but not the only) important effect.
Since this review is concerned with cluster outskirts, we discuss
clumping and several other possible physical and technical
explanations in upcoming Sections, mostly in terms of effects on
temperature and density measurements because the primary focus of this
review is on mass profiles in cluster outskirts.

\subsubsection{Gas/baryon mass fraction}
\label{out:physics:bary}
The gas fraction, $\fg\equiv\mg/\mt$, in clusters typically increases
as a function of both mass and radius (e.g., \citealt{vkf06}; see
Section 7.4 in \citealt{r01} for a discussion of pre-\xmm\ and -\cha\
results).
Going out far enough into the outskirts, the cosmic mean baryon
fraction, e.g., as determined from measurements of the cosmic
microwave background radiation or the primordial
deuterium-to-hydrogen-ratio, should eventually be recovered. It would 
be interesting to measure the characteristic radius at which this
usually happens because this would help constrain the physical
processes relevant for depleting the inner cluster regions of baryons.
Moreover, cosmological applications involving the gas or baryon
fraction of clusters rely on precise estimates of the gas/baryon
depletion factor at a given radius and redshift
\citep[e.g.,][]{emt09,aem11}.

As discussed in Section~\ref{sec:sb},
\ro\ and \cha\ data as well as hydrodynamic simulations mostly indicate a
steepening of the gas density  profile with increasing radius up to
$\sim$$r_{500}$ and beyond, while even further out, \suz\ data seem to
favor a flattening. Taken at face 
value, gas mass fractions in excess of the cosmic mean baryon fraction
are sometimes implied. On the other hand, many different physical
effects, considered in the following Sections, could result in an
artificial trend by affecting either the gas mass or total mass
determination or both. Also, some challenges for measurements
are outlined in the technical Section~\ref{out:tech:X}.

\begin{figure}[htb]
\hspace{1cm}
\includegraphics[angle=0,width=0.8\textwidth]{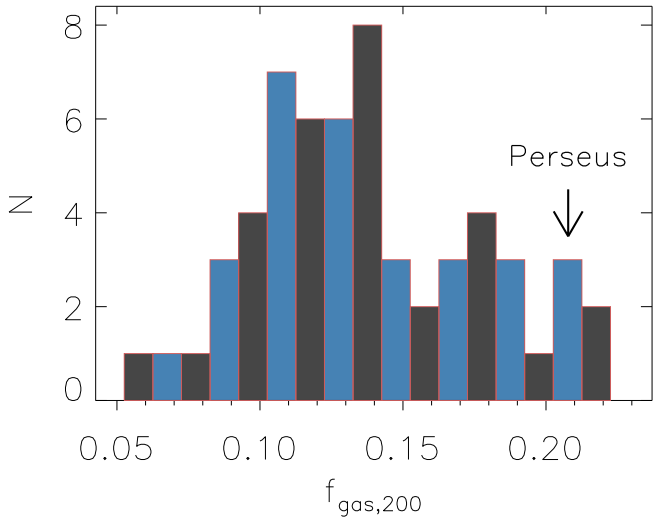}
\vspace{1cm}
\caption{Gas mass fractions within $r_{200}$ based on \ro\ \ps\ data.
Values taken from \citet{r01} and converted to $H_0=70$ km/s/Mpc. Only those 58
clusters are shown for which significant cluster
emission has been detected out to $r>0.8r_{200}$. The gas mass
fraction of the Perseus cluster -- in excellent agreement with the
recent \suz\ measurement \citep{sam11} -- is marked with an arrow.
WMAP9 data indicate a cosmic mean baryon fraction of around (16--17)\%
\citep{hlk13}.}
\label{out:fgas:r01}
\end{figure}

The \ro\ \ps\ was great for measuring gas density profiles out to very
large radius \citep[e.g.,][]{eve12,eem13} because of the very low
particle background level and large field-of-view. For instance,
\citet{r01} measured the gas mass fraction within $r_{200}$ for 106
clusters. For 58 out of these clusters, only a small or no
extrapolation of the measured surface brightness profile was
necessary. The resulting $\fgz$ histogram is shown in Fig.~\ref{out:fgas:r01}. 
One notes that the Perseus cluster, a prominent example for a high
\suz\ gas mass fraction and discussed in terms of gas clumping in
Section~\ref{sec:sb}, 
has one of the highest observed gas mass fractions of all 58 
clusters ($\fgz=0.21$, in excellent agreement with the \suz\
measurement).
While the statistical and systematic uncertainties of the total mass
measurements are large, resulting in a broadening and possibly a
shift of the distribution in Fig.~\ref{out:fgas:r01}, this could be an
indication that the Perseus cluster just happens to lie on the extreme
tail of the intrinsic cluster $\fgz$ distribution. Note that while the
intrinsic dispersion of the $\fgsm$ distribution is small \citep[for
relaxed clusters, e.g.,][]{aem11} it may be larger for $\fgz$ (also
considering the whole cluster population). 
Therefore, more \suz\  (and \emph{Astro-H}) and SZ observations of a
complete sample of clusters may be required for a full understanding
of the typical baryon fraction and possible gas clumping in cluster
outskirts.

\subsection{Structure formation in action}
\label{out:physics:structure}
\begin{figure}[htb]
\includegraphics[angle=270,width=1.0\textwidth]{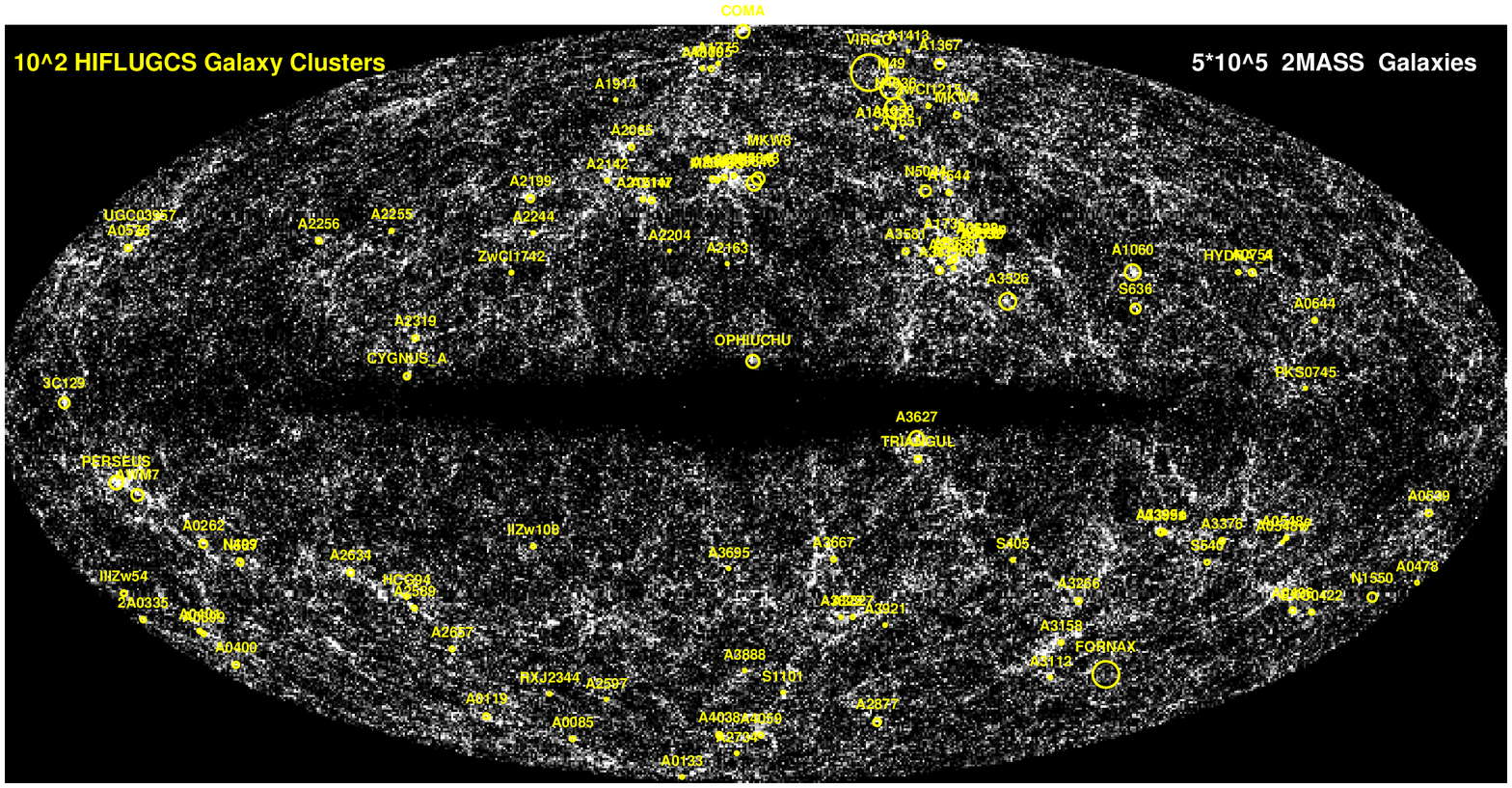}
\caption{Shown as yellow circles are $\sim$100 X-ray bright nearby
galaxy clusters \citep{rb01}. The circle radii indicate the apparent $r_{200}$.
In grey, the distribution of $\sim$0.5 million near-infrared selected
local galaxies from 2MASS is shown \citep{jcc00}. Both populations trace the same
filamentary structure and superclusters. The reason one can see the
same structures with a few X-ray galaxy clusters as with a much larger
number of galaxies is that clusters are strongly biased tracers of
the matter distribution; they are the lighthouses of structure in our
Universe.
Reprinted with permission from \citet{rsb03}.}  
\label{out:physics:structure:2MASS}
\end{figure}

Structure formation simulations show that
galaxy clusters grow through mergers and infall of matter clumps along
filaments \citep[e.g.,][]{bg01,swj05}. Observational evidence of the former is widespread 
\citep[e.g.,][]{mv07}. Filamentary structures have also been seen for
decades, e.g., in the galaxy and galaxy cluster distribution
(e.g., Fig.~\ref{out:physics:structure:2MASS}).
The expected filamentary \emph{gas} distribution \emph{between} clusters;
i.e., the WHIM, likely containing a
significant fraction of all local baryons \citep[e.g.,][]{co99,fp04},
still evades a very significant robust detection in X-rays
\citep[e.g.,][]{klt03,bl06,nme05,wmn06,rkp06,kwd06,b07,bzf09}. Significant
progress has been made at UV wavelengths (tracing the 
lower temperature phase of the WHIM); see \citet{kbs08} for detailed
review articles on the WHIM, in particular \citet{ppk08}.

We are currently entering an era when the region between the
well-observed cluster centers ($\lesssim r_{500}$) and the elusive
WHIM ($\gtrsim 3 r_{200}$) comes within reach of X-ray and SZ
measurements (Section~\ref{out:physics:profiles}).
This is a region where a lot of action related to structure formation
is expected to be happening. For instance,
Fig.~\ref{out:physics:structure:simu} shows infalling clumps of
matter. Typically, these higher density regions are predicted to have
a cooler temperature than their surroundings. Observational
confirmation is now needed to test this picture in detail.
On the other hand, if these clumps are
generally present but remain undetected, e.g., due to poor spatial
resolution, they will bias the X-ray gas density and
temperature measurements in the outskirts (this is described in
Sections~\ref{sec:sb} and \ref{out:physics:multiT}) and,
unless they are in pressure equilibrium with the ambient gas, they
will also bias interpretations of SZ measurements that assume a smooth
distribution. So, quantifying differences of X-ray and SZ
results may allow us to constrain unresolved
clumping \citep[e.g.,][]{gcr01,jbp08,kck12}.

More action in outskirts related to structure formation includes,
e.g., large Mach number accretion shocks and corresponding particle
acceleration. While these phenomena may be traced through the
non-thermal particle population in the radio, hard X-ray, and
$\gamma$-ray bands \citep[e.g.,][]{pes08}, also soft X-ray and SZ
measurements are required for a full understanding of the overall
plasma properties.

\subsection{Hydrostatic equilibrium}
\label{out:physics:hydro}
In the following Sections, we will discuss a few (non-) equilibrium
situations. For the X-ray and SZ mass determination, it is perhaps most
straightforward to see that hydrostatic equilibrium is important since
it is the basis for eq.~(\ref{out:eq:mass:X:hydro2}).

For major cluster mergers, it is obvious that the assumption of
hydrostatic equilibrium is not a good one. Moreover, simulations
suggest that also in less disturbed clusters, turbulence and bulk
motions, e.g., due to infalling clumps, may disrupt hydrostatic
equilibrium at some level \citep[e.g.,][]{nvk07}. This appears to be
more significant the further out one goes in terms of fraction of
overdensity radius \citep[e.g.,][]{lkn09,mrm10}. See also \citet{skk13} for
an alternative interpretation of hydrostatic equilibrium biases as
acceleration term in the Euler equations.

Observationally, direct measurements of/upper limits on ICM turbulent
velocites and bulk motions are currently limited to cluster
cores or to merging subclusters (e.g., \citealt{sfs11}, using the
\xmm\ Reflection Grating Spectrometer (RGS); \citealt{smi08,stn09,thu11}, using \suz\ CCDs).
Claims of an observational detection of such motions in the Centaurus
cluster have been made using \as\ and \cha\ data
\citep[][]{db01,db05,db06} and refuted \citep[][]{off07}
using \suz\ data.
A robust direct measurement of turbulence may need to await future
X-ray missions, like the upcoming \emph{Astro-H} carrying high
spectral resolution micro-calorimeter arrays \citep[e.g.,][]{zck12}.
To reach cluster outskirts, even larger effective areas will be
required, possibly provided through an envisaged \athena-like mission
\citep[e.g.,][]{bbd12,n13}.

In cluster centers, gas ``sloshing'' is widespread, resulting in
spiral-like patterns in the surface brightness distribution
\citep[e.g.,][for a review]{mv07}. It has been suggested based on \ro,
\xmm, and \suz\ observations of the Perseus cluster that features
produced by such motions could extend even out to radii approaching
the virial radius \citep{swu12}.

Another highly sensitive probe of the physical state of the ICM in
cluster outskirts,
in particular turbulent pressure support in high-$z$ ($z\gtrsim 0.5$)
objects,
can be through the Sunyaev-Zel'dovich effect angular
power spectrum, which is the integrated signal from all the unresolved
clusters in the sky \citep{komatsu99,komatsu02}. The SZ
power spectrum is measured as a foreground component of 
the CMB signal with the same spectral dependence as the SZ effect.
Half of the SZ power comes from low-mass clusters and groups ($M_{200}
< 2\times 10^{14} M_{\odot}$), and half of it also comes from
high-redshift ($z>0.5$) systems \citep{komatsu02}.  Therefore, SZ power
might
provide the only method to study these otherwise un-observable
systems with low mass {\it and} high redshift, although only through
their summed contributions.  

Unlike X-ray brightness, the integrated SZ signal of a cluster carries
a significant weight from the volume outside of $r_{500}$, so the
prediction of SZ power is strongly dependent on the pressure profile
in the outskirts. The difficulty lies in detecting the SZ power
itself, which is not the dominant source of foreground anisotropy at
any frequency or angular scale.
Recent SPT  measurements have
constrained the tSZ power at a low amplitude: $3.5 \pm 1.0$ $\mu
\mathrm{K}^2$ at ${\ell} = 3000$ \citep{shirokoff11}. This is lower
than the prediction from X-ray derived cluster pressure models
\citep{shaw10,efstathiou12}, but the difference can be the
result of many different effects, like pressure support from gas bulk
motions, AGN and star-formation feedback, or other non self-similar
evolution of the ICM. Future CMB measurements with better frequency
coverage and angular resolutions are expected to place a tighter
constraint on the thermal SZ power and break some of these
degeneracies. The cosmology dependence of the SZ power can also be
nailed down by other methods, for example an accurate value of
$\sigma_8$ will reduce its degeneracy with the uncertain gas physics.

Another possible, indirect route to determining how strongly X-ray and
SZ hydrostatic mass determinations are affected by turbulence, bulk
motions, and other gas physics effects is through comparison to weak
lensing measurements, which do not rely on the assumption of
hydrostatic equilibrium, in principle. This has been done extensively
for the inner parts of clusters ($\lesssim$$r_{500}$, e.g., Ettori et
al.\ and Hoekstra et al., this volume, for reviews) and it may be
feasible now also for cluster outskirts, although also the weak
lensing mass reconstruction accuracy is more limited in the outer
parts (e.g., Section~\ref{out:tech:WL}).

In addition to merger- or accretion-induced bulk motions, also
\emph{convection} may occur in the ICM. While a gas temperature gradient
in itself does \emph{not} imply a violation of hydrostatic
equilibrium, convection should occur
if the specific entropy decreases significantly with increasing
radius \citep[for a non-magnetic ICM, e.g.,][Sections 4 and V.D.6, respectively]{ll91,s86};
i.e., if
\begin{equation}
\frac{\dif s}{\dif r}<0\,.
\end{equation}
With 
\begin{equation}
s\propto\ln\frac{\tg}{\rog^{\gamma-1}}\,
\end{equation}
and $\gamma=5/3$ this condition becomes
\begin{equation}
-\frac{\dif\ln\tg}{\dif r}>-\frac{2}{3}\frac{\dif\ln\rog}{\dif r}\,.
\end{equation}
Typically, density gradients in clusters are much steeper than
temperature gradients; so, generally no convection is expected.
However, in cluster outskirts, there are some indications that density
profiles may flatten (Section~\ref{sec:sb}), temperature profiles may
steepen (Section~\ref{out:physics:Tr}), and entropy profiles may turn
over (Section~\ref{out:physics:entropy}). For instance,
convectional instability was considered early on for A2163 by \citet{mmi96}.
In the presence of magnetic fields, however, the situation may be more
complex due to other possible instabilities, in particular the
magnetothermal instability \citep[MTI,][]{ba00} and the
heat-flux-driven buoyancy instability \citep[HBI,][]{q08}.

\subsection{Thermal equilibrium, $\tel = \tg$?}
\label{out:physics:TE}

With several tens of millions of degrees the ICM is hot. An important
heating mechanism is shock heating, either in cluster mergers or when
infalling gas passes through an accretion shock.
The ICM consists of electrons and ions. Most of the ions are
protons, so for simplicity we just use the term protons now. In an
accretion shock, primarily protons should be heated since they carry
most of the kinetic energy. After leaving the shock front, the time
scales for protons and electrons to reach Maxwellian velocity
distributions are short; i.e., they will both settle into thermal
equilibrium quickly, albeit at different temperatures. The time scale
is \citep[eq.~5-26]{s56}
\begin{equation}
 t_{\rm eq}(\mathrm{x,x})\approx 1.13\times10^{7}\,\mathrm{yr}\, Z_{\mathrm{x}}^{-4}
 \left(\frac{m_{\mathrm{x}}}{\mpr}\right)^{1/2}
 \left(\frac{\kb T_{\mathrm{x}}}{{\rm 10\,keV}}\right)^{\frac{3}{2}}
 \left(\frac{n_{\mathrm{x}}}{10^{-3}\, {\rm cm^{-3}}}\right)^{-1}
 \left(\frac{\ln{\Lambda_{\mathrm{x}}}}{40}\right)^{-1}\,,
\label{out:eq:physics:teq0a}
\end{equation}
where $Z_{\mathrm{x}}$ is the particle species charge, $m_{\mathrm{x}}$ its mass,
$T_{\mathrm{x}}$ its temperature after reaching equilibrium,
$n_{\mathrm{x}}$ its number density, and $\ln{\Lambda_{\mathrm{x}}}$
its Coulomb logarithm. 
So, in typical cluster regions, this is about $t_{\rm
eq}(\mathrm{p,p})\approx$ 10 Myr for protons and
for electrons about a factor of 43 faster,
\begin{equation}
 t_{\rm eq}(\mathrm{e,e})\approx
 \left(\frac{\mel}{\mpr}\right)^{1/2}t_{\rm eq}(\mathrm{p,p})
 \approx 2.64\times10^{5}\,\mathrm{yr}
 \left(\frac{\kb \tel}{{\rm 10\,keV}}\right)^{\frac{3}{2}}
 \left(\frac{\nel}{10^{-3}\, {\rm cm^{-3}}}\right)^{-1}
 \left(\frac{\ln{\Lambda_{\mathrm{e}}}}{40}\right)^{-1}\,.
\label{out:eq:physics:teq0b}
\end{equation}
If the electrons are indeed heated less efficiently in the shock than
the protons, resulting in $\tel<T_{\mathrm{p}}$ after equilibration,
they will reach equilibrium even faster than implied by the square
root of their mass ratios. 

What takes longer then is raising the electron temperature to the
proton temperature. Assuming this process proceeds primarily through
Coulomb collisions, the corresponding equipartition time scale is
very roughly a factor of 43 longer than the proton equilibration
timescale,
\begin{equation}
 t_{\rm eq}(\mathrm{e,p})\approx
 \left(\frac{\mel}{\mpr}\right)^{-1/2}t_{\rm eq}(\mathrm{p,p})\,,
\label{out:eq:physics:teq0c}
\end{equation}
and is given by \citep[eq.~5-31]{s56} 
\begin{equation}
 t_{\rm eq}(\mathrm{e,p})
 \approx 2.51\times10^{8}\,\mathrm{yr}
 \left(\frac{\kb \tel}{{\rm 10\,keV}}\right)^{\frac{3}{2}}
 \left(\frac{\npr}{10^{-3}\, {\rm cm^{-3}}}\right)^{-1}
 \left(\frac{\ln{\Lambda_{\mathrm{e}}}}{40}\right)^{-1}
 \left(1+\frac{T_{\mathrm{p}}}{\tel}\frac{\mel}{\mpr}\right)^{\frac{3}{2}}\,.
\label{out:eq:physics:teq0d}
\end{equation}
The last two factors are close to 1, so they are ignored in
the following and we also assume $\nel\approx 1.2\npr$. In cluster
outskirts, both ICM densities and temperatures are significantly lower
than in the better observed inner regions. Since the densities drop
much more quickly, the net effect is an increase of the equipartition
timescale. It can come close to one Gyr or even more, 
\begin{equation}
 t_{\rm eq}(\mathrm{e,p})
 \approx 10^{9}\,\mathrm{yr}
 \left(\frac{\kb \tel}{{\rm 1\,keV}}\right)^{\frac{3}{2}}
 \left(\frac{\nel}{10^{-5}\, {\rm cm^{-3}}}\right)^{-1}\,.
\label{out:eq:physics:teq0e}
\end{equation}
So, there the X-ray-emitting electrons may have a cooler temperature
than the invisible protons, resulting in a steeper X-ray temperature
gradient towards the outskirts and, therefore, an underestimate of the
total mass (Fig.~\ref{out:mass:M-Tr}).

The thermal equilibrium/equipartition situation in cluster mergers and
cluster outskirts/WHIM has been
studied theoretically by many authors
\citep[e.g.,][]{fl97,cat98,ef98,t98,t99,ca04,yfh05,rn09,ws09,wsw10}.
It is worth mentioning that the
situation may differ between merger shocks (small Mach
numbers) and accretion shocks (large Mach numbers):
the electron heating efficiency (relative to the one for protons) in a
shock may be anticorrelated to the Mach number squared
\citep[][]{glr07} if other processes in addition to pure Coloumb
heating are considered, resulting in fast relative electron heating in
merger shocks and slow heating in accretion shocks. Indeed, for the
textbook merger shock in the bullet cluster, the equipartition time
scale seems about a factor of five shorter than implied by
eq.~(\ref{out:eq:physics:teq0d}) \citep[][]{mv07} assuming the \cha\ lower
temperature limits from \citet{m06} for the extremely hot post-shock gas to be
robust against systematic calibration uncertainties.

Observational results have also been obtained for cluster outskirt
measurements. For instance, the early \as\ work on Abell 2163 by
\citet{mmi96} triggered some of the theoretical studies mentioned
above. More recently, \suz\ data have been used to constrain the
equipartition state \citep[e.g.,][]{hhs10,ahi11}. With current
instruments, no strong direct constraints have been achieved, though.

Fig.~2 of \citet{rn09} implies that for unrelaxed and relaxed clusters
of similar temperature, the electron temperature in the outskirts
deviates more strongly from the ion temperature for unrelaxed
clusters. Naively, one might conclude from the similarity of
\suz\ profiles (Fig.~\ref{out:physics:Tr:Suz1a}) that no such trend is
observed; i.e., no strong evidence for significant non-equipartition.
However, a proper comparison would need to account for the 
possible difference of intrinsic temperature profiles between relaxed
und unrelaxed clusters, for the inhomogeneous definitions used to
classify clusters as un-/relaxed, and, since the equipartition
timescale depends on temperature (eq.~\ref{out:eq:physics:teq0e}),
for the possible difference in the temperature distributions of the
subsamples.

Since there are not only protons but also ions in the ICM, some of
which are strong X-ray line emitters, there is hope to measure the ion
temperature directly from the line width. This might be feasible with
upcoming missions. For instance, the micro-calorimeter array aboard
\emph{Astro-H} will have an energy resolution of $<$7 eV
\citep[e.g.,][]{tmk12},
possibly just sufficient to detect a thermal broadening of
$\sim$5 eV \citep[e.g.,][in the absence of other more dominant line broadening
effects like bulk motions and turbulence]{rcs08}
but only in the X-ray
bright central regions where 
the collisional equilibration time scale is shorter than the typical
time since the last major merger in any case. An
\athena-like mission could have enough effective area and energy
resolution to measure the
thermal broadening in nearby cluster outskirts, where the ion and
electron temperatures possibly deviate.

\subsection{Ionization equilibrium, $\tx=\tel$?}
\label{out:physics:ioniz}
A very rough timescale for an astrophysical plasma to reach
collisional ionization equilibrium (CIE) is given by
\begin{equation}
 t_{\rm ion-eq}\approx 10^{12}\,
 {\rm s}\left(\frac{\nel}{{\rm
 cm^{-3}}}\right)^{-1}
\label{out:eq:physics:ioneq0}
\end{equation}
with some dependence on the considered element and temperature
\citep{sh10}. Most ions in the ICM are not in an excited state because
radiative deexcitation is very fast.

Recasting (\ref{out:eq:physics:ioneq0}) into units typical of cluster
outskirts yields 
\begin{equation}
 t_{\rm ion-eq}\approx 3\times10^9\,
 {\rm yr}\left(\frac{\nel}{10^{-5}\, {\rm
 cm^{-3}}}\right)^{-1}\,.
\label{out:eq:physics:ioneq}
\end{equation}
Therefore, on the order of a Gyr may be required in cluster outskirts
-- long enough to possibly result in an important, measureable effect
for some clusters.
\begin{figure}[thb]
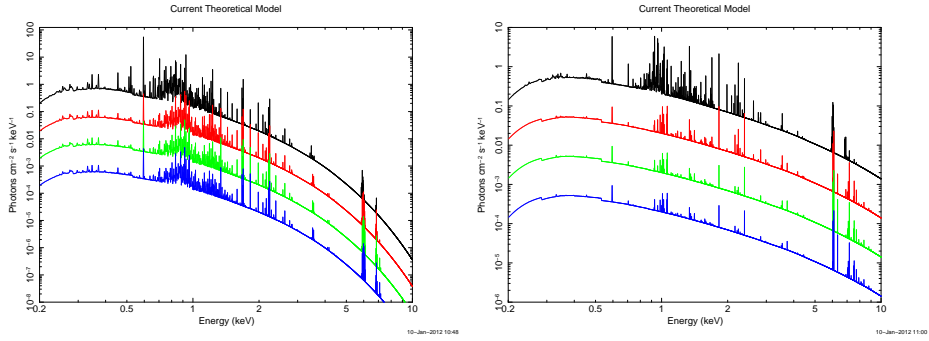

\includegraphics[angle=270,width=0.5\textwidth]{out_ps/nei_1kev-model.ps}
\includegraphics[angle=270,width=0.5\textwidth]{out_ps/nei_5kev-model.ps}
\caption{Model spectra of 1 keV (left) and 5 keV (right) plasmas for
different values of the ionization timescale.
Black: $\tau = 8\times 10^{10}$ s/cm$^3$,
red: $\tau = 3\times 10^{11}$ s/cm$^3$,
green: $\tau = 8\times 10^{11}$ s/cm$^3$, and
blue: $\tau = 3\times 10^{12}$ s/cm$^3$.
The normalizations are scaled for visualization purposes.
Other parameter values are $\nhy=3\times10^{20}$ cm$^{-2}$, metal
abundance = 0.2 solar, redshift = 0.1.}
\label{out:physics:nei_1}
\end{figure}

\begin{figure}[htb]
\hspace{2cm}
\includegraphics[angle=270,width=0.7\textwidth]{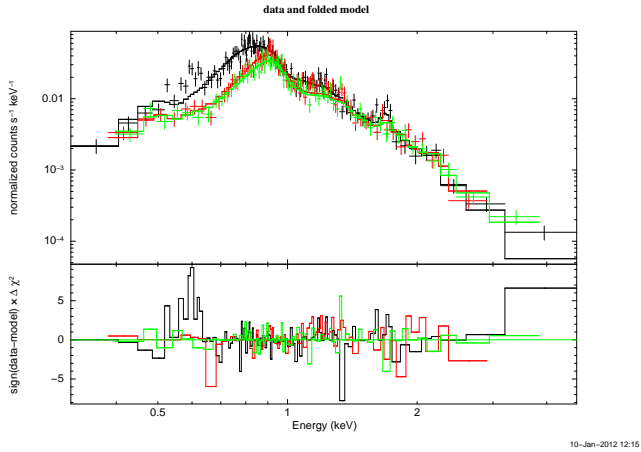}
\caption{Simulated \suz\ observations of a single-temperature
ICM ($\kb T=1$ keV) that is not fully in collisional ionization equilibrium
(2,000--3,000 source photons).
Black: $\tau = 8\times 10^{10}$ s/cm$^3$,
red: $\tau = 3\times 10^{11}$ s/cm$^3$, and
green: $\tau = 3\times 10^{12}$ s/cm$^3$.
} 
\label{out:physics:nei_2}
\end{figure}

\begin{figure}[htb]
\hspace{-0.5cm}
\includegraphics[angle=0,width=0.55\textwidth]{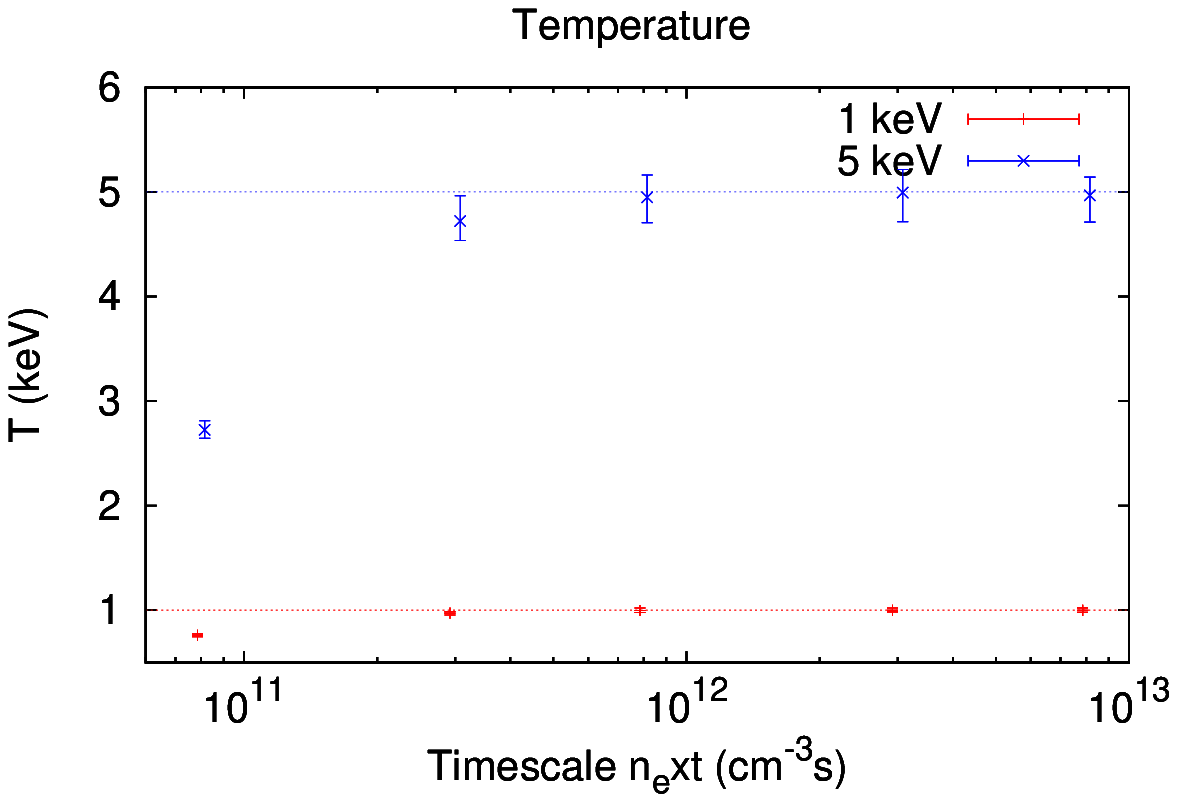}
\hspace{-0.5cm}
\includegraphics[angle=0,width=0.55\textwidth]{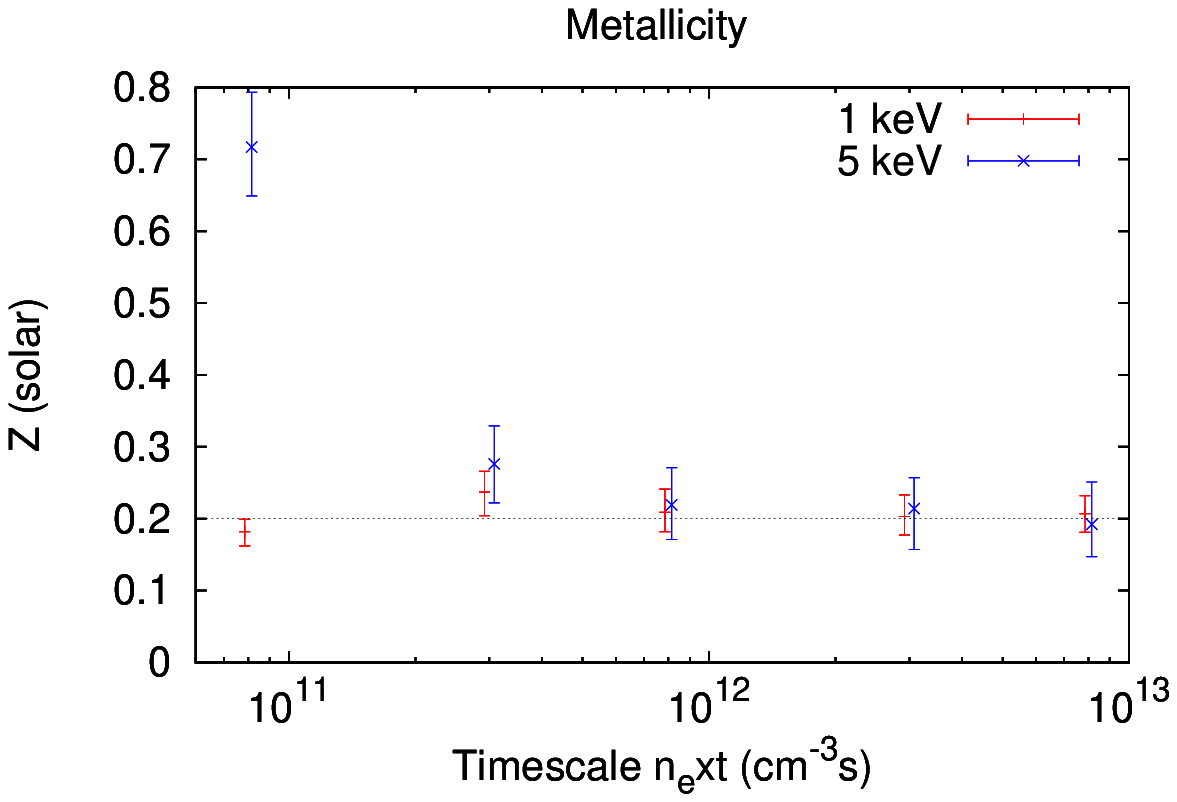}
\caption{Left: Best-fit temperatures when incorrectly assuming collisional
ionization equilibrium as a function of $\tau$; for simulated
\suz\ data, assuming temperatures of 1 keV and 5 keV. Other parameter
values as in Fig.~\ref{out:physics:nei_1}.
Right: Metallicity bias under the same assumptions.
} 
\label{out:physics:nei_2a}
\end{figure}

For a given temperature, how do plasma spectra differ depending on
the ionization state?
Fig.~\ref{out:physics:nei_1} shows that stronger low energy emission
lines are present when the ICM has not yet achieved CIE ($\tau\lesssim
10^{12}$ s/cm$^3$). In typical 
observed spectra with CCD-type energy resolution, this results
in a shift of the peak position of the $\sim$1 keV emission
line complex towards lower energies, especially for a low
temperature plasma (Fig.~\ref{out:physics:nei_2}). This indicates that
one might obtain a biased temperature if 
one wrongly assumes CIE in the fitting process. Due to this shift, as
illustrated in Fig.~\ref{out:physics:multiT:line_complex}, a bias
towards \emph{lower} temperatures may then be expected. Indeed, this
is the case as Fig.~\ref{out:physics:nei_2a} 
(left) demonstrates. In the most extreme non-equilibrium ionization
(NIE) cases simulated, temperatures are underestimated by a factor of
$\sim$2.
Since this bias, if present, will likely be larger for larger radius
where the density is lower, it will translate into a too steep
temperature profile, resulting in an underestimated total mass
(Fig.~\ref{out:mass:M-Tr}).

In Section~\ref{out:chem}, the importance of cluster chemistry
especially in outskirts is described. Here, we show in
Fig.~\ref{out:physics:nei_2a} (right) how the metal abundance
determination is biased if CIE is assumed when it is not yet
established. While there is no strong bias for low temperature
clusters, the metallicity gets severely overestimated for hot clusters
in the most extreme case ($\tau=8\times 10^{10}$ s/cm$^3$). This is
due to enhanced line emission in the NIE case.

Non-equilibrium ionization effects in low density plasmas have been
studied theoretically in detail by several works
\citep[e.g.,][]{ys06,ay10,wsj11}. Some authors have also tried to
estimate these effects observationally
\citep[e.g.,][]{fhn08,fsn10,apk12}. 
Future high-spectral resolution instruments like the upcoming
\emph{Astro-H} or an envisaged \athena-like mission may be able to set
tight constraints on ionization states in the outskirts of galaxy
clusters and thereby constrain merger timescales.

\subsection{How to disentangle multitemperature structure,
$\tx=\tel$?} 
\label{out:physics:multiT}
The widespread presence of temperature gradients in central
\citep[e.g.,][]{asf01,hmr09} and intermediate
\citep[e.g.,][]{vmm05,pbc07} cluster regions, as well as in
cluster outskirts (Section~\ref{out:physics:Tr}) shows that the
ICM is not isothermal. Moreover, gas temperature maps show that even
at a given radius, a wide distribution of temperatures can exist
\citep[e.g.,][]{rsk04,ma09,rcn10,lsk11}, possibly depending on dynamical state
\citep[e.g.,][]{zrf09}. Additionally, in cluster outskirts, emission
from new matter infalling along filaments and possible cooler clumps
might become more important (Fig.~\ref{out:physics:structure:simu}).

\begin{figure}[htb]
\centering
\includegraphics[angle=270,width=0.8\textwidth]{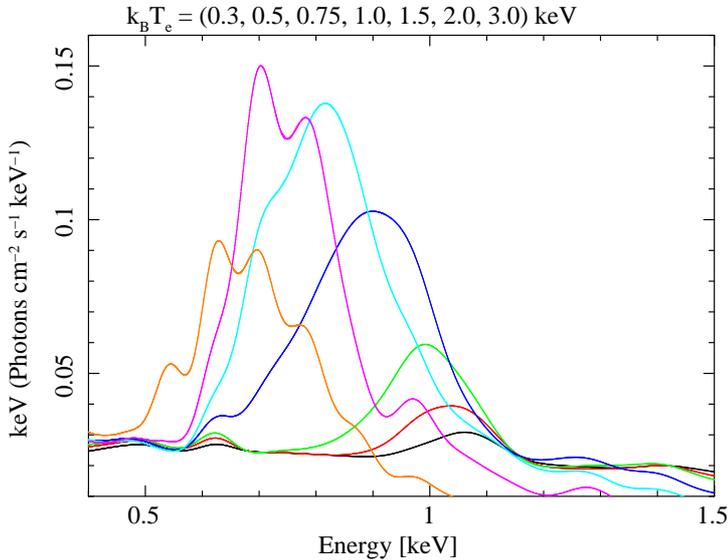}
\caption{The low-energy emission line complex (predominantly Fe and O)
as a function of the intracluster electron temperature assuming a metal
abundance of 0.3 solar, a neutral hydrogen column density of
$N_{\mathrm{H}}=3\times10^{20}\ \mathrm{cm}^{-2}$, and a redshift of $z=0.05$. 
The energy resolution approximates that of current X-ray CCDs. All
spectra assume the same electron density distribution. Notice how the
emission peak shifts to higher energies for increasing temperatures,
from orange (0.3 keV) to black (3 keV).}
\label{out:physics:multiT:line_complex}
\end{figure}

Therefore, in a given spectral extraction region (say, an annulus in
cluster outskirts), emission from gas at multiple temperatures may be
present. The data quality (e.g., number of source photons,
signal-to-noise ratio, energy resolution) on the other hand may not be
sufficient to constrain a multitemperature model. A single temperature
model will then have to be fitted to the multitemperature 
spectrum. What will the best-fit temperature be? It has been shown
that the answer depends on the used X-ray mirror/filter/detector
system \citep[e.g.,][]{me01,mrm04,rmb05,v05}. This is easy to
understand: 
the electron temperature, $\tel$, is usually estimated from fitting a
model, convolved with the instrumental response, to an observed
spectrum. The most temperature-sensitive features in an observed
spectrum with CCD-like energy resolution are (recall
Fig.~\ref{out:mass:apec}) the exponential bremsstrahlung cutoff at
high energies, the slope of the bremsstrahlung emission at
intermediate energies, and the location of the emission line complex
at low energies (Fig.~\ref{out:physics:multiT:line_complex}). An
instrument with more effective area at low energies compared to high
energies -- relative to another instrument -- is more sensitive to the
low energy features and, therefore, will typically give rise to a
lower single temperature estimate.\footnote{Note
that the best-fit single temperature depends on other factors as
well, especially on: (i) the metal abundance because the higher the
abundance the stronger the low-energy emission line complex, (ii) the
hydrogen column density because a higher column density has the same
effect as a decreased effective area at low energies, and (iii) the background
characteristics because, e.g., an instrument with higher particle
background has a poorer signal-to-noise ratio at high energies, which
is similar to a decreased effective area at high energies.}

\begin{figure}[htb]
\includegraphics[angle=0,width=0.55\textwidth]{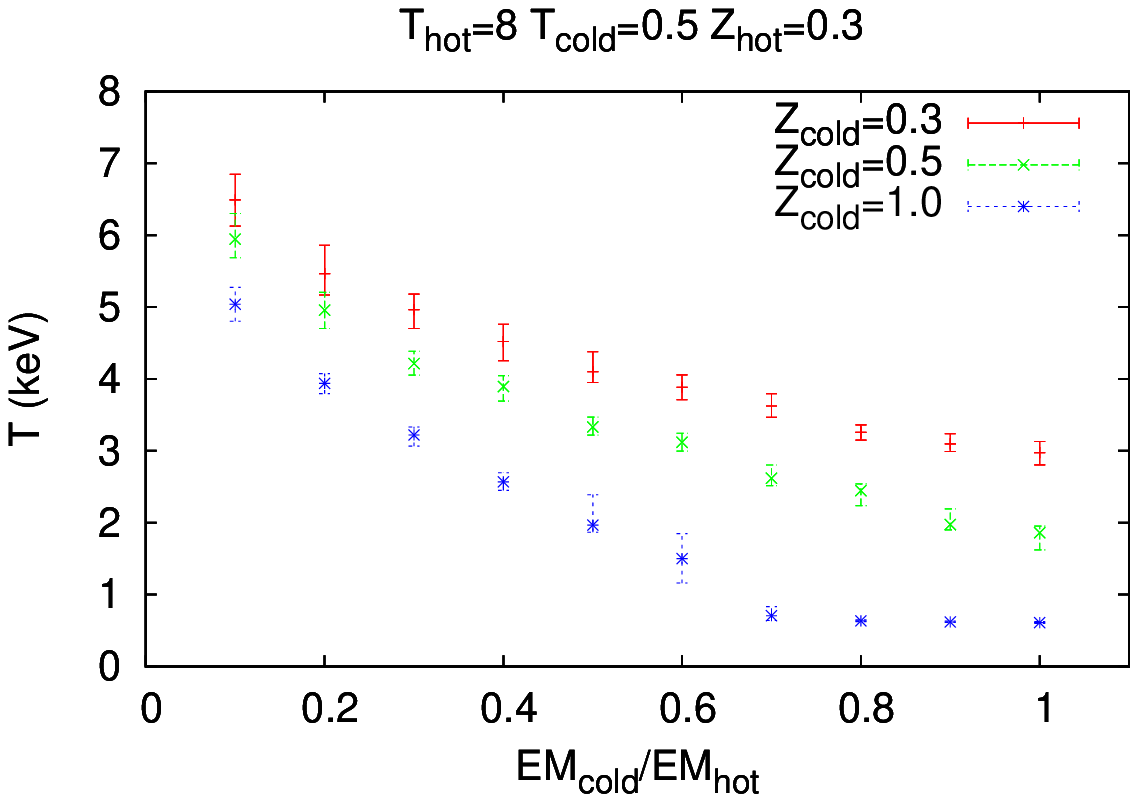}
\hspace{-0.5cm}
\includegraphics[angle=0,width=0.55\textwidth]{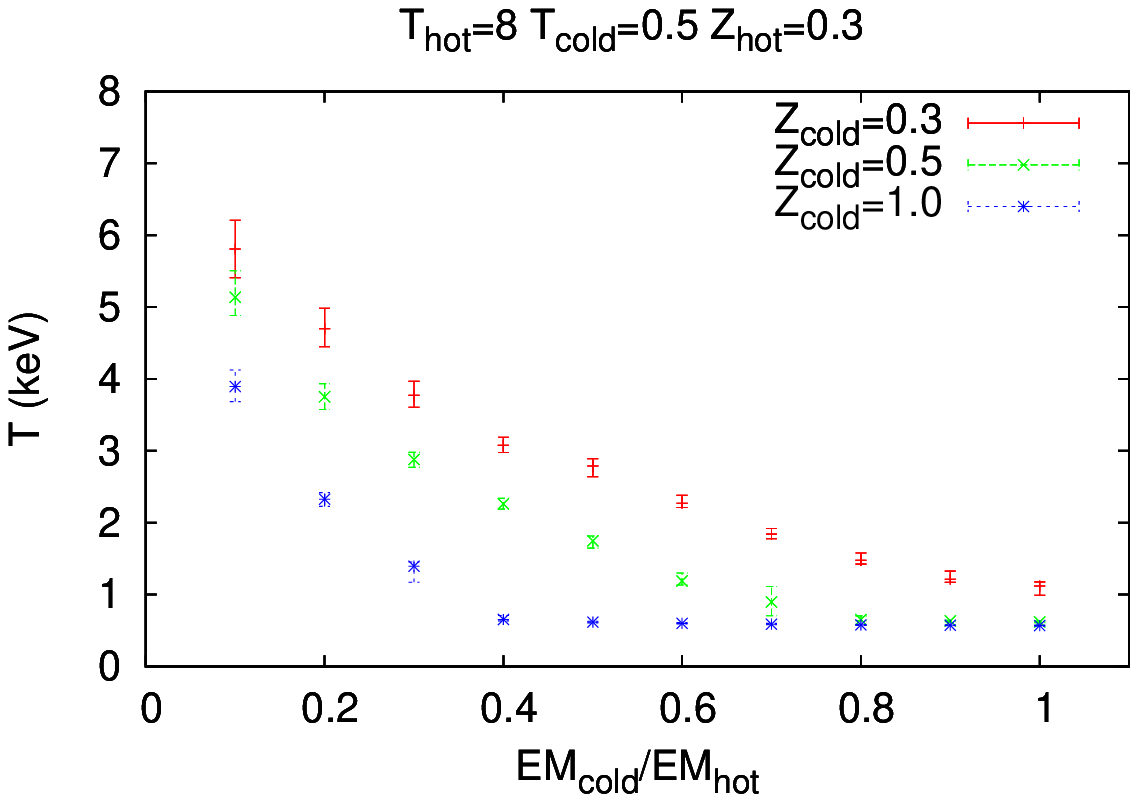}
\includegraphics[angle=0,width=0.55\textwidth]{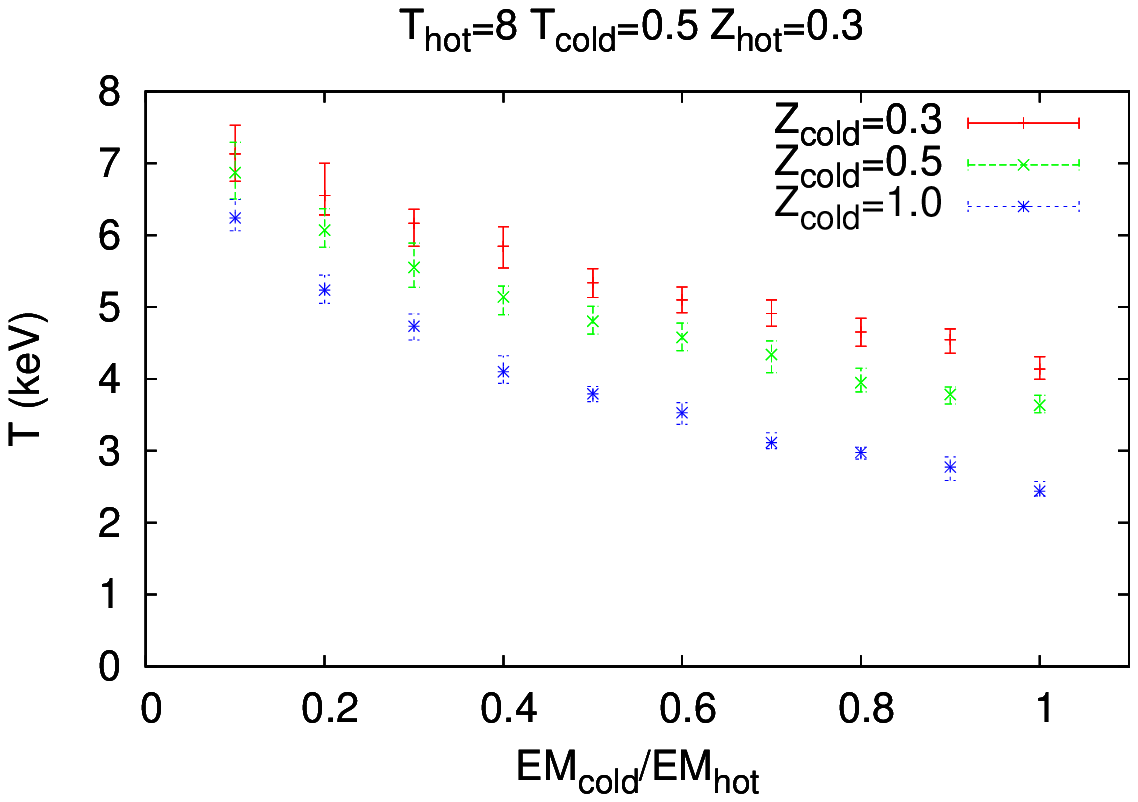}
\hspace{-0.5cm}
\includegraphics[angle=0,width=0.55\textwidth]{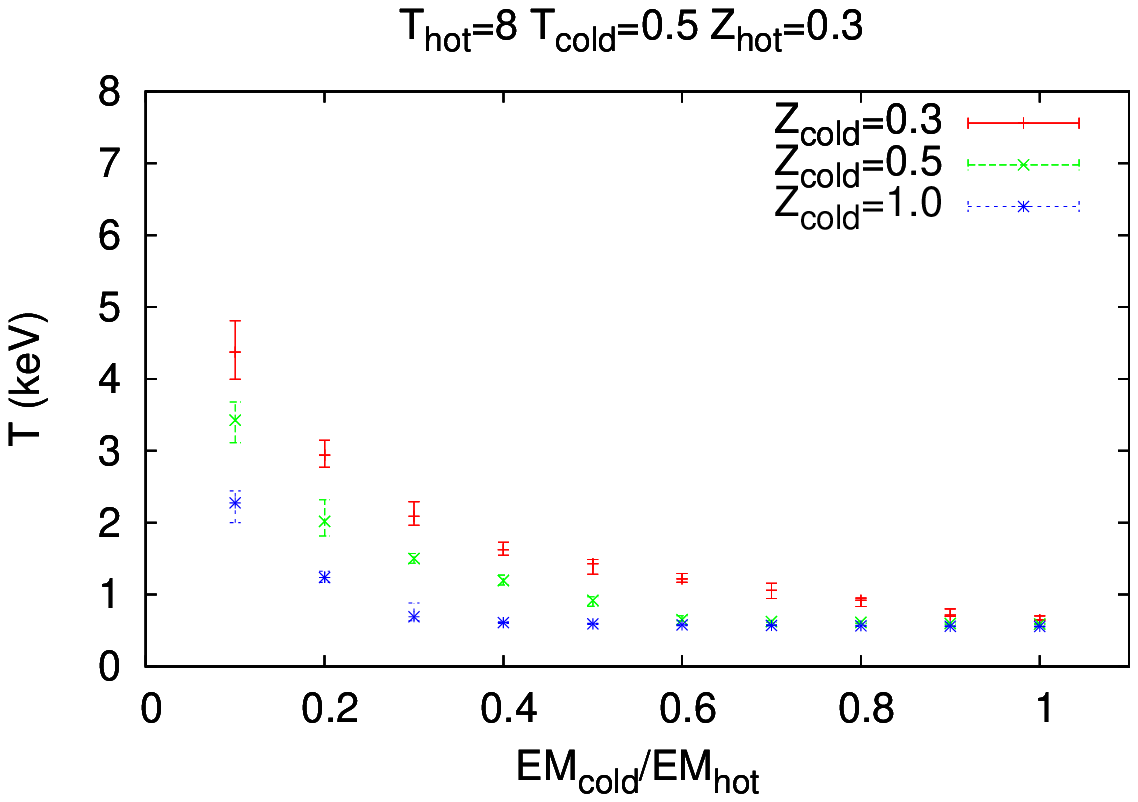}
\caption{Best-fit single temperature of two-temperature plasma
emission as a function of emission measure ratio of the two components,
for different X-ray instruments: \cha\ ACIS-I (upper left), 
\xmm\ EPIC-pn (upper right), \suz\ XIS-FI (lower left), and \rosi\
(lower right). See text for details.}
\label{out:physics:multiT:dets}
\end{figure}

An illustration for the simple case of a two-temperature plasma is
shown in Fig.~\ref{out:physics:multiT:dets} for \cha, \xmm, \suz, and
\rosi. Spectral data are simulated for these instruments assuming emission from plasma at two different temperatures (0.5 keV and 8 keV), with varying contributions (emission measure ratios -- the x-axis) and assuming three different metallicities (0.3, 0.5, and 1 times solar) for the cooler component (the hotter component is always assumed to have a metallicity of 0.3 solar). Then single temperature (and single metallicity) models are fit to the simulated data and the best-fit temperatures are shown in the plots.
The median values of 100
realizations are shown and the 68\% errors are taken from the 
distributions.
The values of parameters not
shown are: hydrogen column density $\nhy=3\times10^{20}$ cm$^{-2}$,
redshift $z=0.1$, number of source photons $=10^4$, and the energy
range used for fitting is (0.5--8.0) keV.
For these illustrations, the source emission is assumed
to dominate over the background at all energies, so no background is
included in the simulations.

The plots show how the best-fit single
temperature decreases with increasing emission measure 
ratio of cold and hot component (from 10\% to equal emission
measure). It also becomes clear how the temperature decreases with
increasing metallicity because the number of low-energy photons
constraining the fit increases in this case. Moreover, the plots
clearly reflect the different sensitivities of the different
instruments, e.g., the relatively hard \suz\ XIS-FI typically returns
much higher temperatures than the relatively soft \rosi.

So, assuming increased multitemperature structure with increasing
radius in cluster outskirts, e.g., due to infalling cold matter,
\rosi\ would see a steeper temperature gradient than \suz\ and,
therefore, would give rise to a lower total mass estimate
(Fig.~\ref{out:mass:M-Tr}) if the different temperature components
cannot be spectrally disentangled.

This seems dramatic; however, we have picked an extreme case
($\kb T_{\mathrm{e,hot}}=8$ keV, $\kb T_{\mathrm{e,cold}}=0.5$ keV)
for illustration purposes. As long as at least one temperature
component is cooler than about 1 keV, the reduced $\chi^2$ is actually
quite bad ($>$1.5) in most cases (for $10^4$ source photons in the
absence of any background); i.e., it is actually clear from the
observed spectrum that a single temperature model is a bad fit. Only
if both temperatures are above about 2 keV provides the single
temperature model an acceptable fit.\footnote{%
Naively, one might draw an immediate conclusion from these very simple
simulations: assuming the reduced $\chi^2$ values of the published
\suz\ temperature profiles in cluster outskirts are close to 1, then
the observed excess surface brightness either would not be due to clumping
or the clumps would not be cool ($<$2 keV, see also Sections~\ref{sec:sb} and
\ref{out:physics:clumping}).
However, this conclusion would be flawed because repeating the
above simulations \emph{with} significant
background, as is the case in cluster outskirts, the reduced $\chi^2$
values actually stay acceptable in most cases (even for $\kb
T_{\mathrm{e,cold}}<2$ keV). This illustrates
that interpretations of observational findings in cluster outskirts
always need to account for the background.}
In the presence of significant background, however, acceptable fits can be obtained also for cooler components.

\begin{figure}[htb]
\centering
\includegraphics[angle=0,width=0.55\textwidth]{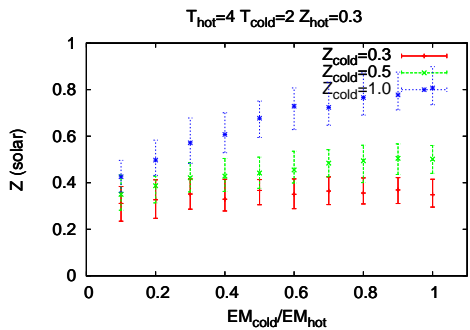}
\caption{Metallicity bias when assuming a single-temperature and
single-metallicity emission while the true emission actually has two
components. Simulated \suz\ data have been used.}
\label{out:physics:multiT:Z}
\end{figure}

Fig.~\ref{out:physics:multiT:Z} illustrates that also the metallicity
determination is biased if the presence of a multitemperature plasma
is ignored. For instance, in case \emph{both} components have a
metallicity of 0.3 solar (the red data points), the resulting best-fit
single component metallicty could be higher for the particular
situation simulated here. For a more detailed
discussion see, e.g., \citet{buo00,geb10}. 

Future instruments (e.g., the upcoming \emph{Astro-H} and an envisaged
\athena-like mission)
will have enough collecting area and spectral resolution to better
distinguish multitemperature structures spectrally. This will be
crucial for a better understanding of the thermodynamics and chemistry
especially in cluster outskirts.

\subsection{Helium sedimentation}
\label{out:physics:helium}
While potentially important for cluster (gas) mass estimates, this
topic is likely more relevant for inner cluster regions; 
therefore, it is only briefly summarized here. More details can be
found, e.g., in \citet{ef06,mar07,pn09,bhb11}, and references therein.

The mean particle weight, $\mu$, enters the hydrostatic
equation~(\ref{out:eq:mass:X:hydro2}); it is inversely proportional
to the 
total mass, $\mt$.\footnote{Recall that for the functional form of
eq.~(\ref{out:eq:mass:X:hydro2}) the radial gradient of $\mu$ has
been assumed to be small.} 
Since the overdensity radius also depends on $\mt$
(eq.~\ref{out:eq:out:o_radius}), the effect on the mass of changing
$\mu$ is amplified, resulting in a ($\sim$1.5 times) stronger
dependence for masses determined within an overdensity radius, e.g.\
$M_{2500}$.
On the other hand, to calculate the ICM mass, $\mg$, from the observed
electron density distribution
(eq.~\ref{out:eq:ff-emiss}), the ratio
$\frac{\rog}{\nel\mpr}$ needs to be known. 
So, the composition of the ICM matters for both the $\mt$ and the $\mg$
determination.

While, e.g., an iron atom is more than ten times heavier than a helium
atom, helium is about 10,000 times more abundant in typical ICM
conditions than iron.
Moreover, diffusion operates faster for lighter elements
\citep[e.g.,][]{gs84,cl04}. 
Therefore, the strongest influence on $\mu$ and
$\frac{\rog}{\nel\mpr}$ should come from variations in the
helium-to-hydrogen ratio, $\frac{n_{\rm He}}{\nh}$.

If a significant fraction of helium ions settled towards
the cluster center over time, then $\frac{n_{\rm He}}{\nh}$ would
increase there. Hydrodynamic simulations indicate, an increase of the
helium-to-hydrogen ratio of a factor of two
might be realistic in some cases (e.g., \citealt{pn09}, with some
dependence on the strength and structure of the cluster magnetic field
and the temperature). Assuming 
$\frac{n_{\rm He}}{\nh}=0.083$ and typical abundances of heavier
elements, $\mu$ is approximately in the range 0.59--0.60, and
$\frac{\rog}{\nel\mpr}$ in the range 1.14--1.15. Doubling the 
helium-to-hydrogen ratio, one obtains about 0.67 and 1.25 for the same
quantities, respectively. Neglecting helium sedimentation; i.e.,
assuming the wrong helium-to-hydrogen ratio (0.083), would in this
case bias high $\mt$ by $\sim$13\% and $\mg$ low by $\sim$8\%, within
a fixed radius. If these were the only effects, then the gas mass
fraction, $\fg\equiv\mg/\mt$, would be biased low by $\sim$19\%
compared to the true fraction.

However, there is another effect because, actually, only the X-ray
flux is directly observed, not the electron number
density. Since the emissivity (eq.~\ref{out:eq:ff-emiss}) depends
on the relative number of ions present, the flux to $\nel$ conversion
is affected by the assumed helium-to-hydrogen ratio. If the true
helium-to-hydrogen ratio is higher by a factor of two than the assumed
one then the electron density will be overestimated by about 5\%
when fitting a typical cluster spectrum (taking into account that the
true electron-to-hydrogen ratio increases from about 1.17 to 1.33 when
increasing the helium abundance).\footnote{Note
that, especially for low temperature systems, the abundance of heavy
elements would be underestimated during a spectral fit due to the
overestimated electron density, possibly giving rise to artificial
observed abundance drops (or less steep increases) towards cluster
centers due to helium sedimentation.
Furthermore, independent of any helium sedimentation, when using the
XSPEC apec or mekal model normalization parameter to determine the
electron density then the result depends on the chosen built-in
abundance table; e.g., for a typical cluster spectrum, the flux in the
(0.1--2.4) keV band decreases by $\sim$7\% for a given normalization
if the \citet{lo03} abundance table is used instead of the 
default \citet{ag89}. This is because the Lodders table assumes
a helium-to-hydrogen ratio of 0.0792 while for Anders \& Grevesse it
is 0.0977. WMAP3 data indicate a primordial value around
$1/(4/0.24819-4)\approx 0.083$ \citep[][]{sbd06} while the final
nine-year WMAP data suggest a higher value around 0.11, although 
consistent with WMAP3 at the 95\% confidence level \citep{hlk13}.}

Therefore, considering all three effects (underestimated $\mu$,
underestimated $\frac{\rog}{\nel\mpr}$, and overestimated $\nel$),
$\fg$ might be very roughly underestimated by $\sim$14\% in some
cluster centers, given these very simplified assumptions. 
Since the true helium-to-hydrogen ratio
in cluster centers is unknown, $\fg$ might be best
determined within large radii, approaching cluster outskirts, where
the helium-to-hydrogen ratio should not be strongly affected by
sedimentation.
Combining X-ray, SZ, and gravitational lensing observations could, in
principle, enable a measurement of the helium sedimentation factor.
In particular, the SZ effect depends only on the electron density and
not on the product of electron and ion density as the X-ray emission.

\subsection{Gas clumping}
\label{out:physics:clumping}
Gas clumping; i.e., inhomogeneities in the gas distribution (and
possibly the temperature distribution), must be present in the
intracluster medium at some level. However, at which 
level, at what size scales, and how it varies with increasing distance from the cluster center is not very clear, yet. Observational and theoretical aspects
of gas clumping have already been discussed in several previous Sections,
including particularly Section~\ref{sec:sb}. This will not be repeated
here. In short, if present, the effect of increasing clumping with
radius would be to overestimate the gas mass, because the gas density
in a given region would be overestimated, and to underestimate the
total mass, because the gas density profile would appear to drop less
steeply (the total mass is proportional to the gas density 
gradient, eq.~\ref{out:eq:mass:X:hydro2}), and because the temperature
profile may appear to drop more steeply if the clumps are cool
(Fig.~\ref{out:mass:M-Tr} shows that a larger gradient in temperature
typically results in a lower total mass estimate). That is, each of the three effects would tend to result in an overestimated gas mass fraction.

If significant clumps are present at large-enough scales, they might
be directly observed in deep X-ray surface brightness images
\citep[e.g.,][]{ves12}.
Alternatively, if clumps are cool, they may be revealed through X-ray
spectroscopy. Furthermore, comparison of X-ray and SZ-derived ICM
profiles may put upper limits on the clumping factor (e.g., references
in Section~\ref{out:physics:structure}).
Also, one may get constraints from considering
whether the already existing slight tension between predicted and
observed SZ angular power spectrum discussed in
Section~\ref{out:physics:hydro} might become more severe if
significant gas clumping were present in cluster outskirts and
resulted in an even more increased SZ signal. 

\section{Chemistry}
\label{out:chem}

\subsection{Relevance of chemical evolution}
Among the several outstanding problems of modern astrophysics, the
chemical enrichment of the Universe is certainly a very topical one. 
In the standard scenario, shortly after the Big Bang, the
element composition of the cosmic gas was very simple -- it was
composed almost entirely of hydrogen and helium, the two lightest
chemical elements. Heavier elements than helium are thought to have
their origin in the interiors and explosions of stars, which formed much later in the
history of the Universe.  It is believed that the first stars
(``Population III'') 
formed few hundred million years after the Big Bang (e.g.,
\citealt{1986MNRAS.221...53C}; \citealt{2002Sci...295...93A}) from
collapse of primordial, metal-free gas
and they have subsequently polluted the surrounding medium by ejecting
the metals produced in their cores (e.g.,
\citealt{1998A&A...335..403G}; \citealt{1999ApJ...527L...5B};
\citealt{2008Sci...321..669Y}; \citealt{2008ApJ...674..644C}).
The minimum amount of gas that may collapse under its own gravity is
approximately given by the Jeans mass, which increases with
temperature as $T^{3/2}$ \citep[e.g.,][]{bl03}.
In a 
primordial environment, the only relevant coolants are H, He and their
derived molecules (e.g., \citealt{1967Natur.216..976S}), which will not cool
below a relatively high temperature of (100--200) K.
Thus, the Population III stars are supposed to be more
massive and short-lived (e.g., \citealt{1995ApJS..101..181W};
\citealt{2002A&A...382...28S}) than the typical stars that form from
metal-rich gas. Once the Population III 
stars had enriched the gas, cooling due to metal line emission became
efficient and gas clouds could cool to lower temperatures and
fragment into smaller pieces. The presence 
of metals probably drove the transition from a top-heavy initial mass
function (IMF) to a ''Salpeter-like'' IMF allowing low-mass stars to
form.

\subsection{Metallicity in cluster outskirts}
The ICM does not only contain primordial elements, but also a
considerable amount of heavy elements. As metals are only
produced in stars, which reside mainly in galaxies, the enriched
material must have been transported from the galaxies into the ICM.
Several processes have been proposed to explain the observed enriched
material: ram-pressure stripping, galactic winds, galaxy-galaxy
interactions, AGN outflows, and intra\-cluster stellar populations (see
\citealt{sd08} for a review).  Each of them has a
different efficiency and pattern distribution (e.g.,
\citealt{2005A&A...435L..25S}; \citealt{2006A&A...452..795D},
\citealt{2006A&A...447..827K}; \citealt{2007A&A...463..513M}).
Simulations show an extended distribution for the metals due
to galactic winds and a more centrally concentrated distribution in case
of ram-pressure stripping. Furthermore, winds seem to be more
efficient in early epochs when the star formation rate was higher,
while ram-pressure starts to play an important role below 
redshift 1--2 (e.g., \citealt{2007A&A...466..813K}). In the past, the
general idea was that in cluster outskirts winds are more
important than ram-pressure stripping because, on the one hand, there is
lower ICM pressure around the galaxies to confine the metals, and, on
the other hand, the ram-pressure stripping increases when the ICM 
density is higher. There is clear evidence nowadays that 
ram-pressure stripping is acting also in the outskirts (e.g.,
\citealt{2005AJ....130...65C}; \citealt{2006ApJ...651..811B};
\citealt{2007MNRAS.376..157C}; \citealt{2007ApJ...671..190S};
\citealt{2009A&A...502..427V}).

The only way to determine the chemical composition of the
intracluster medium is through X-ray observations.
Due to its large effective area, good spectral and spatial resolution,
\xmm\ is currently the best instrument to study the
spatial distribution of metals in the ICM.
Due to its high background
level, it has only been used to determine the abundances in the inner and intermediate
parts of clusters, though, where the emission is brightest
\citep[e.g.,][]{lm08b}.
In 
cluster outskirts, the temperature determination, and so the metal
measurements, are easily affected by the level of subtracted
background \citep[e.g.,][]{dwb06,lm08}.
Thus, a large fraction of the clusters' volume is still unexplored.
 
Recently,
X-ray observations with \suz, thanks to its much lower instrumental
background than that of other X-ray observatories, have pushed
measurements of the ICM metallicity up to the virial
radius for several clusters (e.g., \citealt{fth07};
\citealt{rhz08}; \citealt{bms09};
\citealt{hhs10}; \citealt{sk10},
\citealt{kou10}; \citealt{sam11}, \citealt{ahi11}, \citealt{hbb12},
\citealt{mss13}). 
For instance, \citet{fth07} and \citet{sam11} showed that the outer
regions of clusters are considerably metal-enriched, to a level of
about 0.2--0.3 of the solar metallicity, which is the typical value
observed in intermediate cluster regions. On the other hand,
\citet{uws11} measured with high statistical significance
(5.5$\sigma$) a lower metallicity of 0.11$\pm$0.02 solar at the virial
radius of the Virgo cluster using \xmm\ data.

The physics of the ICM outskirts must be understood to avoid biases in
the chemistry determination (e.g., Figs.~\ref{out:physics:nei_2a},
right, and \ref{out:physics:multiT:Z}). For instance,
gas clumping can lead to an overestimate of the gas density
(Section~\ref{out:physics:clumping}) and an underestimate of the
temperature (Section~\ref{out:physics:multiT}). 
\citet{uws11} argue that unresolved multitemperature structure in the
low temperature outskirts of the Virgo cluster may lead to an
underestimate of the metallicity. Thus, the metal mass inferred from
the metallicity measurement may be a lower limit.
Another source of bias in the metal mass determination can be the
inhomogeneous metallicity distribution. Due to the fact that the
enriched material is not mixed immediately with the ICM, metal blobs and
stripes can be observed. For instance, \citet{lsk11} showed with \xmm\ observations of five cool-core clusters that even in
relaxed clusters the distribution of metals shows significant
inhomogeneities. Using metal maps they evaluate that most previous
metal mass determinations have underestimated the metal mass by
(10--30)$\%$ due to this inhomogeneous distribution.

The spatial distribution of metals is
linked to galactic evolution (e.g., supernova-driven galactic winds, outflows due to AGN, and galaxy-galaxy interactions), galaxy-ICM interactions (e.g., ram-pressure stripping), and gas
dynamics (e.g., gas drifting, turbulence, convection, and mixing).
Important advances in understanding these processes have been made over the past few years.
For instance, \citet{2005MNRAS.359.1041R,2006MNRAS.372.1840R} found that the
optical light profiles are much more peaked than the abundance
profiles, while in absence of mixing they should follow a similar
distribution. This suggests that metals originating  in the central
regions are transported out to large radii by gas motions.
Including AGN
feedback in the simulations leads to iron profiles in better agreement
with observations
 because
they are able to move over a large scale the gas enriched within the
galactic halos \citep[e.g.,][]{2011MNRAS.416..801F}.
The role of AGN in the metal 
transport has been investigated also by \cite{2001ApJ...554..261C}.
They proposed a scenario where radio bubbles, produced by
the jets of the AGN, uplift the cold enriched gas out to large 
distances.
This scenario has been
confirmed observationally by \citet{2008A&A...482...97S} analyzing
\xmm\ data of the cluster Hydra-A.

The ICM is rarely in perfect hydrostatic equilibrium
because large-scale accretion of matter, turbulent motions, gas sloshing, AGN, and galaxy motions are considerably and regularly perturbing the gas
(e.g.,
\citealt{2004A&A...426..387S}, \citealt{2004ApJ...615..181H};
\citealt{2005A&A...434...67V}; \citealt{2006A&A...453..447E};
\citealt{2010MNRAS.402L..11S}; \citealt{2012MNRAS.420.3632R}).
Since both large- and small-scale mixing properties of galaxy clusters
are at work at the same time during the whole cluster evolution, it is
currently difficult to obtain a complete and consistent picture of the evolving
ICM. 
To shed light on the process of metal injection we need to resolve
the metal distribution down to the small scale of mixing, and this
requires the large effective area of an envisaged \athena-like mission.

\section{Technical issues}
\label{out:tech}

\subsection{X-rays}
\label{out:tech:X}

\subsubsection{How to deal with foreground and background
emission in low surface brightness regions?}
\label{out:tech:X:bkg}
The cluster surface brightness decreases fast with increasing radius,
making measurements of gas properties more difficult in cluster
outskirts.
A gas density measurement can be accomplished with much poorer
signal-to-noise data than a temperature measurement. For instance, one
can pick an energy band where the surface brightness, $\sx$, is almost
independent of temperature for hot clusters, e.g., (0.5--2.0) keV
(Fig.~\ref{out:mass:apec}). Then determine the surface brightness, or
rather the number of source photons in this energy band in some 
annulus divided by exposure time and effective area. And finally,
obtain the density by inverting 
\begin{equation}
 \sx\propto\frac{1}{(1+z)^4}\int_{-\infty}^{\infty}\nel^2\,\dif l\,.
\label{out:eq:tech:X:SB}
\end{equation}
Also note the strong redshift dependence due to cosmological surface
brightness dimming.

On the other hand, 
Figures~\ref{out:mass:apec} and \ref{out:physics:multiT:line_complex}
show that measuring the temperature of hot 
($\kb\tel\gtrsim 5$ keV) clusters requires high quality spectral data at
high energies ($E\gtrsim 5$ keV) because of the absence of the Fe L
shell line complex around 1 keV and because the exponential
bremsstrahlung cutoff shifts to high energies. High quality data means
good source photon statistics as well as low and well-determined
fore- and background surface brightness. We will see that, at high
energies, the so-called particle background dominates if times of
so-called soft proton flares have been accurately excluded.

X-ray optics work only for small photon incidence angles (grazing
incidence). The higher the photon energy, the smaller the critical
angle for reflection. For a given focal length, this results in
decreasing effective area with increasing energy as high energy
photons cannot be focused by outer mirror shells anymore. In summary,
for a hypothetical intrinsically flat spectrum X-ray source, the
detected signal decreases rapidly for higher photon energies. This is
also true for astrophysical foreground/background components but not
for other types of background, as dicussed below.

The particle background (events not caused by astrophysical
X-ray sources but indistinguishable from them) spectrum registered on 
the detector is typically fairly flat, so will always become dominant
above some energy. This threshold energy depends strongly on the
instrument setup, including satellite orbit, focal length, presence of
anti-coincidence devices etc. The ambient particle background level is
lower in a low-Earth orbit (LEO, e.g., \ro, \as, \suz, \swi,
{\it Astro-H}) compared to a
highly-elliptic orbit (HEO, e.g., \cha, \xmm) or Lagrange point 2 (L2, e.g.,
\rosi) orbit because satellites are naturally shielded by Earth's magnetosphere.
Particle background also roughly scales with focal length squared and the
number of mirror/detector systems (e.g., \cha\ has a focal length of
10m and one mirror system while \rosi\ has seven mirror systems and a
focal length of only 1.6m). Anti-coincidence devices (e.g., \ro\ \ps)
can currently not be applied very efficiently to X-ray CCDs.
Fluorescent lines due to detector material (e.g., Al, Co) induced by
particle background can be suppressed by applying graded-Z shields
(e.g., \rosi). 
In general, if the particle background is underestimated during
analysis then the resulting temperature profile will often appear to
decline less steeply with radius in cluster outskirts \citep[e.g.,
Fig.~1 in][]{dwb06} because background becomes increasingly important
at larger radii and particle background dominates at high energies,
resulting in an apparent shift of the ICM bremsstrahlungs cutoff
towards higher energies; this would then give rise to an overestimated
total mass (Fig.~\ref{out:mass:M-Tr}).

Satellites in orbit mostly outside Earth's magnetosphere (e.g., \cha\
and \xmm), are occasionally (about 30\% of the time) bombarded with
so-called soft protons. They can hit the focal plane instruments when
entering through the mirrors, increasing the background by a factor of
up to $\sim$100 or more. For analysis of extended sources, times during
such flares need to be removed as far as possible from the data. This
works reasonably well using lightcurves since X-ray CCDs have good
time resolution (analysis is being done on a data-hypercube -- each
individual detected event gets a position, an energy, and an arrival
time assigned). However, low level flares can go undetected in lightcurves and cause
a background signal that needs to be taken care of in subsequent spectral
analyses. Satellites in LEO, like \suz, do not suffer
from such flares.

\begin{figure}[htb]
\vspace{-0.5cm}
\includegraphics[angle=0,width=1.0\textwidth]{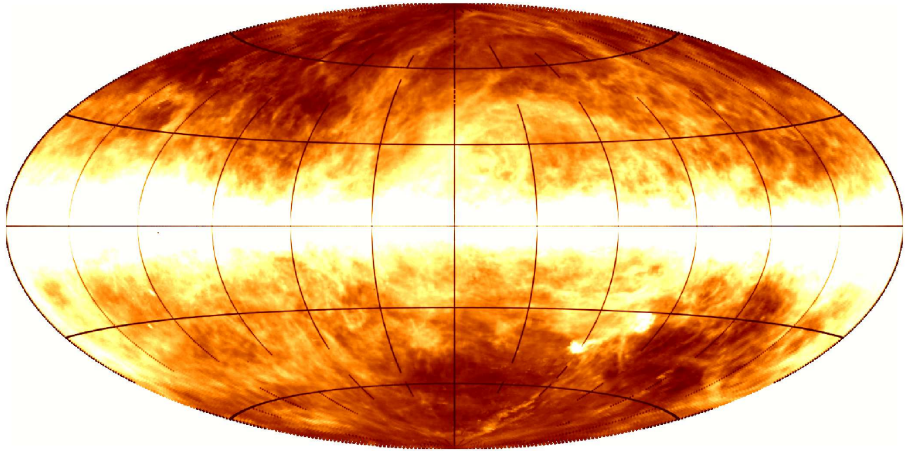}
\caption{The Milky Way seen at a wavelength of 21 cm. Provided by P.
Kalberla and reproduced with permission (Leiden/Argentine/Bonn Survey, \citealt{kbh05}).}
\label{out:tech:LAB}
\end{figure}
\begin{figure}[htb]
\vspace{-0.5cm}
\includegraphics[angle=270,width=1.0\textwidth]{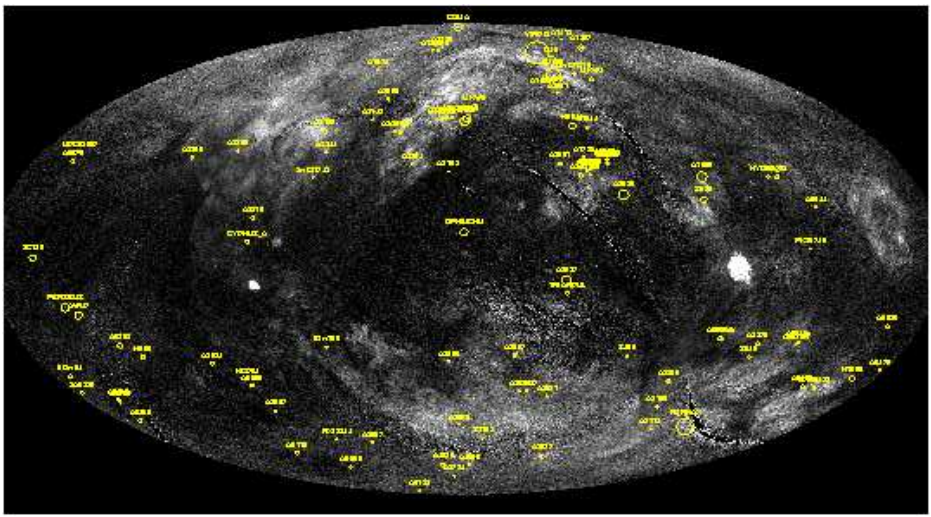}
\caption{The Milky Way as seen in soft X-rays ($\sim$1/4 keV). Notice the
anti-correlation with the Galactic neutral hydrogen map above. Data from the
\ro\ All-Sky Survey (see, e.g., \citealt{smk08}). The yellow circles
indicate the same X-ray bright galaxy clusters as in
Fig.~\ref{out:physics:structure:2MASS}.}
\label{out:tech:RASS}
\end{figure}
The dominant \emph{soft} ($E\lesssim 1$ keV) X-ray background is due
to thermal emission of optically thin gas (bremsstrahlung and
line emission) and charge exchange emission (line emission). The
former varies with pointing direction and is particularly strong at
low Galactic latitudes and the latter varies with time.
The primary components are assumed to be \citep[e.g.,][]{smk08}:
Possibly, emission from the local hot bubble -- hot gas left
       over from one or more long gone supernovae, which exploded in
       the vicinity of the Sun ($\kb\tx\approx 0.1$ keV). 
Galactic emission from the disk and Galactic halo
       ($\kb\tx\approx$ (0.1--0.2) keV).
Possibly, local group emission ($\kb\tx\approx 0.2$ keV).
Emission from large-scale Galactic objects like supernova remnants and
from a large number of unresolved stars at low Galactic latitudes.
And, last not least, time variable solar wind charge exchange emission from the
       geocorona and/or heliosphere, which may be responsible for a
       significant fraction of the emission previously attributed to
       the local hot bubble \citep[e.g.,][]{kse11}.

Furthermore, there is a cooler interstellar medium (ISM) in our
Galaxy, as traced, e.g., by neutral hydrogen. Since the ISM has been
enriched by metals, when there is hydrogen then there
are also heavy elements, causing X-ray absorption. So, hydrogen maps
can be used to estimate Galactic X-ray absorption for distant sources.
The anti-correlation seen in Figs.~\ref{out:tech:LAB} and
\ref{out:tech:RASS} indicates that a significant fraction of the soft
X-rays does \emph{not} originate in the local hot bubble or from solar
wind charge exchange but comes from farther away
\citep[e.g.,][]{pkk98}.

Last not least, at intermediate energies, above $E\gtrsim 1$ keV and
below the particle background threshold energy, the sum of emission
from distant undetected AGN dominates. This component can be reduced
if the instrument has good spatial resolution (e.g., \cha), which
allows to detect point sources down to very low fluxes.

All of these fore- and background components have to be subtracted or
modelled with increasing accuracy the further out one goes into the
cluster outskirts. E.g., Fig.~2 (right) of \citet{rhz08} shows that
in an annulus approaching $r_{200}$ of the cluster A2204, the cluster
emission is subdominant at almost every energy and different
fore- and background components dominate different parts of the
spectrum. Methods need to be optimized for each instrument
and observational situation to obtain robust results, particularly on
temperature measurements. The characterization of the X-ray emission
from outskirts of clusters at low Galactic latitudes is especially
challenging. A useful approach is also to combine \cha\ and \xmm\
observations for point source detection with \suz\ observations for
the faint cluster emission. However, since AGN are variable sources,
such observations would ideally be performed simultaneously.
Moreover, a future X-ray mission could be optimized for cluster
outskirts studies by combining good spatial and spectral resolution
over a large field-of-view with large effective area and short focal
length in a low-Earth orbit.

\subsubsection{How to deal with the \suz\ point-spread-function
and stray light?}
\label{out:tech:X:PSF}
Currently, \suz\ is the best X-ray instrument to determine the
temperature and metal abundance of the ICM in cluster outskirts. In
addition to the general fore- and background issues discussed
above, the particular challenges for the \suz\ analysis include PSF
and stray light correction. While \suz's PSF does not strongly depend
on photon energy -- a significant advantage over \as\ -- it is broad,
about 2 arcmin half-power-diameter, resulting in contamination of the
observed spectrum in a given annulus from neighboring shells, e.g.,
from a bright cluster core. This can be corrected for by ray tracing
simulations \citep[e.g., using xissim,][]{imf07} given a finely binned cluster
surface brightness and temperature profile. I.e., if one knows exactly
how the cluster looks like, one can accurately correct for PSF
effects. So, one has to stick in what one actually wants to measure.
Therefore, an iterative approach needs to be taken and, again, additional
\cha\ or \xmm\ data to model the inner regions are very helpful.
Furthermore, the PSF limits the determination of reasonably resolved
temperature profiles to low redshift clusters ($z\lesssim 0.3$).

Also stray light can be an issue; i.e., single reflections of photons
from sources outside the field-of-view that reach the detector. This
is particularly true for very low redshift clusters that require multiple
\suz\ pointings to cover a significant radial range in terms of
$r_{200}$, but any observation can be contaminated if there is a
bright source nearby. This can, in principle, be corrected for with
ray tracing simulations but this approach is limited by the accuracy of
the underlying calibration. Moreover, the observing strategy can be
adjusted to minimize stray light effects. Off-axis observations of
the Crab nebula have shown this to be the case if the CCD \emph{corners} point
towards the bright source \citep{tah12}.

\subsection{SZ}
\label{out:tech:SZ}

\subsubsection{Direct measurement of the cluster pressure profile}
As seen in eq.~(\ref{eq:sz}) the SZ signal is proportional to the product of electron
density and temperature integrated along the line-of-sight; therefore,
deprojection of a resolved SZ profile will directly yield the
three-dimensional pressure profile for a cluster. In practice, such
direct inversion of an SZ image is difficult to perform due to the
poor resolution of the current SZ experiments, and also due to the
difficulty in obtaining filtering-corrected SZ profiles out to large
radius without assuming some kind of model. A majority of studies to
date have, therefore, made use of  simultaneous fitting of SZ and X-ray
data using some parametric model for the gas density and temperature
distribution  (e.g., \citealt{mahdavi07},
\citealt{mbc09},
\citealt{bhb12}). The generalized NFW model is currently
considered to be the best description of the shape of the
intracluster gas pressure, and is constrained with high precision
from X-ray data inside $r_{500}$
\citep{app10}.
Some early attempts
at non-parametric comparison (e.g.,
\citealt{jbp08},
\citealt{bnp09})
also found generally good agreement between the X-ray prediction of
pressure and the SZ measurements. It must be mentioned that data from
single clusters have too little constraining power to find systematic
differences between X-ray and SZ measurements at large radius, and a
more meaningful comparison can be obtained after stacking the SZ
signal from several clusters. Early stacking results do not show any
statistically significant deviation (Section~\ref{out:physics:P} and
\citealt{pba10}, \citealt{paa13}), although such
comparisons are still in their early stages.

\begin{figure*}[t]
\centering
  \includegraphics[width=0.7\columnwidth]{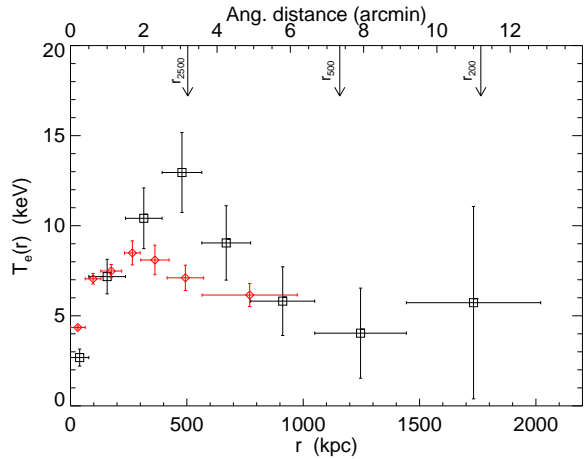}
  \caption{
  Temperature profile for the cluster Abell 2204 obtained
  through a joint X-ray/SZ analysis (black squares). Reprinted with permission from
  \citet{bnp09}. X-ray 
  information includes only the surface brightness, no spectra. The
  red diamonds result from the spectral \xmm\ analysis of the same
  cluster by \citet{zhang08}.}
  \label{fig:sz}
\end{figure*}

\subsubsection{Density and temperature profiles from joint X-ray/SZ
modelling}

A more direct analysis of the state of the intracluster medium at
large radius is possible if X-ray and SZ observations are used
together to model the density and temperature profiles
simultaneously. Since these two observables have a different dependence
on $\nel$ and $\tel$, it is in principle straightforward to solve for
them using X-ray and SZ data. Such joint modelling was originally
proposed by \citet{sw78} and has been applied to simulated data and
analytical models \citep[e.g.,][]{ys99,pb06,abp09}
but application to real cluster data
have been very few (\citealt{kitayama04}; \citealt{nord09};
\citealt{bnp09}). The major difficulty lies in obtaining high-quality
SZ data with sufficient resolution, and to compensate SZ
flux loss due to filtering.  Once an SZ profile is obtained, one can simply
use the X-ray surface brightness from the soft energy band (roughly
0.5--2 keV) to model first the gas density, where the X-ray cooling
function is practically insensitive to gas temperature in a hot
cluster (Fig.~\ref{out:mass:apec}), and then use it to obtain the gas temperatures from the SZ
data. The potential advantage lies in the fact that the X-ray surface
brightness profile can be obtained reliably with a smaller number of
photons than required for an accurate spectral measurement of the gas
temperature. 
This technique, therefore, opens a new window to the cluster outskirts
by measuring the temperature (and mass) of the ICM independently of
X-ray spectroscopy (e.g., Fig.~\ref{fig:sz}).

One difficulty with the above method is that any biases incurred in
the X-ray brightness profile measurement will affect the SZ-derived
temperature measurement with opposite bias. The SZ data alone cannot
constrain the gas temperature, as it measures only the integrated
pressure at different scales. The density normalization must come from
X-ray brightness profile measurements. Therefore, if the X-ray
emission is boosted, e.g., due to the presence of gas clumping, the
temperature profile inferred from X-ray/SZ joint analysis will show a
steeper decline. This will complicate the comparison between
SZ-derived temperature and X-ray spectroscopic temperature,
particularly in the cluster outskirts where clumping effects may be
significant (Section~\ref{sec:sb}).

\subsubsection{How to measure profiles with poor spatial resolution?}
As already mentioned, current comparisons between the X-ray and SZ
measurements are mainly limited by the low angular resolutions of  the SZ 
experiments. SZ survey telescopes generally
operate with a beam size of $\sim$1 arcmin, and the \pla\ 
surveyor has an effective beam size of 7 arcmin for SZ science.
Massive clusters at very low redshifts ($z\lesssim 0.1$) are not suitable for
imaging with ground based SZ instruments due to their large angular
dimensions, as the limited field-of-view of bolometer cameras make the removal  
of atmospheric noise difficult. Interferometers act like a high-pass filter of 
the sky signal and are generally unsuitable for imaging large objects. 
In addition, single-frequency SZ experiments are affected by the confusion 
from primordial CMB fluctuations in their reconstruction of the 
SZ signal at large radii. 

On the other hand, most X-ray
spectroscopic studies of cluster outskirts have been done with very
low redshift targets (Section~\ref{out:physics:Tr}), either to have
sufficient signal-to-noise over X-ray fore- and backgrounds
(Section~\ref{out:tech:X:bkg}), or to avoid the complicated PSF
smearing effect of the {\it Suzaku} satellite
(Section~\ref{out:tech:X:PSF}). Some of the low redshift
cluster profiles have now been measured by the \pla\ 
satellite with sufficient accuracy (Section~\ref{out:physics:P}) and
can be compared or combined with low-redshift X-ray data
(Section~\ref{out:physics:entropy}).
But the low resolution \pla\  results need to be tested against 
finer resolution ground-based instruments for possible systematic biases in their 
beam deconvolution or multi-frequency matched filtering techniques.  
Future high resolution SZ instruments like CARMA and the Cerro
Chajnantor Atacama Telescope (CCAT)
will provide sufficient angular resolution for intermediate
redshift clusters, and CCAT will also have a large field-of-view to image 
some of these $z\lesssim 0.1$ objects.

The other aspect of reliable SZ imaging at large angular scales is the
development of effective deconvolution methods, to account for the
loss of signal due to the multiple high-pass filters employed on the data 
to remove atmospheric noise and the CMB contributions. While fitting a model to 
the data such filtering renders the model sensitive to only a very limited spatial 
range of the cluster SZ signal, typically near its central region. 
Thus computing the integrated SZ signal or measuring the pressure gradient 
in the outskirts is done from model extrapolation into 
the domain where the actual measurements have no impact. 
We remind that the situation used to be somewhat similar in
X-ray-based analyses of cluster pressure profiles
\citep[e.g.,][]{app10} where the pressure 
slope beyond $r_{500}$ was fixed by simulations. With \suz, this is
now changing.
Model independent (non-parametric) reconstruction of  the full SZ signal has
so far been attempted by two different methods: by making a direct
deconvolution of the filtered map \citep{nord09,sayers11}
and by an analytic inversion of the radial profile assuming spherical
symmetry \citep{pba10}. These studies are currently limited by  the  
SZ instrument sensitivities and low number statistics, although rapid
progress is being made and future SZ/X-ray comparisons of the cluster
outskirts hold much potential.

\subsection{Weak lensing}
\subsubsection{How to deal with projection effects?}
\label{out:tech:WL}
Wide-field imaging cameras allow us, in principle, to trace the weak
lensing signal of a
typical cluster out to several virial radii. However, the weak lensing signal of any
individual cluster blends into the cosmic shear background arising from the
combined shear of the intervening large-scale structure at much smaller
clustercentric separations than technically accessible.
From the early simulation-based work by \citet{2003MNRAS.339.1155H}, a 
rule-of-thumb could be drawn that the effect of projected large-scale structure becomes a 
considerable source of error for separations $\gtrsim\!10$ arcmin from the
cluster center, even for massive clusters at redshifts yielding a high lensing
efficiency.

Recently, due to the improved capabilities for ray tracing through cluster
$N$-body simulations, this issue has found more widespread attention
\citep[e.g.,][]{2011MNRAS.412.2095H,2011ApJ...740...25B,2011MNRAS.416.1392G}.
Based on ray tracing results from the Millennium simulation 
\citep{swj05,2009A&A...499...31H}, \citet{2011MNRAS.412.2095H}
confirmed their earlier results \citep{2001A&A...370..743H,2003MNRAS.339.1155H} 
that accounting for large-scale structure lensing is crucial when studying the mass profile in 
cluster outskirts.
Starting with a simulation by \citet{tkk08}, 
\citet{2011ApJ...740...25B} investigated the biases and uncertainties arising 
from large-scale structure uncorrelated to the cluster (and also from cluster triaxiality) 
and find large-scale structure to add significant scatter to weak lensing mass estimates obtained from 
profile fitting. 
They arrive at an additional uncertainty induced by large-scale structure consistent with the
findings of \citet{2003MNRAS.339.1155H} and \citet{2011MNRAS.412.2095H}.
\citet{2011MNRAS.416.1392G}, who draw attention to the
influence of structure \emph{correlated} to the cluster of interest,
also corroborate this picture: shear profiles are increasingly
dominated by large-scale structure scatter beyond a separation of $\sim$10 arcmin. 

Again using ray tracing simulations \citep{2009ApJ...701..945S}, 
\citet{2011MNRAS.414.1851O} considered the outer cluster profile
itself, which they find best described by a modified NFW profile with
a smoothed 
cut-off \citep{2009JCAP...01..015B}, motivated by the halo model.
Furthermore, if a \citet{2009JCAP...01..015B} profile were fitted with
an NFW function to an outer boundary of ($\sim$10--30) arcmin, the
consequence  would be a (5--10)\% underestimate of the cluster mass, 
while its concentration would be overestimated. 

How can one incorporate these simulation results in practice? For
cluster masses obtained by fitting a parametric function (e.g., NFW)
to the shear profile, shear by large-scale structure can be taken into
account as an additional component in the error estimation. 
\citet{ier10,ier12} provide a practical
example for this method, using an estimation of large-scale structure contribution
from the simulations of \citet{2003MNRAS.339.1155H}.

Weak lensing tomography aims at direct disentanglement of lensing
structures at different lens redshifts \citep[e.g.,][for
applications]{2007ApJS..172..239M,2010A&A...516A..63S}.
So, in principle, contributions from uncorrelated projected large
scale-scale structure could be measured and individual cluster mass
profiles corrected for them.
Accurate photometric redshifts requiring deep imaging in several bands is
necessary to achieve resolution in the redshift dimension (although a poorer one
than in the two ``celestial'' dimensions).

Due to the inevitable lensing by projected, unrelated structures, an
accurate weak lensing mass measurement is currently limited to
massive ($M\!\gtrsim\!10^{14}\,\mathrm{M}_{\odot}$) individual
clusters, though. 
\emph{Stacking techniques}, however, make a wider range of the cluster mass
function accessible to weak lensing. More importantly, cluster
stacking averages out the effects of substructure and departure from
spherical symmetry, thus defining a ``typical'' (outer) mass profile
\citep{2006MNRAS.372..758M,2009ApJ...703.2217S,2010PASJ...62..811O}.
Now that an axisymmetric mass distribution can be assumed, the quantity
\begin{equation}
\Delta\Sigma(\theta) = \Sigma_{\mathrm{crit}}\gamma_{\mathrm{t}}(\theta)
= \bar{\Sigma}(<\!\theta) - \Sigma(\theta)\,,
\end{equation}
with $\bar{\Sigma}(<\!\theta)$ the mean projected density within the radius 
$\theta$, provides an estimate of the projected surface mass density 
$\Sigma(\theta)$.
However, when stacking clusters, special care has to be taken to the 
determination of the cluster center.

Combining weak lensing shear and magnification measurements for some of the
most massive clusters in the Universe, 
\citet{2011ApJ...729..127U,2011ApJ...738...41U} compute mass profiles, and 
average them in a stacked profile. Using their Bayesian method, which
breaks the mass-sheet degeneracy by including lensing magnification
data, these authors find the logarithmic slope of the mass profile to
steepen towards the center,
over a large radial range, consistent with the NFW profile.

\section{Outlook}
\label{out:outlook}

As illustrated in this review, new instruments and improved
theoretical modelling have resulted in significant progress in
understanding the physics and chemistry of the intracluster medium as
well as cluster mass profiles beyond $r_{500}$. Also, new questions
emerged. With upcoming SZ (e.g., Atacama Large
Millimeter/submillimeter Array, ALMA; CCAT) and X-ray instruments
(e.g., {\it Astro-H}; \rosi), it will likely be possible to clearly
disentangle, which of the suggested physical and enrichment processes
dominate in cluster outskirts, which will in turn inform modelling of
cluster and galaxy evolution. Still, we are far from reaching the really
interesting regime, the region beyond the cluster virial radius
($>$$r_{100}$), diving into the expected filamentary structure. A
stacking analysis of \rosi\ data at cluster positions in mass and 
redshift bins seems promising. Still, for the gaseous component, new
optimized instruments will likely need to be constructed to study
individual systems. In the X-ray regime, breakthroughs could be
achieved with a large effective area, \athena-like, mission. For a more
accurate weak lensing determination of 
cluster mass profiles out to the filamentary regime, there is hope
that with ongoing and upcoming deep, large-area
imaging surveys (e.g., Kilo Degree Survey, KiDS;
Dark Energy Survey, DES; Euclid; Large Synoptic Survey Telescope,
LSST), detailed stacking analyses can be performed. 
This will constrain structure formation models and possibly also the
nature of dark matter. Let's move on to an exciting future.

\begin{acknowledgements}
We would like to thank both referees, in particular R. Schmidt, for
providing feedback that helped improve the presentation of this paper.
We acknowledge H. Akamatsu, D. Nagai, and C. Sarazin for useful
discussions and 
H. Akamatsu,
A. Fabian,
A. Hoshino,
P. Humphrey,
M. Kawaharada,
E. Miller,
D. Nagai,
K. Sato,
T. Sato,
A. Simionescu, and
S. Walker
for making simulated or observed temperature profiles available
electronically.
We acknowledge the XSPEC12 software \citep{da01,a96}, which has been used
to create several figures. 
THR acknowledges support
by the German Research Association (DFG) through
Heisenberg grant RE 1462/5 and grant RE 1462/6.
LL acknowledges support by the German Aerospace Agency (DLR)
with funds from the Ministry of Economy and Technology
(BMWi) through grant 50 OR 1102 and by DFG grant RE 1462/6.
SE and MR acknowledge the financial contribution from contracts 
ASI-INAF I/023/05/0 and I/088/06/0. MR acknowledges also the support
of the contracts ASI I/016/07/0 COFIS, ASI Euclid-DUNE I/064/08/0, ASI
Uni Bologna-Astronomy Dept. Euclid-NIS I/039/10/0, and PRIN MIUR Dark
energy and cosmology with large galaxy survey.
EP acknowledges the financial support of grant ANR-11-BS56-015.
\end{acknowledgements}

\hyphenation{Post-Script Sprin-ger}

\end{document}